\documentclass{aa}  
\usepackage{amsmath}
\usepackage{mathtools}
\usepackage{placeins}
\usepackage{graphicx}
\usepackage{txfonts}
\usepackage{float}
\makeatletter
\DeclareRobustCommand{\ion}[2]{%
  \text{#1\,\check@mathfonts\fontsize\sf@size\z@\selectfont #2}%
}
\makeatother

\usepackage{etoolbox}

\newcommand{\kms}{km\,s$^{-1}$}
\newcommand{\lya}{Ly$\alpha$}
\newcommand{\hi}{H~{\sc i}}
\newcommand{\nhi}{$N_{\rm HI}$}
\newcommand{\bhi}{$b_{\rm HI}$}
\newcommand{\linetools}{{\sc linetools}}

\usepackage[breaklinks, colorlinks, citecolor=blue, linkcolor=blue]{hyperref}

\begin{document} 

\title{A positive correlation between broad \hi\ \lya\ absorptions and local overdensities of galaxies} 

\titlerunning{A positive correlation between broad \hi\ \lya\ absorptions and local overdensities of galaxies}
\subtitle{}
\author{Ismael Pessa
          \inst{1}
          \and
          Nicolas Tejos\inst{2}
          \and
          Karen Martinez-Acosta\inst{2}
          \and
          Sebastian Lopez\inst{3}
          \and 
          Jessica Werk\inst{4}
          \and \\
          J. Xavier Prochaska\inst{5,6,7,8}
          }

\institute{Leibniz-Institut für Astrophysik Potsdam (AIP), An der Sternwarte 16, 14482, Potsdam, Germany. 
\email{ipessa@aip.de}
\and
Instituto de F\'isica, Pontificia Universidad Cat\'olica de Valpara\'iso, Casilla 4059, Valpara\'iso, Chile. 
\email{nicolas.tejos@pucv.cl}
\and
Departamento de Astronom\'ia, Universidad de Chile, Casilla 36-D, Santiago, Chile. 
\and
Department of Astronomy, University of Washington, Seattle, WA 98195, USA
\and
Department of Astronomy and Astrophysics, University of California, Santa Cruz, 1156 High St., Santa Cruz, CA 95064, USA
\and
Kavli IPMU (WPI), UTIAS, The University of Tokyo, Kashiwa, Chiba 277-8583, Japan
\and
Division of Science, National Astronomical Observatory of Japan, 2-21-1 Osawa, Mitaka, Tokyo 181-8588, Japan
\and
Simons Pivot Fellow
             }
             \date{Received XX; accepted XX}

  \abstract
 {A large fraction of the baryon budget at $z<1$ resides in large-scale filaments in the form of diffuse intergalactic gas, and numerous studies have reported a significant correlation between the strength of the absorptions produced by this gas in the spectra of bright background sources, and impact parameter to cosmic filaments intersected by these sightlines. However, a similar relation is harder to determine for the warm-hot phase of the intergalactic gas, since its higher Doppler parameter and significantly lower neutral gas fraction makes this gas difficult to detect in absorption. We use a sample of 13 broad Ly$\alpha$ absorbers (BLAs) detected in the HST/COS spectrum of a single QSO ($z\sim0.2685$), whose sightline intersects several inter-cluster axes, to study the relation between BLAs and the large-scale structure of the Universe. Given their Doppler parameters of $b>40$\,\kms, BLAs are good tracers of warm-hot intergalactic gas. We use VLT/MUSE and VLT/VIMOS data to infer local overdensities of galaxies at the redshifts of the BLAs, and to assess the potential association of the BLAs with nearby galaxies. We find that out of the 13 BLAs in our sample, four are associated with a strong overdensity of galaxies, and four with tentative overdensities. The remaining five are located at redshifts where we do not identify any excess of galaxies. We find that these overdensities of galaxies at the redshift of BLAs ($\pm 1000$\,\kms) are local, and they vanish when larger cosmic volumes are considered, in terms of a larger velocity offset to the BLA or larger impact parameter to the QSO sightline. Finally, we find a positive correlation between the total hydrogen column densities inferred from the BLAs, and the relative excess of galaxies at the same redshifts, consistent with the picture where warm-hot gas resides deep within the gravitational potential well of cosmic filaments.}

\maketitle

\section{Introduction}

The cosmic web, an intricate structure of Mpc-scales filaments and cluster, defines the Universe large-scale structure \citep{Bond1996,GonzalezPadilla2010, AragonCalvo2010}. Within this cosmic lattice, the Warm-Hot Intergalactic Medium (WHIM) is thought to hold a significant portion of the cosmic baryons \citep{Cen1999,Danforth2010, Shull2012, Nicastro2018}. According to different hydrodynamical cosmological simulations based on $\Lambda$CDM cosmology, a fraction between $25\%$ and $50\%$ of the total baryons at low redshift resides in the WHIM \citep{Shull2012, Martizzi2019,Tuominen2021}. Thus, investigating the WHIM nature, properties, and distribution is vital for understanding cosmic structure and galactic evolution \citep{Martizzi2019}. The WHIM temperature lies between $10^5$ - $10^7$\,K, and it is located mostly within the filaments of the cosmic web \citep{Dave2001, Nicastro2018}. These high temperatures result from the shock heating produced by the falling of the intergalactic gas into the potential well set by the large-scale structure of the Universe. As a consequence of the high temperatures, the neutral gas fraction of the WHIM is extremely low \citep{Richter2004, Tripp2006}, which, together with the overall low densities of the WHIM, makes this gas hard to study in absorption, because of the very shallow and broad absorption features produced by this gas in the spectra of background sources, thermally broadened to Doppler parameters $b$ > 40\,\kms\ \citep{Richter2006a, Richter2006b, Lehner2007}. Additionally, non-thermal line broadening processes, line blends, and noise features can mimic broad spectral features \citep{Richter2006b, Garzilli2015}, further complicating the study of the WHIM in absorption. This contrasts with the X-ray emission produced by the WHIM, detected between galaxy clusters \citep[see, e.g.,][]{Kull1999, Werner2008, Reiprich2021}.

As a consequence, some studies have attempted to target cosmic filaments inferred from the distribution of galaxy clusters to detect and characterize WHIM properties \citep{Tejos2016, Pessa2018,Vernstrom2021}. In particular, \citet{Tejos2016} and \citet{Pessa2018} studied the sightline of a QSO that intersects 7 independent inter-cluster axes, at impact parameters <3 Mpc in the redshift range $0.1 \leq z \leq 0.5$, which shows an excess of broad ($>40$\,\kms) Lyman alpha absorbers (BLAs) with respect to the random field expectations; these authors conclude that this excess of BLAs is likely tracing WHIM within cosmic filaments. Detecting BLAs associated with warm gas absorbers is crucial for determining the physical properties of the WHIM and, fundamentally, its total baryon content \citep{Savage2011}.

An alternative approach to studying the baryon content consists of using the dispersion measure (DM) of fast radio bursts (FRBs) \citep{Mcquinn2014,Macquart2020}. However, one limitation of this approach is the difficulty of separating the contribution to FRB DMs from the cosmic web and that from other components in their lines-of-sight, such as that from the Milky Way, intervening galaxy halos, and the FRB galaxy host itself \citep{Simha2020, Simha2021, Lee2022, Simha2023, Khrykin2024}. Furthermore, FRBs DMs are insensitive to the temperature of the ionized gas. Another limitation is the degeneracy between cosmological parameters and the fraction and location of baryons in the intergalactic medium \citep[IGM;][]{Bao2023, Baptista2023}. 

There are several works that have focused on connecting absorptions in the spectra of background sources to the properties of galaxy filaments (i.e. cosmic filaments traced by the distribution of galaxies). \citet{Wakker2015} use Ly$\alpha$ absorption lines in the \textit{HST} spectra (resolution of about $\sim20$\,\kms) of 15 active galactic nuclei (AGN) to probe the properties of the gas within an individual galaxy filament. They find that Ly$\alpha$ absorbers tend to be closer to the filament axis, and that there is a negative correlation between the equivalent width of the absorbers and the filament impact parameter, meaning that the gas is preferentially concentrated towards the inner regions of the filament, where the gravitational potential is deeper. However, their sample of absorbers contains only a few BLAs, making it difficult to extend their conclusions to warm-hot gas. In a similar study, \citet{Bouma2021} use a large sample of 32 QSO sightlines to probe the gas within five previously mapped \citep{Courtois2013} local filaments, and find a higher number density of Ly$\alpha$ absorbers per unit redshift near galaxy filaments, compared to the general population of Ly$\alpha$ absorbers, which most likely reflects the overdensity of matter within them. 

On the other hand, \citet{Burchett2020} use the galaxy distribution inferred from the SDSS data to model the geometry and density of the cosmic web, and then use \textit{HST} spectra to compare the strength of the measured absorption versus the modeled gas density in the filaments, and find progressively stronger absorptions towards the inner and denser regions of the cosmic web. However, after reaching a maximum, they find a sharp drop in the absorption signal in the densest regions of filaments. This is likely due to the difficulty to detect in absorption the warm-hot gas located in the deepest parts of the potential of the cosmic web, because to its low neutral fraction. Altogether, while studies of Ly$\alpha$ absorbers have shown that the intergalactic medium is denser in the inner regions of filaments, these results have not been extended to the warm-hot phase of the intergalactic medium yet, due to the difficulty of detecting broad Ly$\alpha$ absorptions, as this requires high resolution and high S/N UV spectra of background sources. In this line, \citet{Parchat2016} find evidence of metal-enriched warm-hot gas in a QSO sightline that could be intercepting a large-scale filament connecting two groups of galaxies, supporting the idea that broad absorbers lie in cosmic filaments.

In this paper, we search for observational evidence of the connection between BLAs and the densest parts of the large-scale structure of the Universe traced by cosmological filaments. To do it, we proceed in a similar fashion as in \citet{Tejos2016} and \citet{Pessa2018}. That is, targeting a single QSO (SDSSJ161940.56+254323.0), whose sightline intersects several inter-cluster axes, where theoretical models predict a higher probability of finding a large-scale filamentary structure \citep{AragonCalvo2010, GonzalezPadilla2010}. We detect $13$ BLAs candidates in the sightline of the QSO, as well 21 narrow Ly$\alpha$ absorptions (NLA) at redshifts $\sim0.01-0.26$. Out of this sample of absorbers, we deem as reliable the $13$ BLAs and $12$ NLAs, based on their equivalent widths and equivalent width uncertainties. We use then VLT/MUSE and VLT/VIMOS data to infer local overdensities of galaxies at the redshift of the absorptions, which allows us to study the potential association of the BLAs with nearby galaxies and cosmic structures. While 13 BLAs is not a particularly large number, compared to other studies, it is still significantly higher than the random field expectation reported by \citet{Danforth2010} of $5\pm3$, for a single sightline at $z\sim0.27$, for which our combined dataset allows us to explore correlations between the properties of the absorbing gas and its local environment. Furthermore, previous works with larger samples of absorbers do not necessarily have a larger number of BLAs, due to the technical challenges that their detection represents \citep[see, e.g.,][]{Wakker2015, Tejos2016}.

Our paper is structured as follows: In Sec.~\ref{sec:data} we describe the datasets used in our analyses, in Sec.~\ref{sec:results} we present our results on the association between BLAs and local overdensities of galaxies, and we discuss them in Sec.~\ref{sec:discussion}. Finally, we summarize our findings in Sec.~\ref{sec:summary}. For our analysis, we assume a $\Lambda$CDM cosmology based on the results of the \citet{Planck2016}. All the distances quoted in this paper refer to comoving distance.

\section{Data and selection of the sightline}
\label{sec:data}
\subsection{Selection of the QSO SDSSJ161940.56+254323.0}
\label{sec:selection}

We selected SDSSJ161940.56+254323.0 in a similar fashion as the selection of SDSSJ141038.39+230447.1 described in \citet{Tejos2016}, that is, maximizing the number of inter-cluster filaments intersected. The QSO SDSSJ161940.56+254323.0 was selected from the QSO catalog published by \citet{Schneider2010}, based on SDSS DR7 data. This catalog comprises $\sim$ 100\,000 QSOs with well-known magnitudes and spectroscopic redshifts. We searched for bright ($r < 17.5$ mag) QSOs whose sightlines pass through the maximum number of individual inter-cluster axes, connecting cluster pairs at lower $z$ than that of the QSO and above $z>0.1$ (the details of the identification of the inter-cluster axes are presented in Sec.~\ref{sec:gal_clust}). The motivation behind this selection criteria is that in the $\Lambda$CDM cosmological paradigm, galaxy clusters correspond to the nodes of the cosmic web, where several filamentary structures intersect. We also searched in the Galaxy Evolution Explorer \citep[GALEX][]{Martin2005} database and prioritized those QSOs with high FUV fluxes, enabling a signal-to-noise ratio (S/N) $\sim10$ spectra to be observed in a relatively short exposure time (no larger than 15 \textit{HST} orbits). 

We found that SDSSJ161940.56+254323.0 \citep[together with SDSSJ141038.39+230447.1, see][]{Tejos2016, Pessa2018} satisfied the imposed criteria, maximizing the number of independent cluster-pair structures, for the minimum observing time.

\subsection{Galaxy clusters}
\label{sec:gal_clust}
As described in Sec.~\ref{sec:selection}, we selected SDSSJ161940.56+254323.0 because its sightline is expected to intersect the largest number of inter-cluster axes, within some maximum impact parameter. These inter-cluster axes are defined as straight lines in the plane of the clusters connecting pairs of galaxy clusters close (in the projected distance) to the QSO sightline, and they are expected to coincide with the presence of large-scale cosmic web filaments \citep{GonzalezPadilla2010}.  To identify cluster pairs close to the QSO sightline and define the inter-cluster filaments, we used the Gaussian Mixture Brightest Cluster Galaxy (GMBCG) catalog \citep[][]{Hao2010}. The GMBCG is based on SDSS DR7, and consists of over 55000 rich clusters across the redshift range from $0.1 < z < 0.55$.

The sample of cluster pairs has been defined imposing the following criteria (see figure 1 of \citealt{Tejos2016} for an illustration):

\begin{itemize}
    \item[(i)] The rest-frame velocity difference between the clusters has to be $<1000$\,\kms.
    \item[(ii)] Both clusters must be at $z < z_{\mathrm{QSO}}$.
    \item[(iii)] At least one of the two members of a cluster pair must have a spectroscopic redshift determination\footnote{Only one cluster from all the final selection of cluster pairs near the SDSSJ161940.56+254323.0 sightline does not have spectroscopic redshift.}.
    \item[(iv)] The transverse separation between the cluster centers has to be $< 30$ Mpc.
    \item[(v)] The impact parameter between the inter-cluster axis and the candidate QSO sightline has to be $<5$ Mpc.
\end{itemize}

Throughout this paper, we use 1000\,\kms\ as a conservative velocity limit typically used to characterize galaxy filaments \citep[e.g.,][]{Wakker2015, Bouma2021}. This value also accounts for the typical velocity dispersion of galaxy clusters on the order of $\sim1000$ \kms\ \citep{Struble1999}. Nevertheless, In Sec.~\ref{sec:overdensity_larger_scales}, we explore the impact of using a different velocity limit. The 30 Mpc maximum separation between clusters in a cluster pair was motivated by theoretical results from N-body simulations in $\Lambda$CDM universes, which predict a high probability of having coherent large-scale filamentary structures between nearby galaxy clusters 
\citep[][]{GonzalezPadilla2010}. Finally, the choice for the maximum impact parameter of $\sim5$ Mpc is somewhat higher than the typical width of filaments of $\sim3$ Mpc \citep[][]{Wakker2015, AragonCalvo2010}. We decided to include larger impact parameters as the shape of the filaments is often not a straight line, and thus, a larger impact parameter to the inter-cluster axis does not necessarily imply a larger impact parameter to the actual galaxy filament. 

Imposing these criteria, we find a total of eight cluster pairs close to the SDSSJ161940.56+254323.0 sightline, whose properties are summarized in Table~\ref{tab:cluster_pairs_props}. We use these cluster pairs to study the presence of an inter-cluster axis at the redshift of any of the BLAs described in Sec.~\ref{sec:blas_survey}, and overdensities of galaxies in our VLT/VIMOS survey (see Sec.~\ref{sec:overdensity_blas}).

\begin{table}

\begin{tabular}{ccccc}
\hline
ID & $z_{\mathrm{sys}}$ & separation & $\Delta v$ & Impact param. \\
(1)&          (2)       &    (3)     &      (4)    &    (5)    \\
& & Mpc &\,\kms & Mpc \\
\hline
1 & 0.14688 & 14.3 & 875 & 2.2 \\
2 & 0.16065 & 17.3 & 990 & 3.3 \\
3 & 0.17194 & 17.7 & 673 & 3.5 \\
4 & 0.18279 & 22.5 & 756 & 4.3 \\
5 & 0.18819 & 29.5 & 910 & 1.0 \\
6 & 0.18908 & 26.0 & 766 & 2.8 \\
7 & 0.18923 & 16.0 & 687 & 4.3 \\
8 & 0.22924 & 29.9 & 556 & 4.4 \\
\hline
\end{tabular}
\caption{\label{tab:cluster_pairs_props} Properties of the cluster pairs described in Sec.~\ref{sec:gal_clust}. Column (2) shows the mean redshift of the two pair members, column (3) indicates their transverse separation, column (4) shows their respective velocity difference along the line of sight, and column (5) corresponds to the impact parameter from the QSO sightline to the inter-cluster axis, at the redshift of the cluster pair.}
\end{table}

\subsection{HST/COS FUV spectroscopy}

We obtained HST/COS FUV spectroscopy of QSO SDSSJ161940.56+254323.0 as part of a HST GO program (ID 13832, PI Tejos). The observations were carried out in August 2015. We employed the G130M and G160M gratings, and using four Fixed Pattern Noise Positions (FP-POS) with two central wavelengths for G130M (1291 and 1309 \AA) and one central wavelength (1577 \AA) for G160M. The inclusion of multiple FP-POS positions is necessary to mitigate the impact of fixed-pattern noise inherent to the COS instrument. Our setup provides spectral coverage between $1134-1751$ \AA\ with resolving power $R \sim 20\,000$.

We retrieved the individual exposures of SDSSJ161940.56+254323.0 spectra from the Mikulski Archive for Space Telescopes (MAST). The downloaded data reduced files {\tt x1d.fits} had been previously uniformly processed by the HST/CalCOS \citep{Danforth2016}. CalCOS is a processing system designed to calibrate COS data. It achieves this by rectifying instrumental effects, producing a wavelength calibration tailored to each exposure, and finally extracting and producing a one$-$dimensional spectrum that is calibrated for flux, for the entire observation \citep{James2022}. We co-added all exposures corresponding to each grating. Subsequently, we used \linetools\ \citep{Prochaska2016} to rebin the original spectra into a single linear wavelength scale of $0.0395$\,\AA\,pixel$^{-1}$, which corresponds roughly to two pixels per resolution element. In the spectral range covered by both gratings, we used the data from the G160M grating due to its higher resolution and S/N.
 
This approach yielded a final spectrum characterized by a S/N$\sim$12 per resolution element. This process was done with \linetools\, which ensures the conservation of total flux in the process.

\begin{figure*}
\centering
\includegraphics[width = 0.8\textwidth]{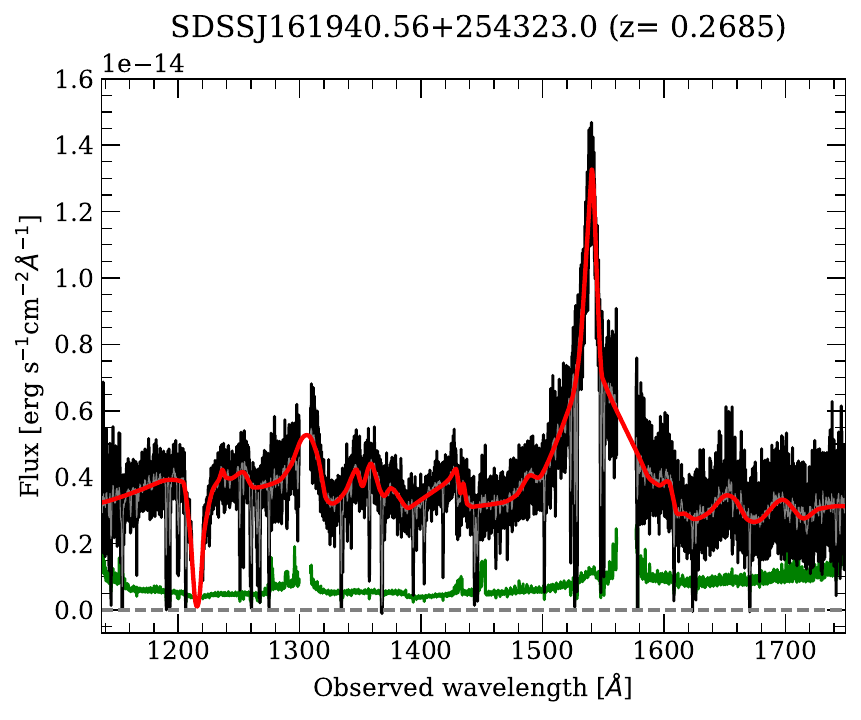}
\caption{ Observed HST/COS FUV spectrum of QSO SDSSJ161940.56+254323.0 in black. The green line shows the $1-\sigma$ uncertainty of the spectrum. The red line corresponds to the modeled pseudo-continuum. The grey line shows the HST spectrum convolved with a Gaussian kernel with a FWHM of $1 \AA$. Figure~\ref{fig:BLAs_sample} shows a zoom-in of the QSO spectrum at their wavelengths.}     
\label{fig:spectrum_HST}
\end{figure*}

The calibrated one-dimensional final spectrum was further analyzed using the \linetools \,{\tt lt\_continuumfit} script to fit a pseudo-continuum to it\footnote{i.e. including the intrinsic broad emission lines and the Galaxy’s damped Ly$\alpha$ system wings \citep{Tejos2016}.}. Figure~\ref{fig:spectrum_HST} illustrates the reduced SDSSJ161940.56+254323.0 spectrum, alongside its associated uncertainty, and our pseudo-continuum fit model.

\subsection{VLT/MUSE Integral field spectroscopy}
\label{sect:muse_obs}

\begin{figure}
\centering
\includegraphics[width = \columnwidth]{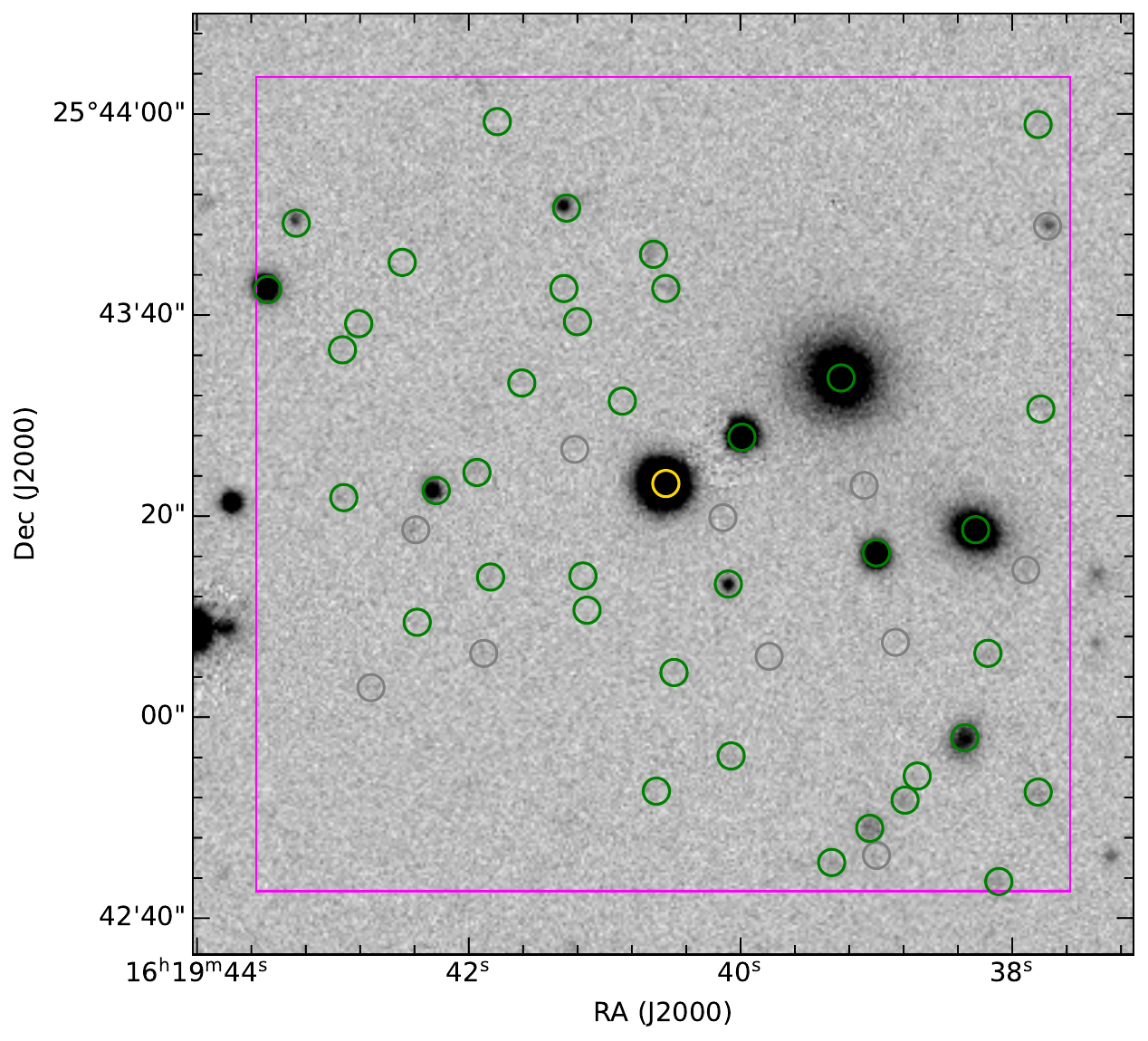}
\caption{Combined FoV of our VLT/MUSE observations. The magenta square shows approximately the size of the combined FoV. The background image corresponds to an $r$-image from The Dark Energy Spectroscopic Instrument \citep[DESI;][]{Dey2019} Legacy Surveys. The green and grey circles show the sources identified in the field, with and without a redshift measurement, respectively (see Sec.~\ref{sec:MUSE_survey}). The central yellow circle corresponds to SDSSJ161940.56+254323.0.}     
\label{fig:MUSE_FOV}
\end{figure}

We obtained VLT/MUSE \citep{Bacon2010} data of a $\sim 1.3\times 1.3$\,arcmin$^2$ field-of-view (FoV) containing SDSSJ161940.56+254323.0 as part of the ESO programme 096.A-0426 (PI Tejos), which is in coordination with the HST/COS observations. The observations were taken in Service Mode with a seeing of $\sim$1$"$, sampled at $0.2\times0.2$\,arcsec$^2$, with a spectral range from $4750-9350$\,\AA, and a resolving power of $ R\sim 1770-3590$. A total of $6$ exposures of $18$\,min each were used. Table~\ref{tab:muse_point} summarizes these observations, including the pointings and position angles (PA) of each individual exposure and Fig.~\ref{fig:MUSE_FOV} shows the approximated combined FoV of our VLT/MUSE data overlaid on an $r$-image from The Dark Energy Spectroscopic Instrument \citep[DESI;][]{Dey2019} Legacy Surveys.

\begin{table}{\large VLT/MUSE observations}
\centering
\begin{tabular}{lccr}
\hline
\multicolumn{2}{c}{Pointings}&Exposure time&PA\\
          RA&DEC & (s) & (deg) \\
\hline
16:19:41.30&+25:43:33.17&$2 \times 1057$&0\\

16:19:39.82&+25:43:33.20&$2 \times 1057$&90\\

16:19:39.82&+25:43:13.19&$1057$&180\\

16:19:41.30&+25:43:13.19&$1057$&270\\
\hline
\end{tabular}
\caption{Summary of our observation described in Section~\ref{sect:muse_obs}.}
\label{tab:muse_point}
\end{table}

Datacubes have been reduced and combined using the standard MUSE pipeline version muse-1.6.1 (see \url{http://www.eso.org/observing/dfo/quality/PHOENIX/MUSE/processing.html} for details). 
As a postprocessing optimizaton, we used the ZAP software \citep{Soto2016} to improve the sky subtraction. The survey of galaxies assembled from the VLT/MUSE data is described in detail in Sec.~\ref{sec:MUSE_survey}.

\subsection{VLT/VIMOS Multi-object spectroscopy}
\label{sect:vimos_mos_obs}

We obtained VLT/VIMOS \citep{LeFevre2003} multi-object-spectroscopy (MOS) to complement the VLT/MUSE observations as part of ESO programmes 097.A-0560 and 099.A-0197 (PI Tejos). 
Its field of view (FoV) of $\approx 15 \times 16$\,arcmin$^2$ allows us to survey galaxies around the QSO SDSSJ161940.56+254323.0 at much larger scales, up to several Mpc, depending on the redshift considered. Note that the VLT/MUSE FoV lies within the separation of the four quadrants of the VLT/VIMOS FoV, that is, there is no overlap between both datasets.

We used the low-resolution blue grism ($R=220$) with the OS Blue filter, which covers a wavelength range of $4000 - 6700$\,\AA. Observations were performed in Service Mode with a seeing of $\sim 1.1$ arcsec on the night of UT 2017-08-16. The VIMOS data has been reduced using the standard ESO pipeline version vimos-3.1.9 (see \url{https://www.eso.org/sci/software/pipelines/vimos/} for details).

We use $R$ and $I$ pre-imaging observations for designing the MOS masks. In order to avoid observing galaxies at $z>z_{\mathrm{QSO}}\approx0.3$, we applied a $R-I<0.7$ color criteria for placing slits into objects. By using data from the SDSS DR16 as a reference, we expect that $\approx80\%$ of these galaxies will lie at $z<0.3$. From the total of objects satisfying the color criteria, we placed slits randomly on galaxies that also satisfy a brightness criterion of $R\leq 22.5$, aiming for a minimum S/N $\approx5$ per resolution element at the continuum level, in order to facilitate the redshift identification (although some galaxies fainter than this limit are included in our sample as fillers).

The total exposure time for the MOS observation was $1.9$ hours, from $6$ individual exposures of $1135$\,s each, with a position angle of PA$=90^{\circ}$. These observations are summarized in the Table~\ref{tab:vimos_point}. The survey of galaxies assembled from the VLT/VIMOS data is described in detail in Sec.~\ref{sec:VIMOS_survey}
\begin{table*}
\centering
\large VLT/VIMOS observations
\\
\begin{tabular}{lccccr}
\hline
\multicolumn{2}{c}{Pointing}&Exposure time&Grism&Filter&PA\\
          RA&DEC & ( s)&&&deg\\
\hline
16:19:40.54&+25:43:25.57&$150$& --& R & 90\\
16:19:40.54&+25:43:25.57&$120$& --& I & 90\\
16:19:40.54&+25:43:25.57&$6 \times 1135$&LR Blue&OS Blue&90\\
\hline
\end{tabular}
\caption{Summary of our VLT/VIMOS observations.}
\label{tab:vimos_point}
\end{table*}

\subsection{Galaxy survey}
In this section, we describe our methodology for building a survey of galaxies near the sightline of SDSSJ161940.56+254323.0, that are identified in our VLT/MUSE and VLT/VIMOS datasets. We use the VLT/MUSE dataset to study the small-scale (CGM-scales) environment around the QSO sightline, where the VLT/VIMOS data is not available, and look for potential intervening galaxies associated with the absorptions detected in the HST/COS spectrum, whereas the VLT/VIMOS data is used to study the environment around the sightline on scales of a few Mpc, and evaluate if the number of galaxies detected is consistent with an under- or an over-dense region of the Universe. Further in the paper, we use this survey to evaluate the existence of overdensities of galaxies at the redshift of a given BLA identified in the HST/COS spectrum of SDSSJ161940.56+254323.0, as well as to identify potential associations between a BLA and a galaxy near the SDSSJ161940.56+254323.0 sightline. 
\subsubsection{VLT/MUSE survey}
\label{sec:MUSE_survey}
We have built this VLT/MUSE survey in the same fashion as in \cite{Pessa2018}. In summary, the source identification has been done in two steps, first, we use {\sc SExtractor} \citep{Sextractor1996} in the `white' image to detect the visible sources in the field. Then, we use MUSELET \citep{MPDAF} to search for additional emission-lines-only galaxies, and we added $2$ sources. \\
We ended up with a sample of 51 sources in the VLT/MUSE field of view, and for each one of them, we extracted a 1-D spectrum.
For the sources detected by {\sc SExtractor} in the `white' image, we estimate the apparent $r$ magnitude by constructing an $r$ image, convolving the cube with the SDSS $r$ transmission curve. The zero-point calibration of this photometry has been done with a cross-match using sources in our survey with available SDSS $r$ photometry. 
\\
The redshift of each source was measured using the  {\sc Redmonster} software \citep{Hutchinson2016}, with a posterior visual inspection. We adopt the same reliability scheme defined in \cite{Pessa2018} for classifying the level of confidence in the $z$ characterization of each source: `a' for the best characterized sources (at least two well-identified spectral features), `b' for the reasonably well characterized sources (one well-identified spectral feature and at least one other tentative identification), `c' for the most uncertainly classified sources (only one clear spectral feature) and `d' for the sources that could not be assigned a $z$. Stars safely identified by their continuum shape and/or characteristic stellar absorptions are labeled as `a'. Out of the $51$ sources, we successfully obtained the redshifts for $40$ of them (see Table~\ref{tab:muse_ids} in Appendix~\ref{sec:appendix_muse_sources}). These include 35 obtained with {\sc Redmonster} and 5 from a posterior visual inspection of strong emission lines. We note that in our VLT/MUSE survey, only one galaxy is labeled as `c', and it is at a redshift higher than the central QSO, and therefore, it does not affect our results. The absolute $r$ magnitude $M_{r}$ and the spectral classification included in the table were determined following \cite{Pessa2018}. The remaining 11 sources that were not characterized with a redshift are listed in Table~\ref{tab:muse_noids} for completeness. Figure~\ref{fig:MUSE_FOV} shows the location of the sources identified in the combined VLT/MUSE FoV, with and without a redshift measurement, with green and grey circles, respectively. While the VLT/MUSE FoV is not large enough to be useful for probing the large-scale structure of the Universe, it is useful to study in detail the small-scale environment of the QSO sightline, and investigate the presence of galaxy halos potentially associated with the absorptions identified in the QSO spectrum.

\subsubsection{Completeness of our VLT/MUSE survey}
\begin{figure*}
        \begin{minipage}[b]{0.33\textwidth}
                \includegraphics[width=\columnwidth]{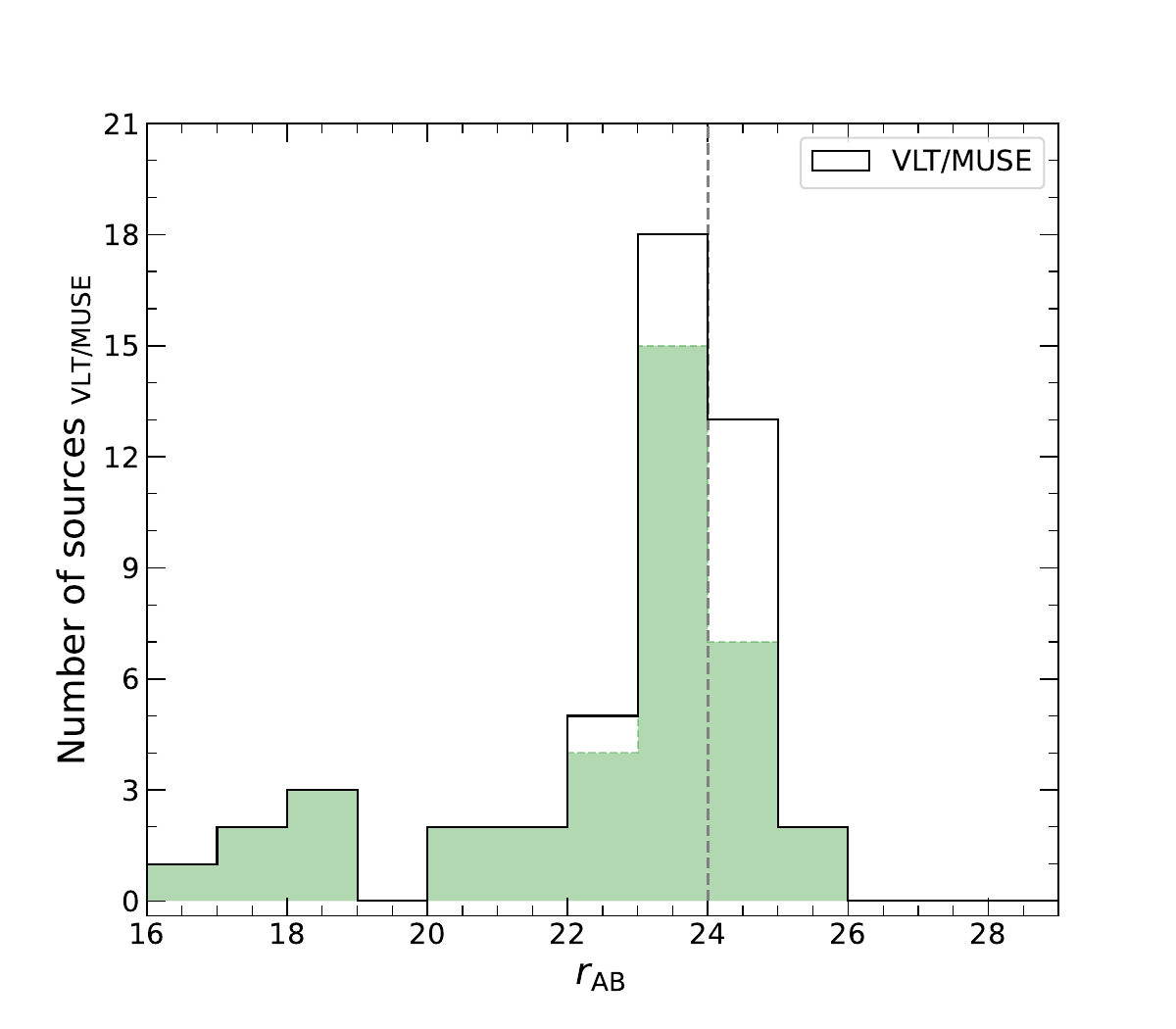}

        \end{minipage}%
        \begin{minipage}[b]{0.33\textwidth}
                \includegraphics[width=\columnwidth]{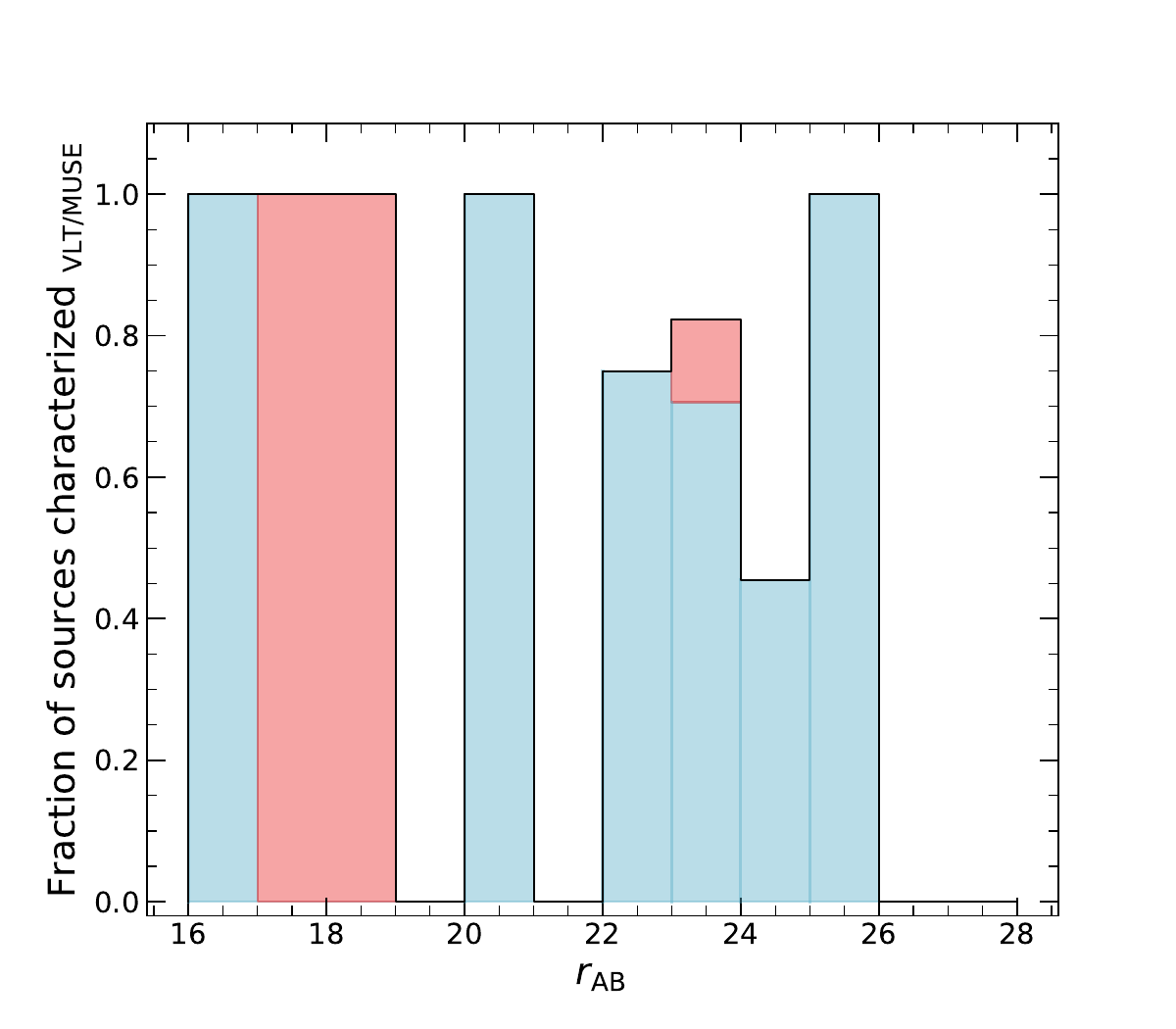}

        \end{minipage}%
        \begin{minipage}[b]{0.33\textwidth}
                \includegraphics[width=\columnwidth]{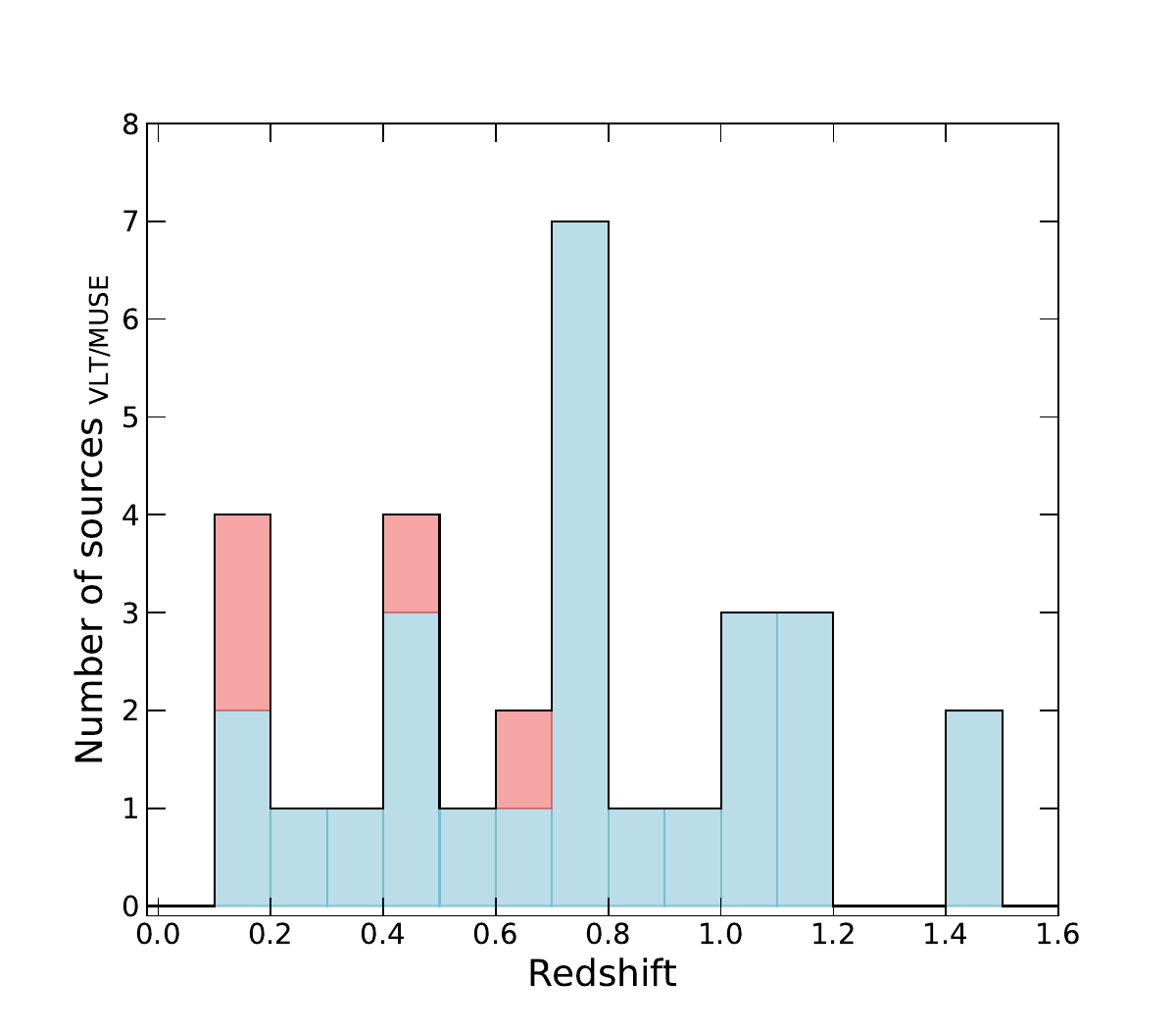}

        \end{minipage}%

        \caption{\textit{Left}: Survey histogram colored in green for our VLT/MUSE sample with measured redshifts. The black line shows the distribution for the whole sample detected by SExtractor. The distribution suggest a detection threshold of around $r\sim 24$ mag, indicated with the vertical dashed line. \textit{Center}: completeness fraction of the redshift survey. The star forming galaxies and non-star forming galaxies are shown in blue and red, respectively (stars are not included). We reach to $\sim78\%$ successful characterization fraction at an apparent magnitude $r_{\rm AB}=24$ mag. \textit{Right}: shows the redshift distribution of our full characterized galaxy sample colored by star forming and non-star forming fraction same as the central panel.}
        \label{fig:muse_survey_hists}
\end{figure*}

In order to characterize the completeness of our VLT/MUSE survey, we use the apparent $r$ magnitude. The left panel of the Fig.~\ref{fig:muse_survey_hists} shows the $r$ magnitude distribution of the sources detected by {\sc SExtractor} in our VLT/MUSE survey. The sources characterized with a $z$ are represented in green, and the complete sample is represented by the black solid line. This histogram suggests a detection threshold of $r\sim24$ mag. An apparent magnitude of $r=24$ mag corresponds to a luminosity of $1.9 \times 10^{8}$ L$_{\odot}$  $\approx 0.006$ L$_{*}$ at $z=0.2$. The central panel of  Fig.~\ref{fig:muse_survey_hists} shows the fraction of sources characterized with a redshift per magnitude bin. Our characterization rate reaches $\sim78$\% down to $r$ magnitude $r=$ 24 mag (it reaches $100$\% at fainter magnitudes, but with only two detected sources, both classified as star-forming galaxies due to the presence of emission lines). Finally, the right panel shows the redshift distribution of the sources in our VLT/MUSE survey. The central and right panels are colored by spectral type, blue and red for star-forming and non-star-forming galaxies, respectively. Most galaxies are at $z > z_{\mathrm{QSO}}$, with only $14$ at $z < z_{\mathrm{QSO}}$ (of which 10 are foreground stars, and are not included in the central and right panels of Fig.~\ref{fig:muse_survey_hists}).

\subsubsection{VLT/VIMOS survey}
\label{sec:VIMOS_survey}
As explained above, in Sec.~\ref{sect:vimos_mos_obs}, the target selection was made imposing a $R-I$ color criterion in order to select galaxies in the redshift range of interest, i.e., those galaxies at $z<z_{\mathrm{QSO}}$. We use {\sc esorex} routines to reduce and extract the spectra from a total of 297 sources, and for each one of them, we obtain its apparent $r$ and $i$ magnitude from the SDSS DR16 catalog. We then use {\sc Redmonster} to calculate the redshift of these sources. The left panel of Fig.~\ref{fig:vimos_compl_z} shows the $r$ magnitude distribution of our VLT/VIMOS survey, for the complete sample, and for sources for which we could determine a redshift. This distribution suggests a detection threshold of $r=22.5$ mag. The redshift distribution of the sources in our VLT/VIMOS survey is shown in the right panel of Fig.~\ref{fig:vimos_compl_z}.
\\
We successfully measured the redshifts of 244 sources out of the total sample, including 53 stars, used to check the {\sc ESOREX} wavelength solution. We find this wavelength calibration to be satisfactory, with a dispersion of about $\sim200$\,\kms\ around zero velocity. We use the same reliability scheme as in the VLT/MUSE survey. Table~\ref{tab:VIMOS_survey_table} summarizes the properties of the sources in our VLT/VIMOS survey. Out of the 244 sources characterized with a redshift, 111 are at a lower redshift than the QSO (including stars), of which only 5 are labeled with an uncertain redshift (category `c'). Thus, while these possible misidentifications could weaken any putative real correlation, our results are unlikely to be driven by them. Furthermore, we have verified that the removal of these galaxies from our survey does not impact our main results.
\subsubsection{Completeness of our VLT/VIMOS survey}
\label{sect:vimos_compl}
In order to characterize the completeness of our VIMOS survey, we take into account two independent biases:
\begin{enumerate}
\item We are not able to measure the redshift of 100\% of the sources in our VIMOS survey.
\item The real number of sources with apparent $r$ magnitude $< 22.5$ mag in the VIMOS field of view that satisfies the imposed color criteria is much higher than the total number of sources in our VIMOS sample.
\end{enumerate}
To correct by these effects, we:
\begin{enumerate}
\item Study the successful redshift characterization ratio, in order to estimate how many galaxies we are missing in our own sample due to our limited ability to measure redshifts. Figure~\ref{fig:vimos_compl_z} shows this effect. The distribution of sources characterized by a redshift is represented in green, and the distribution of the full targeted sample is indicated with a black solid line. The central panel shows the fraction of sources characterized with a redshift per magnitude bin. 

On average, we recover about 84\% of the galaxy redshifts in the $17.5<r<22.5$ interval, where most of the sources in our survey lie. 

\begin{figure*}
        \begin{minipage}[b]{0.33\textwidth}
                \includegraphics[width=\columnwidth]{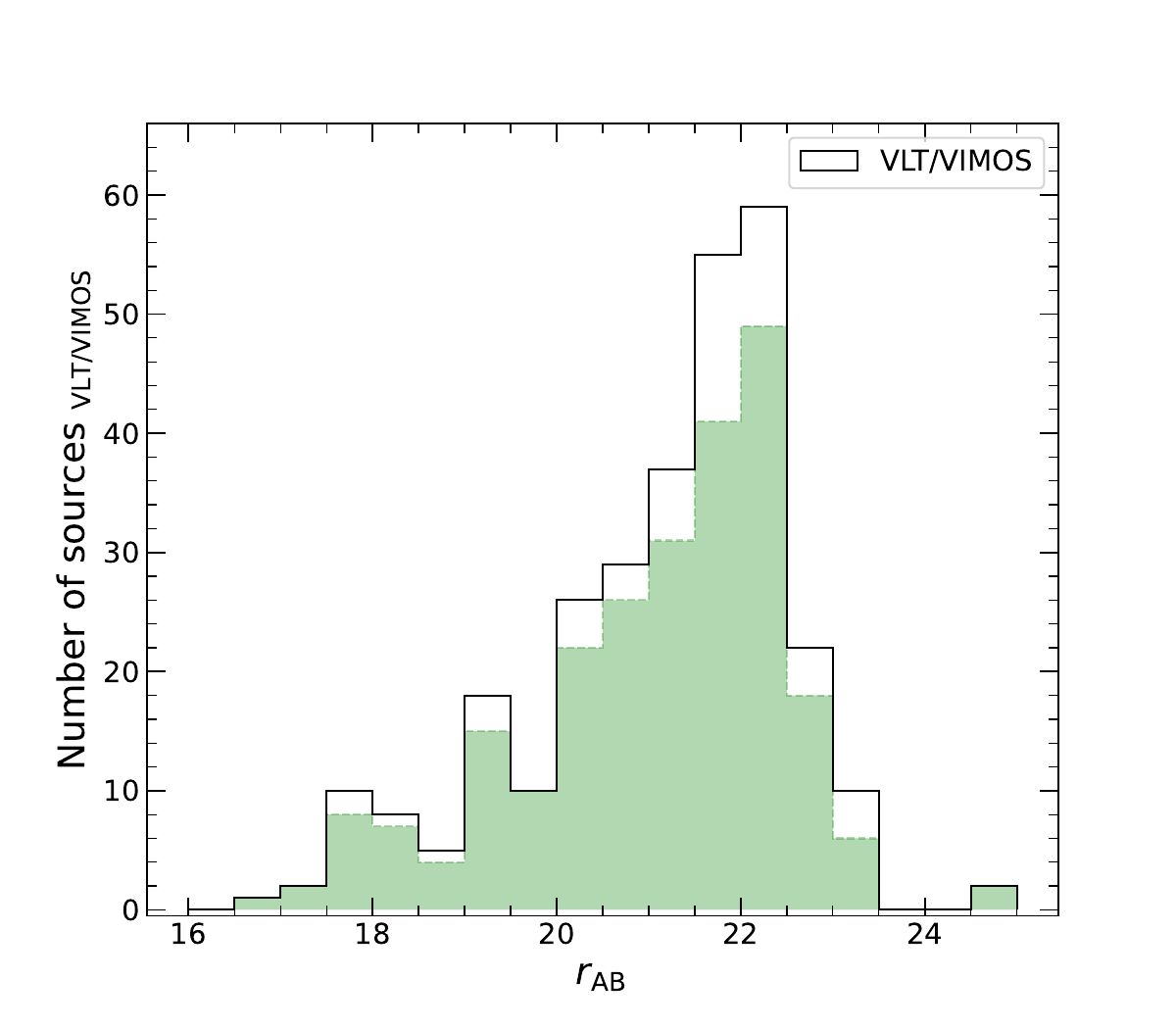}

        \end{minipage}%
        \begin{minipage}[b]{0.33\textwidth}
                \includegraphics[width=\columnwidth]{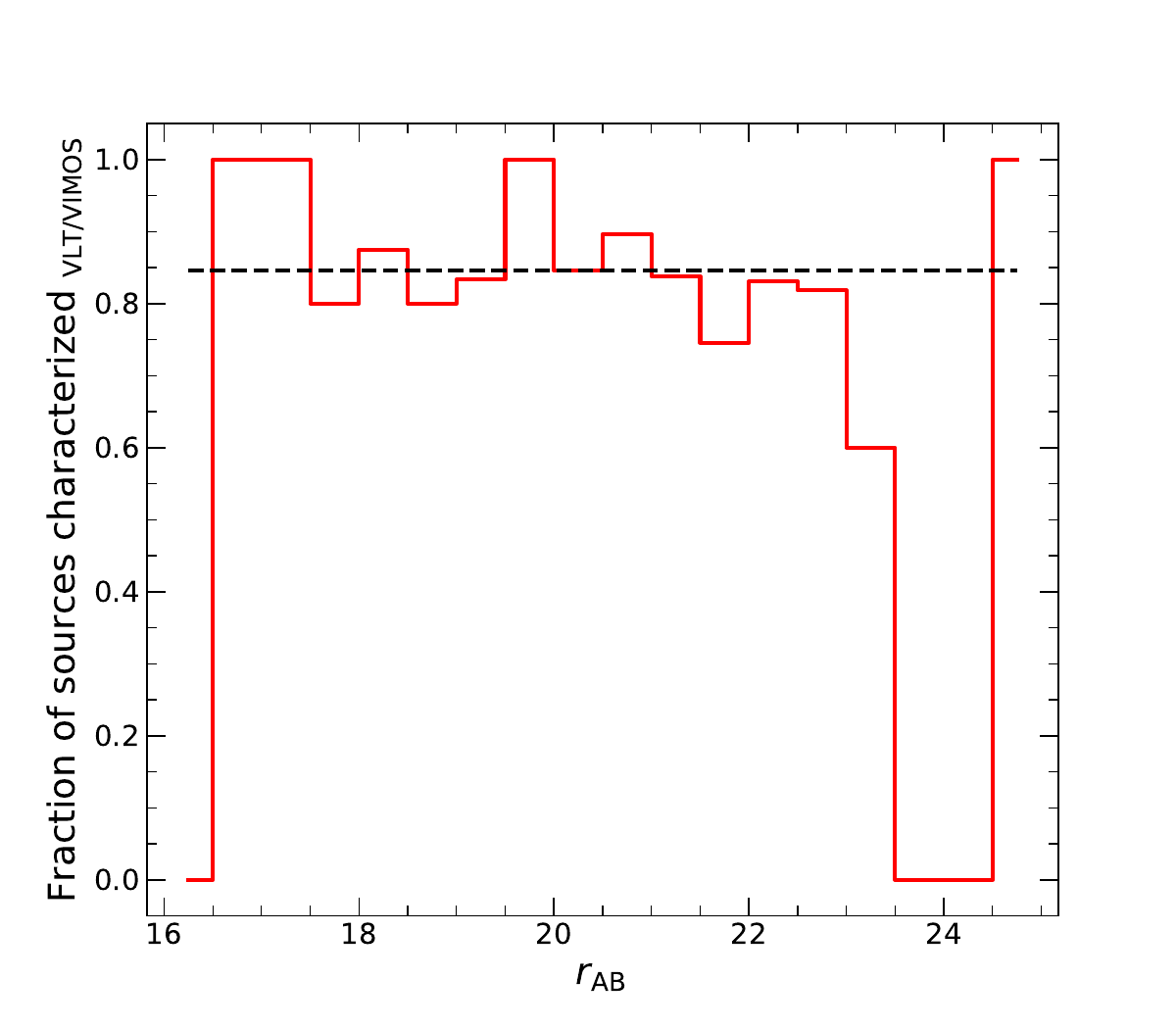}

        \end{minipage}%
        \begin{minipage}[b]{0.33\textwidth}
                \includegraphics[width=\columnwidth]{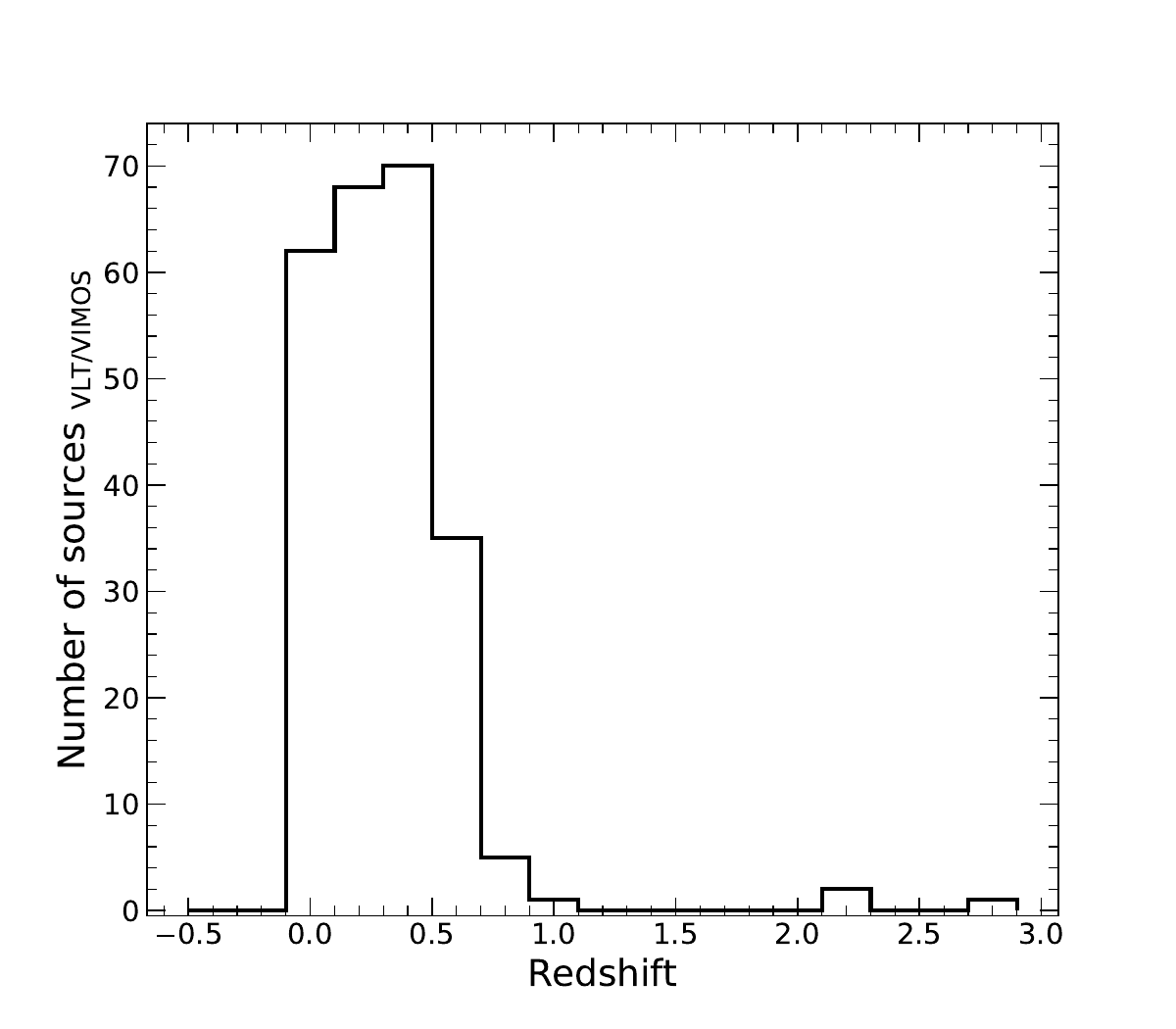}

        \end{minipage}%

        \caption{\textit{Left}: $r$ magnitude distribution of the sources in our VIMOS survey. The sources characterized with a redshift are represented in green, and the complete sample is represented by the black solid line. \textit{Center}: Fraction of sources characterized with a redshift per magnitude bin. The mean of this fraction considering only the $17.5<r<22.5$ interval is marked by the black dashed line. On average, we have measured the redshift of $\sim$84\% of the sources in our VIMOS survey. \textit{Left}: Redshift distribution of the sources in our VLT/VIMOS survey.}
        \label{fig:vimos_compl_z}
\end{figure*}

\item Use the SDSS DR16 photometric catalog to estimate the total number of galaxies brighter than $r$ = 22.5 mag within the VLT/VIMOS FoV that satisfy the imposed color criteria of $R-I<0.7$. The left panel of Fig.~\ref{fig:vimos_compl_sdss}, shows the total number of sources in the SDSS data and the number of sources in our VIMOS survey per $r_{\mathrm{AB}}$ magnitude bin, in blue and black respectively. The right panel shows the fraction between these two quantities and the black dashed line shows the mean of this fraction, considering only the $17.5<r<22.5$ interval. On average, our VIMOS target selection recovers around $\sim$27\% of the sources in the field. The median $5\sigma$ depth for SDSS photometric observations is $r = 22.70$, so it provides an appropriate reference to evaluate the completeness of our VLT/VIMOS survey. 

\begin{figure*}
        \begin{minipage}[b]{0.48\textwidth}
                \includegraphics[width=\columnwidth]{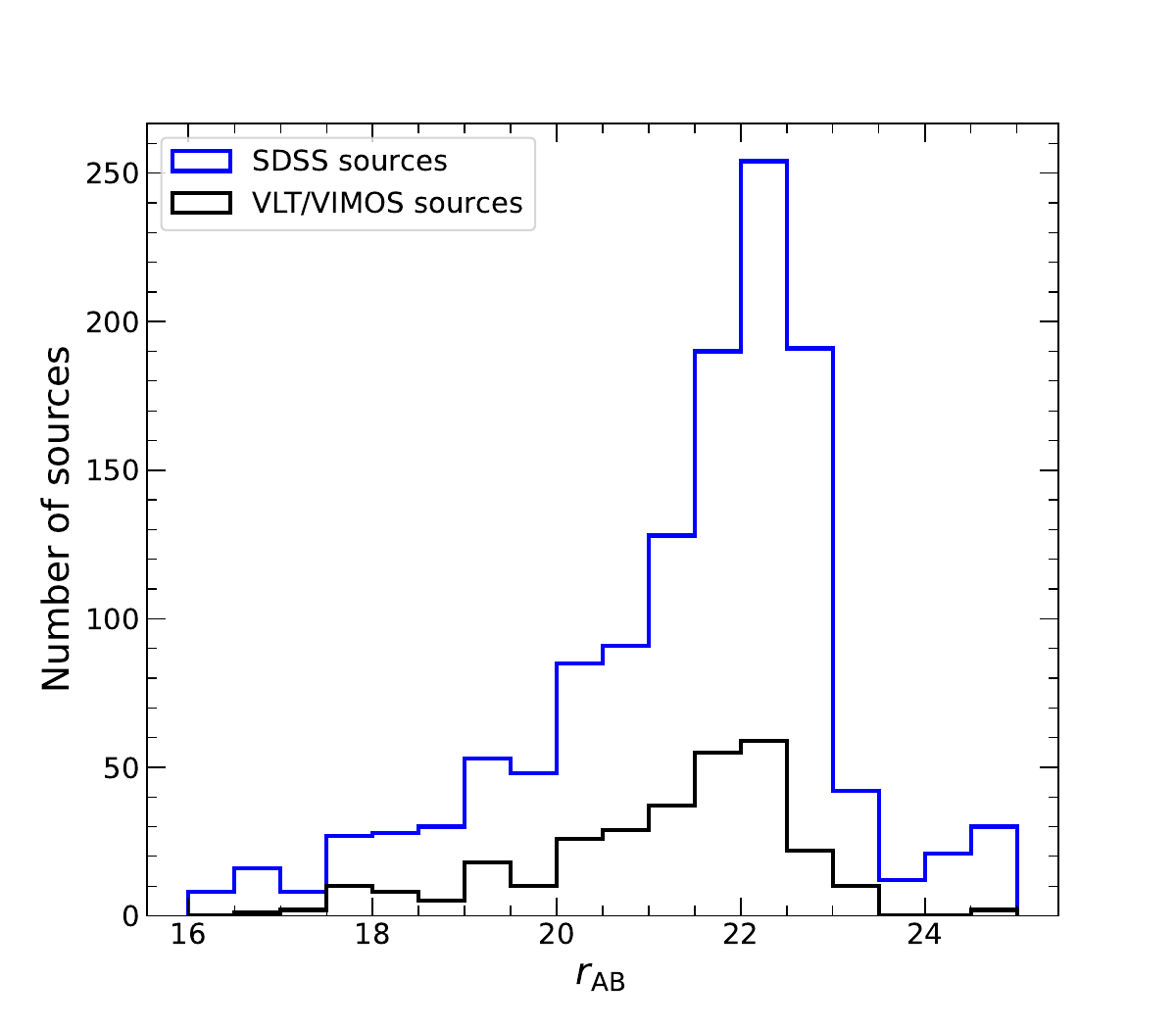}

        \end{minipage}%
        \begin{minipage}[b]{0.48\textwidth}
                \includegraphics[width=\columnwidth]{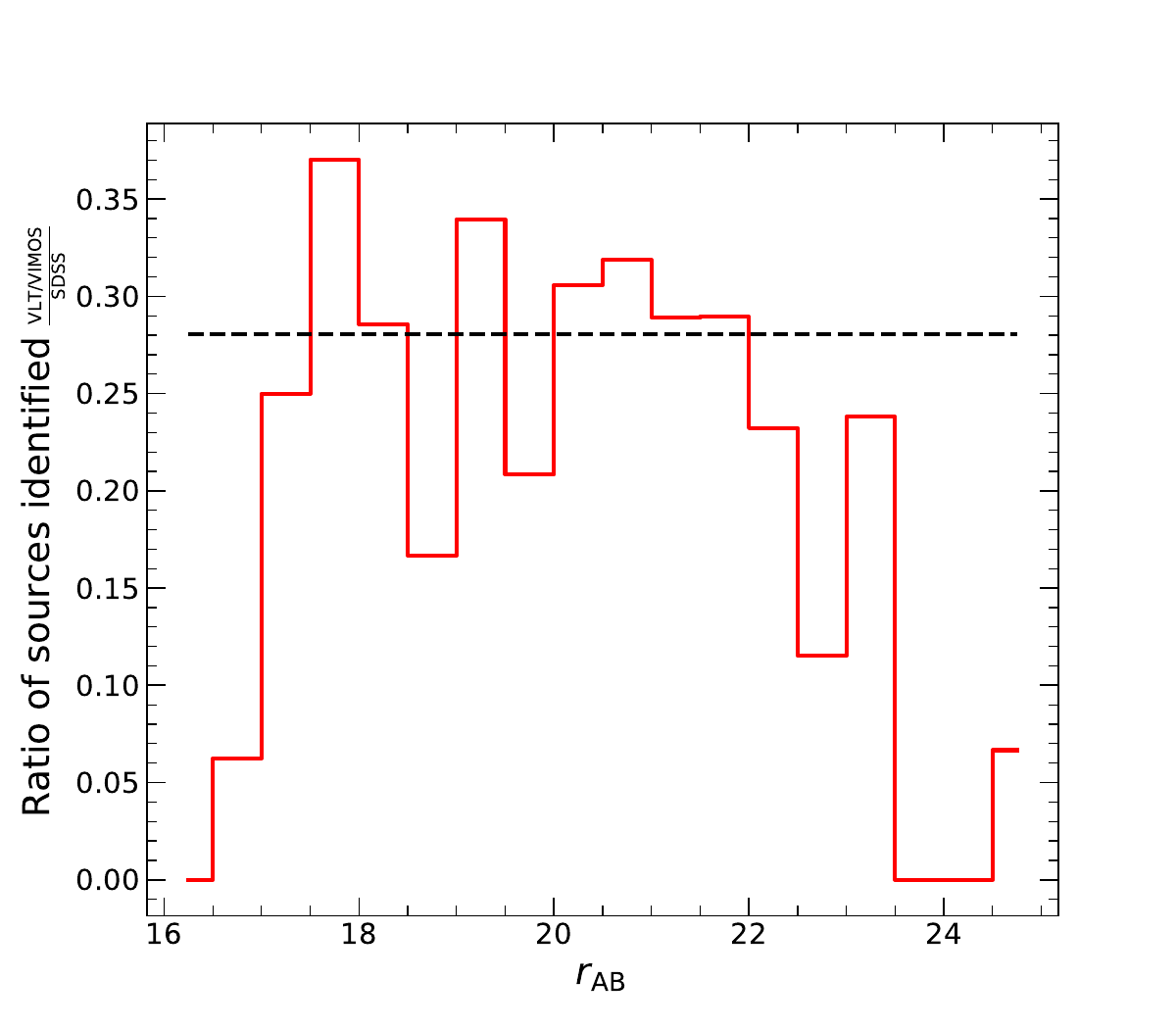}

        \end{minipage}%

        \caption{\label{fig:vimos_compl_sdss} \textit{Left}: Total number of sources in the SDSS DR16 data that satisfies the imposed color criteria of $R-I<0.7$ (blue) and number of sources in our VIMOS survey (black) per magnitude bin. \textit{Right}: Ratio between the total number of sources in the SDSS DR16 data and our VIMOS survey per magnitude bin. The black dashed line mark the mean of this fraction, for the $17.5<r<22.5$ interval. On average, we have recovered $\sim$27\% of the sources present on the field with our target selection.}
\end{figure*}

\end{enumerate}

Considering these two effects, we divide the number of galaxies effectively found in our VLT/VIMOS survey (within a given redshift window) by $0.84$ and $0.27$ to estimate the real underlying number of galaxies in the same cosmic volume. While our absolute recovery fraction of sources in the VLT/VIMOS FoV is relatively low, our understanding of the selection function allows us to correct for this incompleteness. We use the Poissonian errors calculated following \citet{Gehrels1986} to estimate the uncertainty of this correction.

\subsection{BLAs survey}
\label{sec:blas_survey}
In this section, we present the methodology employed for the identification of absorption lines in the SDSSJ161940.56+254323.0 normalized spectrum, and describe how we deal with possible ambiguities. The characterization of the absorption lines in the spectrum of SDSSJ161940.56+254323.0 of the HST/COS was carried out throughout the full line-of-sight, not only limiting ourselves to the regions where known structures exist. In this way, we minimize possible misidentifications and better handle possible interloper absorption.

We employ the same criteria to identify and characterize these absorption features as in \citet{Tejos2016} (see their section~4 for details). For this, we used the custom software {\tt IGM\_guesses}\footnote{\url{https://github.com/pyigm/pyigm}.}. This software facilitates the visualization and Gaussian fitting of absorption features, ultimately yielding initial parameters for subsequent Voigt profile fitting (see below); these include redshift ($z$), column density ($N$), and Doppler parameter ($b$) for the ions identified.

In summary, we first identify all possible absorption components (\hi\ and metals) at redshift $z = 0$, with the categories `a' (reliable), `b' (possible), or  `c' (uncertain) as appropriate (this reliability scheme is independent of that used to classify the redshift measurement of galaxies, although for simplicity we use the same naming convention for the reliability categories). In cases where the component shows at least two well-resolved transitions, we assign them to the category `a', and in the rest of the cases, we assign them to the category `b'. Similarly, we identify all possible absorption components (\hi\ and metals) at redshift $z = z_{\rm QSO}$ and label them in the respective categories. We then identify the absorption components of \hi\ that show at least two transitions from $z = z_{\rm QSO}$ to $z = 0$, with the category `a'. In the redshifts of these components, we identify all the possible metal absorptions, and we assign them to the respective category (given the prior of having a well-identified \hi\ absorption). Finally, we consider that all unidentified absorption features are \hi\ Ly$\alpha$ in category `b', and their corresponding associated metal absorption are identified (if any). If metal ions are present, we reassign the \hi\ component to the category `a'; otherwise, we maintain the component in category `b'. Some components in category `b' could move to category `c', based on the equivalent width ($W_r$) significance criterion: $W_r/\delta W_r < 3$, where $\delta W_r$ is the uncertainty of the $W_r$ measurement. However, none of the BLAs in our sample fall in the `c' category. 

Later we used these initial guesses as inputs for an automatic Voigt profiling process using the software {\sc Veeper} \footnote{\url{https://github.com/jnburchett/veeper}.}, taking into account the non-Gaussian COS line spread function (LSF) and restricting the sample to absorption lines having $W_r>0.01$\,\AA.

\begin{figure*}
\centering
\includegraphics[width = \textwidth]{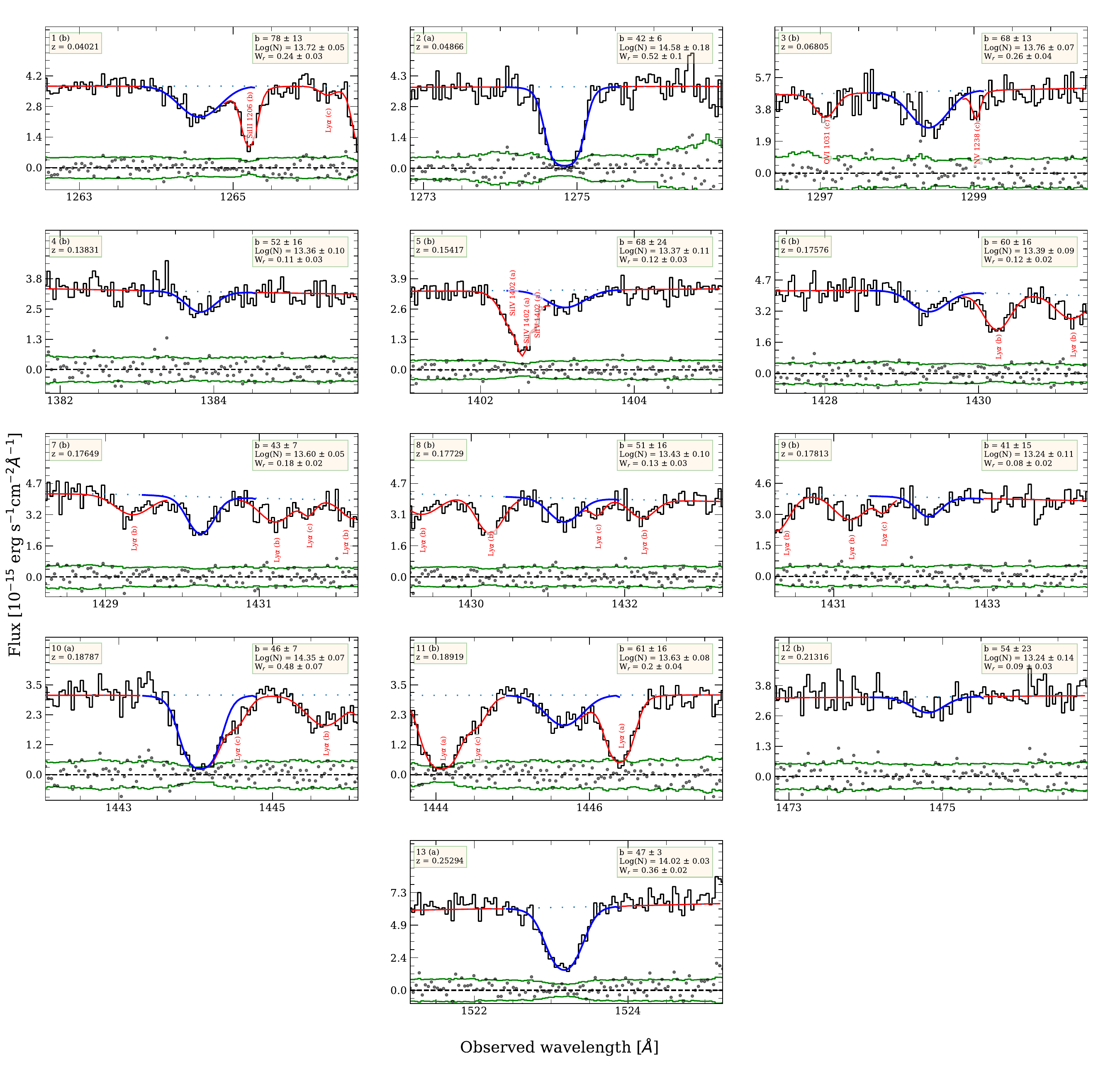}
\caption{\label{fig:BLAs_sample} Identified BLAs in the HST/COS spectrum of SDSSJ161940.56+254323.0. The panels show the best-fitting Voigt profile for each BLA in blue and the full model of the QSO spectrum in red. The dotted line shows the continuum level. Additional modeled transitions in the same wavelength range are labeled in red. The derived observational parameters column density log($N$/cm$^{-2}$), Doppler parameter in \kms, equivalent width in \AA, and redshift are indicated in each panel, together with the reliability of the identification (see Sec.~\ref{sec:blas_survey} for details). For saturated lines, our method leads to unrealistically large $W_r$ uncertainties due to the poor column density constraint.}     
\end{figure*}

From this analysis, we obtained a sample of $13$ identified reliable BLAs for which we measured their redshift, neutral hydrogen column density \nhi, and Doppler parameter \bhi. In this work, we identify as BLA any hydrogen absorption with an observed Doppler parameter larger than $40$\,\kms\ \citep[see, e.g.,][]{Lehner2007, Danforth2010, Danforth2016}, however, we acknowledge that due to the presence of non-thermal broadening, a $b>40$\,\kms\ does not necessarily always imply gas at $T \gtrsim 10^{5}$\,K \citep[see, e.g.,][]{Sameer2024}. These BLAs are shown in Fig.~\ref{fig:BLAs_sample}. When characterizing absorption lines, there is always the possibility of introducing biases and/or systematics. We discuss possible caveats in the identification of our BLA sample in Section~\ref{sec:blas_caveats}.

From the Voigt profile fit parameters of these BLAs ($z$, \nhi, \bhi), we have then followed \citet{Pessa2018} to infer the temperature ($T$), ionization fraction ($f_{\mathrm{ion}}$) and ionized gas column density ($N_{\mathrm{HII}}$). In summary, we split the observed Doppler parameter ($b_{\rm obs}$) into two different components, thermal ($b_{\rm th}$) and non-thermal ($b_{\rm non-th}$),

\begin{equation}
b_{\rm obs} = \sqrt[]{b_{\rm th}^{2} + b_{\rm non-th}^{2}}
\end{equation}

\noindent The thermal broadening only depends on the temperature ($T$) of the gas,

\begin{equation}
b_{\rm th}=\sqrt[]{\frac{2k_{b}T}{m}} \approx0.129\sqrt[]{\frac{T}{A}}\,\mathrm{km \, s}^{-1}
\label{eq:bth}
\end{equation}

\noindent where $k_{b}$ is the Boltzmann constant, $m$ is the gas particle mass and $A$ is the atomic weight of the element. For \hi\ equation~\ref{eq:bth} follows \citep[e.g.,][]{Richter2006b}:

\begin{equation}
T \approx 60 \left(\frac{b_{\rm th}}{\mathrm{km}\,  \mathrm{s}^{-1}}\right)^{2}\, \mathrm{K}
\label{eq:temp}
\end{equation}

On the other hand, non-thermal broadening mechanisms include turbulence, line blending, etc. In overdensities like galaxy filaments we expect that turbulence dominates \citep{Dolag2008}, and we parametrize it as being proportional to the thermal broadening, such that $b_{non-th} \approx b_{turb} \approx \alpha b_{th}$. This would imply: 
\begin{equation}
b^2_{\rm obs} \approx b_{\rm th}^2 + b_{\rm turb}^2 \approx b_{\rm th}^2 (1 + \alpha^2)
\label{eq:btot}
\end{equation}

In principle, to calculate the value of $\alpha$ for a BLA, it is needed an aligned metal component to separate the thermal and non-thermal broadening components \citep[see e.g.,][]{Savage2014, Stocke2014}. This relies upon the underlying assumption that each absorption component maps a spatially isolated `cloud' structure that has well-defined properties. However, \citet{Marra2024} uses hydrodynamical cosmological simulations of two $z = 1$ galaxies and synthetic quasar absorption-lines spectra of their CGM, to track the location of the gas associated with a given absorption component, and find that generally, a single absorption component is produced by gas on multiple different locations along the line of sight, due to the complex line of sight velocity distribution of the CGM. This means that even when aligned metal components are available, the inferred value of $\alpha$ is highly uncertain. Nevertheless, in order to provide a first-order empirical estimation, we calculate the $\alpha$ value for the absorption components of aligned\footnote{$\Delta v < 10$\,\kms\ between both components} \hi\ and O~{\sc vi} absorbers reported by \citet{Savage2014} and \citet{Stocke2014} (for those systems where $b_{\mathrm{H}~{\rm I}} > b_{\mathrm{O}~{\rm VI}}$). For the \citet{Savage2014} sample, we calculate a median $\alpha = 0.7$, with a dispersion of 0.8. For the \citet{Stocke2014}, we calculate $\alpha = 0.6\pm0.3$. If we consider only those absorptions with $b_{\mathrm{H}~{\rm I}} \geq 40$, we obtain $\alpha\ = 0.6\pm0.4$ and $\alpha = 0.5\pm0.2$, respectively, with most of the values within the $0-1.5$ range. For the following calculations, we choose a fiducial value of $\alpha = 0.7$, but we quantify the systematic uncertainty introduced by this assumption by also evaluating our results for $\alpha = 0$ and $\alpha = 1.5$, which encompasses most of the different measurements and also corresponds to the fiducial value plus the largest dispersion found among the different samples (0.8). We acknowledge that while this choice is likely not accurate for individual absorption lines, it should be appropriate for statistical inferences (as we do here), and an accurate measurement of this quantity is hard to obtain with the current observational methods \citep{Marra2024}. Moreover, we note that the additional uncertainty introduced by this assumption is generally comparable to our statistical uncertainties (see Sec.~\ref{sec:correlation_galaxies_BLA_gas}).

Given an $\alpha$ value, we can then estimate the temperature of the gas using equations~\ref{eq:temp} and ~\ref{eq:btot}. The ionized gas column density is then given by the ionization fraction of the gas, i.e., the number of ionized hydrogen per neutral ones, defined as:

\begin{equation}
f_{\rm ion}\equiv\frac{N_{\rm HI}+N_{\rm HII}}{N_{\rm HI}}\approx \frac{N_{\rm HII}}{N_{\rm HI}}
\label{eq:fion}
\end{equation}

\noindent and the combined UV-background photoionization plus collisional ionization model from \cite{Richter2006a} suggests a linear relation between $\log(f_{\rm ion})$ and $\log(T)$, such that:

\begin{equation}
\mathrm{log}(f_{\rm ion}) \approx -0.75+1.25\mathrm{log}\left(\frac{T}{\rm K}\right)
\label{eq:fion_richter}
\end{equation}

We finally calculate the ionized gas column density $N_{\rm H} \approx N_{\rm HII}$ for each individual BLA simply as $N_{\rm H} \approx f_{\rm ion} N_{\rm HI}$. Table~\ref{tab:blas_params} summarizes the physical properties of the gas inferred from each one of the BLAs in our sample. While in this paper we focus primarily on BLAs, for completeness and comparison purposes we also proceed in the same manner with the detected NLAs. Table~\ref{tab:nlas_params} summarizes the physical properties of the gas derived for the NLAs. In total, ten NLAs (and zero BLAs) fall in the low-reliability category `c'. These transitions are excluded from further analyses in the following sections. Note that equation~\ref{eq:fion_richter} is defined in \cite{Richter2006a} for a temperature range of $4 < \log(T/\mathrm{K}) < 7$, where low-temperature absorption features are primarily photoionized, rather than collisionally ionized. Since most of the NLAs listed in Table~\ref{tab:nlas_params} exhibit temperatures $\log(T/\mathrm{K}) > 4$ (even in the case of $\alpha = 1.5$, especially those with reliabilities `a' and `b'), we find that using the same model to derive the ionization fraction of the NLAs is appropriate and avoids introducing inhomogeneities into our analyses. However, for the single NLA with a reliability higher than `c' that shows a temperature lower than $10^{4}$ K (ID 2), this calculation represents an extrapolation of the model, and we caution the reader about this caveat (this particular NLA has no impact in any of the conclusions of this paper).

\setlength{\tabcolsep}{1.5pt}
\begin{table*}
\centering
\begin{tabular}{cccccccccc}
\hline

ID & $z$ & $b$/km s$^{-1}$ & log($N_{\mathrm{HI}}$ / cm$^{-2}$) & log($N_{\mathrm{HII}}$ / cm$^{-2}$) & $\Delta\,\mathrm{log(N}_{\mathrm{HII}}$/ cm$^{-2}$) & $\Delta\log (T$/K) & $\Delta\log f_{\mathrm{ion}}$ & Reliability & Metals\\
(1) & (2) & (3) & (4) & (5) & (6) & (7) & (8) & (9) & (10) \\
\hline
1 & 0.04021 & 78$\pm$13 & 13.72$\pm$0.05 & 19.71$\pm$0.18 & 19.93 - 19.29 & 5.6 - 5.1 & 6.21 - 5.57 & b& -  \\
2 & 0.04866 & 42$\pm$6 & 14.58$\pm$0.18 & 19.90$\pm$0.24 & 20.11 - 19.47 & 5.0 - 4.5  & 5.53 - 4.89 & a & \ion{Si}{III}, \ion{Si}{IV}, \ion{C}{IV}  \\
3 & 0.06805 & 68$\pm$13 & 13.76$\pm$0.07 & 19.60$\pm$0.22 & 19.82 - 19.18 &  5.4 - 4.9  & 6.06 - 5.42 & b & -  \\
4 & 0.13831 & 52$\pm$16 & 13.36$\pm$0.10 & 18.90$\pm$0.36 & 19.11 - 18.47 &  5.2 - 4.7  & 5.75 - 5.11 & b & -  \\
5 & 0.15417 & 68$\pm$24 & 13.37$\pm$0.11 & 19.20$\pm$0.40 & 19.42 - 18.78 & 5.4 - 4.9 & 6.05 - 5.41 & b & - \\
6 & 0.17576 & 60$\pm$16 & 13.39$\pm$0.09 & 19.09$\pm$0.30 & 19.31 - 18.67 &  5.3 - 4.8  & 5.92 - 5.28 & b & -  \\
7 & 0.17649 & 43$\pm$7 & 13.60$\pm$0.05 & 18.93$\pm$0.18 & 19.15 - 18.51  &  5.0 - 4.5  & 5.55 - 4.91 & b & -  \\
8 & 0.17729 & 51$\pm$16 & 13.43$\pm$0.10 & 18.97$\pm$0.36 & 19.18 - 18.54 &  5.2 - 4.7 & 5.75 - 5.11 & b & -  \\
9 & 0.17813 & 41$\pm$15 & 13.24$\pm$0.11 & 18.54$\pm$0.41 & 18.75 - 18.11 & 5.0 - 4.5  & 5.51 - 4.87 & b & - \\
10 & 0.18787 & 46$\pm$7 & 14.35$\pm$0.07 & 19.75$\pm$0.17 & 19.97 - 19.33 &  5.1 - 4.6  & 5.62 - 4.98 & a & \ion{O}{VI}  \\
11 & 0.18919 & 61$\pm$16 & 13.63$\pm$0.08 & 19.35$\pm$0.30 & 19.57 - 18.93 &  5.4 - 4.8 & 5.94 - 5.30 & b & -  \\
12 & 0.21316 & 54$\pm$23 & 13.24$\pm$0.14 & 18.82$\pm$0.48 & 19.03 - 18.39 & 5.2 - 4.7 & 5.79 - 5.15 & b & -  \\
13 & 0.25294 & 47$\pm$3 & 14.02$\pm$0.03 & 19.47$\pm$0.08 & 19.68 - 19.04 &  5.1 - 4.6 & 5.66 - 5.02 & a & -  \\
\hline
\end{tabular}
\caption{\label{tab:blas_params} Physical properties of the gas inferred from each one of the BLAs in our sample, as described in Sec.~\ref{sec:blas_survey}. Column (9) shows the reliability of the Voigt profile fitting, and column (10) indicates the metal transitions found at the same redshift of each BLA. The ionized hydrogen column densities in column (5) have been calculated using a constant $\alpha$ (turbulent-to-thermal Doppler contribution ratio) of 0.7. The uncertainties in column (5) reflect only the statistical uncertainties propagated from the line fitting. However, the inferred ionized hydrogen column densities are subject to systematic uncertainties due to the assumptions made to compute them. Columns (6), (7), and (8) show the range of variation of the determined log($N_{\mathrm{HII}}$), log($T$), and log $f_{\mathrm{ion}}$, respectively, when using $\alpha$ values in the range [$\alpha_{\rm low}$ -  $\alpha_{\rm high}$], with $\alpha_{\rm low}$ = 0, and  $\alpha_{\rm high}$ = 1.5, as detailed in Sec.~\ref{sec:blas_survey}.}
\end{table*}
\setlength{\tabcolsep}{6pt}

\setlength{\tabcolsep}{3pt}
\begin{table*}
\centering
\begin{tabular}{ccccccccc}
\hline
ID & $z$ & $b$/km s$^{-1}$ & log($N_{\mathrm{HI}}$ / cm$^{-2}$) & log($N_{\mathrm{HII}}$ / cm$^{-2}$) & $\Delta\,\mathrm{log(N}_{\mathrm{HII}}$/ cm$^{-2}$) & $\Delta\log (T$/K) & $\Delta\log f_{\mathrm{ion}}$ & Reliability \\
(1) & (2) & (3) & (4) & (5) & (6) & (7) & (8) & (9) \\
\hline
1 & 0.00869 & 35$\pm$24 & 12.97$\pm$0.19 & 18.08$\pm$0.77 & 18.29 - 17.65 & 4.9 - 4.3 & 5.32 - 4.68 & c \\
2 & 0.01702 & 13$\pm$16 & 12.74$\pm$0.21 & 16.79$\pm$1.31 & 17.01 - 16.37 & 4.0 - 3.5 & 4.27 - 3.63 & a \\
3 & 0.01982 & 13$\pm$18 & 12.67$\pm$0.24 & 16.70$\pm$1.56 & 16.92 - 16.28 & 4.0 - 3.5 & 4.25 - 3.61 & c \\
4 & 0.02893 & 26$\pm$3 & 14.08$\pm$0.08 & 18.88$\pm$0.15 & 19.09 - 18.45 & 4.6 - 4.1 & 5.01 - 4.37 & b \\
5 & 0.04157 & 28$\pm$32 & 12.70$\pm$0.31 & 17.56$\pm$1.31 & 17.78 - 17.14 & 4.7 - 4.2 & 5.08 - 4.44 & c \\
6 & 0.04197 & 29$\pm$4 & 14.17$\pm$0.09 & 19.09$\pm$0.16 & 19.31 - 18.67 & 4.7 - 4.2 & 5.14 - 4.50 & b \\
7 & 0.04246 & 30$\pm$4 & 14.33$\pm$0.13 & 19.27$\pm$0.21 & 19.48 - 18.84 & 4.7 - 4.2 & 5.15 - 4.51 & b \\
8 & 0.08051 & 20$\pm$15 & 12.93$\pm$0.18 & 17.43$\pm$0.84 & 17.64 - 17.00 & 4.4 - 3.9 & 4.71 - 4.07 & c \\
9 & 0.12490 & 25$\pm$9 & 16.91$\pm$2.14 & 21.65$\pm$2.18 & 21.86 - 21.22 & 4.6 - 4.1 & 4.95 - 4.31 & a \\
10 & 0.12506 & 19$\pm$53 & 17.06$\pm$2.24 & 21.52$\pm$3.74 & 21.74 - 21.10 & 4.3 - 3.8 & 4.68 - 4.04 & a \\
11 & 0.12529 & 32$\pm$29 & 14.79$\pm$1.30 & 19.80$\pm$1.64 & 20.02 - 19.38 & 4.8 - 4.3 & 5.23 - 4.59 & a \\
12 & 0.14246 & 26$\pm$16 & 12.94$\pm$0.17 & 17.74$\pm$0.69 & 17.95 - 17.31 & 4.6 - 4.1 & 5.01 - 4.37 & c \\
13 & 0.14850 & 27$\pm$6 & 13.42$\pm$0.06 & 18.26$\pm$0.25 & 18.47 - 17.83 & 4.6 - 4.1 & 5.05 - 4.41 & b \\
14 & 0.17764 & 20$\pm$18 & 12.90$\pm$0.26 & 17.43$\pm$1.01 & 17.65 - 17.01 & 4.4 - 3.9 & 4.75 - 4.11 & c \\
15 & 0.18494 & 32$\pm$22 & 13.00$\pm$0.19 & 18.04$\pm$0.77 & 18.25 - 17.61 & 4.8 - 4.3 & 5.25 - 4.61 & c \\
16 & 0.18824 & 35$\pm$22 & 13.38$\pm$0.28 & 18.48$\pm$0.75 & 18.70 - 18.06 & 4.9 - 4.3 & 5.32 - 4.68 & c \\
17 & 0.18978 & 36$\pm$5 & 14.10$\pm$0.07 & 19.25$\pm$0.17 & 19.47 - 18.83 & 4.9 - 4.4 & 5.37 - 4.73 & a \\
18 & 0.19203 & 29$\pm$30 & 13.10$\pm$0.31 & 18.00$\pm$1.17 & 18.22 - 17.58 & 4.7 - 4.2 & 5.12 - 4.48 & c \\
19 & 0.24255 & 21$\pm$7 & 13.27$\pm$0.08 & 17.83$\pm$0.36 & 18.05 - 17.41 & 4.4 - 3.9 & 4.78 - 4.14 & b \\
20 & 0.24922 & 8$\pm$9 & 12.96$\pm$0.16 & 16.45$\pm$1.19 & 16.67 - 16.03 & 3.6 - 3.1 & 3.71 - 3.07 & c \\
21 & 0.25696 & 28$\pm$4 & 13.94$\pm$0.05 & 18.82$\pm$0.17 & 19.04 - 18.40 & 4.7 - 4.2 & 5.10 - 4.46 & a \\
22 & 0.25728 & 32$\pm$4 & 14.04$\pm$0.04 & 19.07$\pm$0.14 & 19.29 - 18.65 & 4.8 - 4.3 & 5.25 - 4.61 & b \\
\hline
\end{tabular}
\caption{\label{tab:nlas_params} Same as Table~\ref{tab:blas_params}, for the NLAs ($b < 40$\,\kms) detected in the QSO sightline (except for the last column, which is not included since we do not discuss NLAs case-by-case, as for BLAs).}
\end{table*}
\setlength{\tabcolsep}{6pt}

\subsection{Potential caveats in the identification of BLAs}
\label{sec:blas_caveats}

As explained in the previous section, we identify as a BLA any hydrogen absorption with Doppler parameter larger than 40\,\kms, as this generally implies gas temperature higher than $10^{5}$\,K \citep{Richter2006a, Richter2006b}. However, noise and line blends can potentially mimic broad and shallow absorption features \citep{Richter2006b, Garzilli2015, Tejos2016}. The impact of this effect depends on the S/N of the data, and eventually higher S/N enables a better characterization of the kinematic structure of the absorption features \citep[e.g.,][]{Richter2006b, Danforth2010}. This limitation is intrinsic to the absorption-line technique and affects, to some extent, any absorption-base study. In order to minimize the possible impact of blends, we introduce an additional component for absorption features that showed asymmetric profiles (e.g., ID 10 in Table~\ref{tab:blas_params}, see also Fig.~\ref{fig:BLAs_sample}).

Additionally, since constraining $b$ and $N$ becomes degenerate for saturated lines ($N_{\mathrm{\ion{H}{I}}} \gtrsim 10^{14}$\,cm$^{-2}$), we present here a closer examination of our fits for the saturated BLAs in our sample (IDs 2, 10, and 13 in Table~\ref{tab:blas_params} and Fig.~\ref{fig:BLAs_sample}), as well as the fits of their associated metals and/or other\,\hi\, absorption found at the same redshift, if any (figures are provided in Appendix~\ref{sec:appendix_plots_SaturatedBLAs_metals}). We also discuss possible alternative best-fitting solutions for some of the BLAs in our sample (IDs 1, 3, 5, 6).

The BLA \#1 ($z\sim0.04021$) could be alternatively modeled as three separate narrow components. Naturally, this could potentially reproduce the data better than our one-component fitting. Thus, to choose the best model, while avoiding overfitting, we use the Bayesian information criterion (BIC), which introduces an additional penalty term for the number of parameters in the model \citep{BIC} \footnote{The BIC is also sensitive to the improvement of the fitting, and the goodness of fit will generally increase when more parameters are added to the model. In that sense, in principle, an arbitrarily high number of components could be added to fit any given BLA; thus, here we use the BIC to determine up to which point the data support a more complex model.}. For this BLA, the BIC is significantly lower for the one-component fitting, and thus, a one-component model is the preferred solution.

The BLA \#2 ($z\sim0.04866$), shown in Fig.~\ref{fig:blas_metals_2}, exhibits \ion{Si}{III}, \ion{Si}{IV} and \ion{C}{IV} absorption at the same velocity, none of which show any sign of asymmetry, therefore we find it unlikely that this BLA is an artifact produced by the blending of narrower absorptions. Unfortunately, Ly$\beta$ is not available at this redshift, however, fitting the absorption profile with a narrower Doppler parameter (1-$\sigma$ lower than best-fitting value) produces a qualitatively worse fit that does not reproduce the spectral profile of the BLA. Thus, we conclude that the identification of this absorption as BLA is not a result of an unconstrained Doppler parameter.

The BLA \#3 ($z\sim0.06805$) could also be modeled as two separate narrower components. However, the measured BIC value is significantly lower for the one-component model, and thus, a one-component model is the preferred solution.

The BLA \#5 ($z\sim0.15417$) is blended with a multi-component Galactic \ion{Si}{IV} absorption feature, which makes it difficult to infer its Doppler parameter confidently. The BLA fitting depends on the modeling of the Galactic absorption. We report here the best-fitting value obtained after carefully modeling the Galactic \ion{Si}{IV} 1402 absorption, using three \ion{Si}{IV} components. Using fewer components could lead to a broader BLA. However, the three-component fitting is supported by additional Galactic transitions present in the QSO spectrum (e.g., \ion{Si}{IV} 1393, \ion{C}{IV}). Since in any case, this absorption would still qualify as a BLA, and we find that three components are the most consistent model with what is seen in the HST data, we keep it in our sample.

The BLA \#6 ($z\sim0.17576$) could be alternatively modeled as two separate narrow components. However, similar to BLA \#3, the BIC is significantly lower for the one-component fitting, and thus, we keep a one-component model as the preferred solution.

The BLA \#10 ($z\sim0.18787$) is shown in Fig.~\ref{fig:blas_metals_10}. This \hi\ absorption feature presents an asymmetry that we model with an additional narrower component. Ly$\beta$ and Ly$\gamma$ coverage are available at this redshift, and the spectra show consistency with absorption (although the latter is extremely shallow), further constraining the Doppler parameter and column density of the BLA. The detection of a \ion{O}{VI} absorption is also consistent with the presence of warm-hot gas at this redshift. Overall, we believe that although there is a blend, the detection of this BLA is robust, and the parameters of the Voigt profile are well-constrained. Indeed, we find that fitting the BLA with a $1$-$\sigma$ narrower $b$ produces a model that is not consistent with the observed absorption profile.

The BLA \#13 ($z\sim0.25294$) is shown in Fig.~\ref{fig:blas_metals_13}. This BLA does not present any sign of asymmetry in its spectral profile. Ly$\beta$ and Ly$\gamma$ are also available at this redshift, although similar to the BLA \#10, the latter is very shallow (Ly$\beta$, on the other hand, represents a more clear detection). We also find that fitting the BLA with a $1$-$\sigma$ narrower $b$ produces a model that is not consistent with the data. Altogether, we find that this BLA is a robust identification and that the parameters of the Voigt profile are well-constrained.

\section{Results}
\label{sec:results}
\subsection{Cross-correlating the presence of BLAs with overdensities of galaxies}
\label{sec:overdensity_blas}

\begin{figure*}
        \begin{minipage}[b]{\columnwidth}
                \includegraphics[width=2\columnwidth]{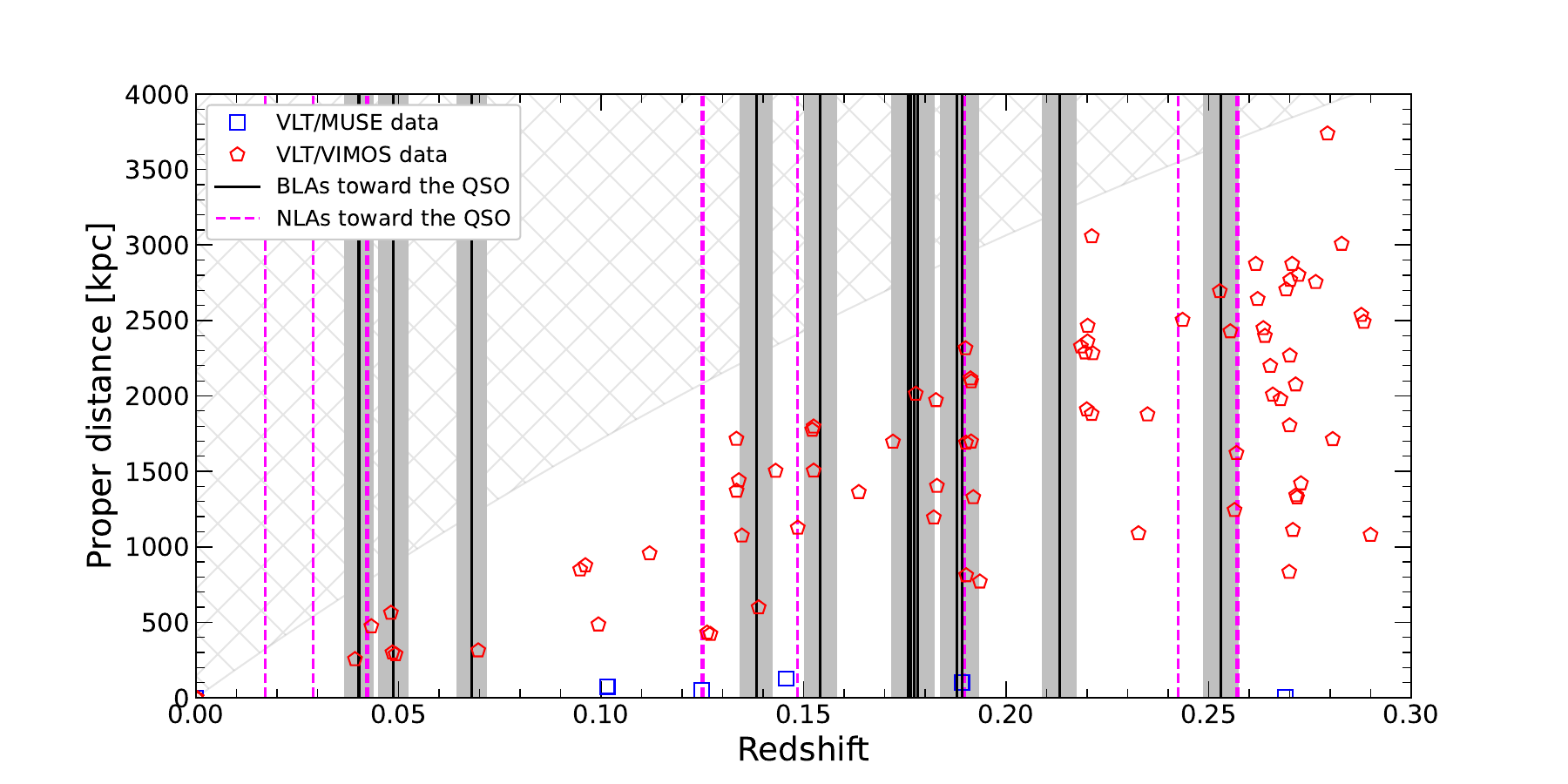}

        \end{minipage}%
        \\
        \begin{minipage}[b]{\columnwidth}
                \includegraphics[width=2\columnwidth]{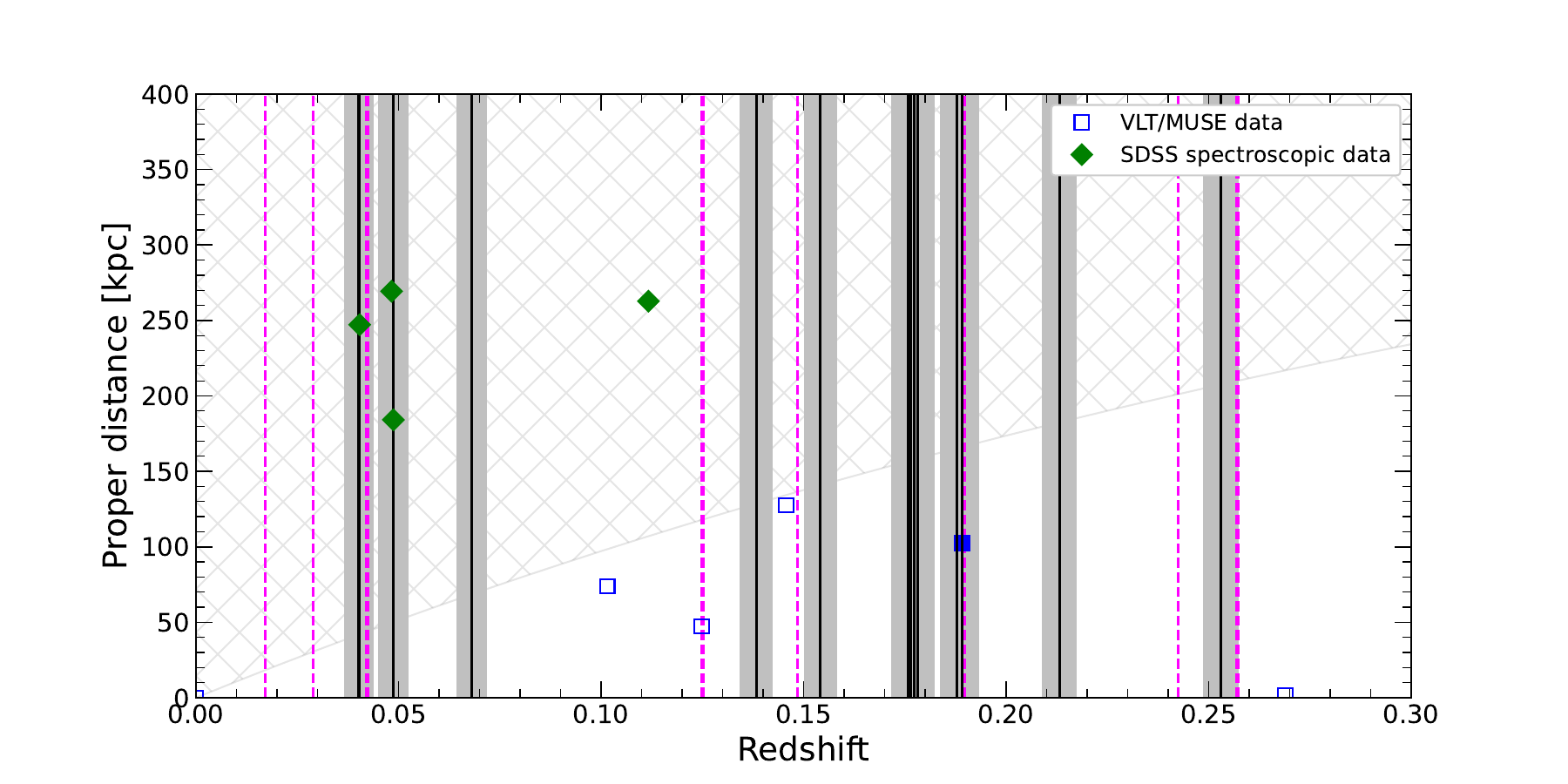}

        \end{minipage}%
        \caption{\label{fig:muse_vimos_ipz} Impact parameter to the QSO sightline as a function of the redshift of each source in our survey. \textit{Top:} Sources up to a projected distance of 4 Mpc from the QSO sightline. Objects in our VLT/MUSE and VLT/VIMOS surveys are represented with blue squares and red pentagons, respectively. The vertical black solid lines and their background grey area mark the redshifts of the reported BLAs $\pm 1000$ \kms. The hatched area shows a 15 arcminute distance from the central QSO, which would correspond to approximately the edges of the VLT/VIMOS field of view at a given redshift. The vertical dashed magenta lines indicate the redshift of the NLAs detected in the spectrum of the QSO. \textit{Bottom:} Sources closer to the QSO sightlines in our VLT/MUSE survey, at impact parameters up to $\sim150$ kpc (blue squares). The filled blue square represents a galaxy found to be very close in velocity (within $\pm1000$ \kms) to the BLAs at $z\approx0.18787$ and $z\approx0.18919$. The hatched area shows the approximate edges of the VLT/MUSE field of view. Additionally, we include data from the SDSS DR16 (green diamonds), in order to look for nearby galaxies that we could be missing due to our lack of impact parameter coverage at low redshift. }
\end{figure*}

\begin{figure*}
\centering
                \includegraphics[width=2\columnwidth]{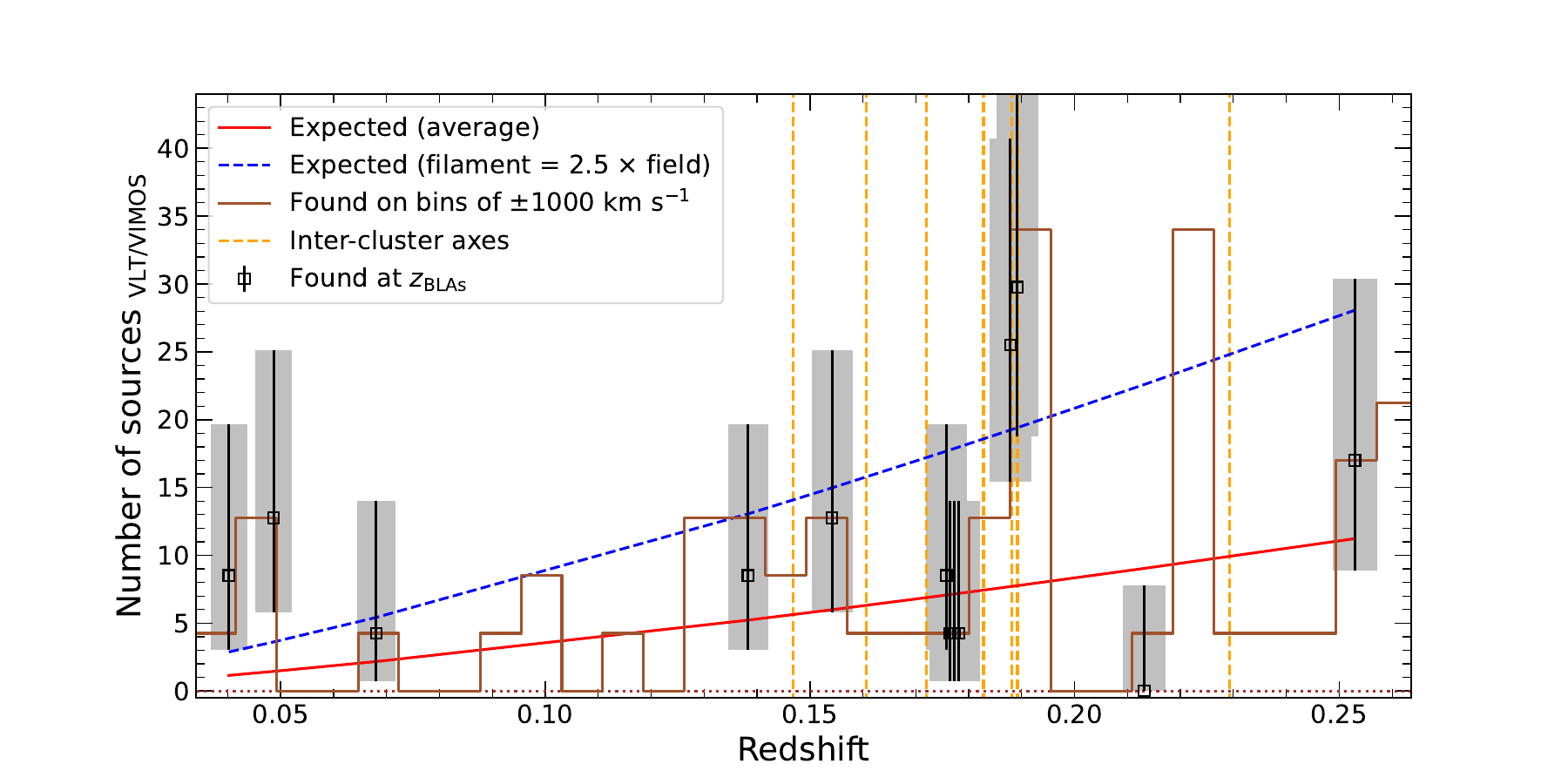}
\caption{\label{fig:bla_excess} Comparison between the number of galaxies randomly expected at the redshift of the BLAs reported in Section~\ref{sec:blas_survey}, within a velocity window of $\pm1000$ \kms, inside the physical area enclosed by the total VLT/VIMOS FoV at a given redshfit, with the number of galaxies actually `found' at that redshift range (corrected by the completeness effects described in Section~\ref{sect:vimos_compl}). The black squares represent this corrected number of detected galaxies with their respective Poissonian uncertainty at the redshift of the BLAs (reliability `a' or `b', see Sec.~\ref{sec:blas_survey}). The Poissonian error bars have been calculated following \citet{Gehrels1986}. The width of the grey area marks the $\pm$1000 \kms\ region around the BLAs. The brown solid line shows the completeness-corrected number of galaxies found in bins of $\pm1000$ \kms, regardless of the presence of a BLA. The brown dotted line shows the zero level. 
The red line represents the randomly expected number of galaxies and the blue dashed line represents the number of galaxies expected in an environment $2.5$ times denser, such as the filaments. At the redshift of 4 of the BLAs ($\approx0.04021, \approx0.04866, \approx 0.18787$,  $\approx 0.18919$), we find a clear excess of galaxies, consistent with an overdense region. The vertical orange dashed lines mark the redshift at which we find a galaxy cluster pair whose inter-cluster axis passes near the LoS of the QSO (see Sec.~\ref{sec:gal_clust}).}     
\end{figure*}

In this section, we look for overdensities of galaxies in our galaxy survey, at the redshifts of the BLAs reported in Sec.~\ref{sec:blas_survey}. These overdensities would suggest that there is a large-scale structure, and therefore, a positive correlation would mean that the BLAs may be preferentially tracing the large-scale filamentary structure of the Universe \citep[the much larger volume fraction occupied by cosmic filaments, compared by cosmic nodes make this the most likely scenario, see, e.g.,][]{Cautun2014}. We compare the number of galaxies detected at the redshift of each BLA with the field expectation to quantify the strength of the galaxy overdensity.

The top panel of Fig.~\ref{fig:muse_vimos_ipz} shows the impact parameter to the QSO sightline as a function of the redshift for the sources in our VLT/MUSE and VLT/VIMOS data. The VLT/VIMOS survey reaches larger impact parameters than that of VLT/MUSE because of its much larger field of view. However, VLT/MUSE is much more complete at impact parameters close to the QSO sightline. Thus, we use our VLT/VIMOS survey to study the large-scale distribution of galaxies, and our VLT/MUSE survey to look for galaxy halos closer to the QSO sightline that could be associated with some of the absorption features detected. The figure also shows the redshift of the reported BLAs (and NLAs). We find at least one galaxy close to the redshift of every BLA, except at $z\approx0.21316$ where we did not find any nearby galaxy. To quantify the significance of this number of galaxies, and determine if it does represent an excess with respect to the field expectation, consistent with a large-scale structure, we use the code presented in \cite{Reed2007} to study the Halo Mass Function (HMF) at low redshift. This code uses cosmological parameters as an input to generate a modeled HMF. In particular, it generates the number of haloes more massive than an arbitrary mass value expected per cubic Mpc h$^{-1}$, that is, the number density of haloes per mass bin. Comparing this randomly expected number of galaxies with the number of galaxies `found' (corrected by the two incompleteness effects mentioned in Sec.~\ref{sect:vimos_compl}) within $\pm1000$\,\kms\ from the redshift of the BLAs reported in Sec.~\ref{sec:blas_survey}, permit us to know how significant is this number of found galaxies and if it does actually represent an overdensity, where the size of the velocity window determines the depth of the cosmic volume spanned by the VLT/VIMOS FoV. In Sec.~\ref{sec:overdensity_larger_scales}, we explore the impact of using a different velocity limit on the inferred density of galaxies.

Figure~\ref{fig:bla_excess} shows this comparison. The black squares represent the corrected number of galaxies `found' at the redshift of each BLA with its Poissonian error. Recall that the number of galaxies found in our VLT/VIMOS survey at a given redshift is corrected by the completeness of our survey, as detailed in Sec~\ref{sect:vimos_compl}, thus, the number of sources in the $y$-axis of Fig.~\ref{fig:bla_excess} is larger than the number of red symbols shown in the top panel of Fig.~\ref{fig:muse_vimos_ipz}.
The red line represents the number of galaxies randomly expected to be found in the VLT/VIMOS field of view within $\pm1000$\,\kms\ of a given redshift. This value comes from the modeled HMF mentioned earlier, and corresponds to the number of haloes more massive than a given mass lower limit expected in the VLT/VIMOS field of view. We used as mass lower limit the stellar mass that corresponds to our detection threshold of magnitude $r = 22.5$ mag (at any given redshift), and assuming a $\frac{M*}{M_{\odot}} \propto \frac{L}{L_{\odot}}$ relation \citep[see, e.g.,][]{Bell2001, Kauffmann2003, Schombert2022}. We use then the relation provided by \citet{Moster2010} to transform this stellar mass into a threshold halo mass, that is, the lower integration limit for the HMF. 

According to \cite{GonzalezPadilla2010}, large-scale filamentary structures are $2-3$ times denser than the average density of the Universe. Therefore, the blue dashed line corresponds to the expected number of galaxies found in an overdense region, $2.5$ $\times$ denser than average density. Thus, at the redshift of 4 of the BLAs, we found a tentative excess of galaxies, consistent with an overdense region: the BLAs at $z\approx0.04021$, $z\approx0.04866$, and the two BLAs at (average) $z\approx0.18853$. We also found a milder excess of galaxies at the BLAs at $z\approx0.06805$, $z\approx0.13831$, $z\approx0.15417$, and $z\approx0.25294$, although for these redshifts, the number of galaxies found is also consistent with the field expectation given the large uncertainties. 

By randomly selecting $2000$ sets of 13 redshifts in the range 0-$z_{\mathrm{QSO}}$, we have determined that given the redshift distribution of our VLT/VIMOS survey, the probability of, by chance, coinciding with at least 4 overdense regions (above our threshold of 2.5 times the expected number of galaxies), and 4 mildly overdense regions (above the expected density) is of about $\sim3\%$, meaning that the BLAs are indeed found preferentially at redshift that exhibit an excess of galaxies. This also suggests that our BLA sample is likely not dominated by noise features (see Sec.~\ref{sec:blas_caveats} for a more comprehensive discussion about the caveats on the BLA identification).

\subsection{Galaxies close to the QSO sightline at the redshift of BLAs}
\label{sec:close_galaxies}
In Sec.~\ref{sec:overdensity_blas} we studied the significance of having observed a certain number of galaxies at the redshift of each BLA, and we found that for the BLAs at $z\approx0.04021$, $z\approx0.04866$, $z\approx0.18787$ and $z\approx0.18919$, the number of galaxies observed is consistent with an overdense region of the Universe. Here, we study the distribution of galaxies nearby to the QSO sightline, in order to get more insights about the possible origin of these BLAs. 

The bottom panel of Fig.~\ref{fig:muse_vimos_ipz} shows the impact parameter to the QSO sightline of nearby galaxies as a function of its redshift. The vertical lines mark the redshifts of the BLAs. The blue squares represent sources from our VLT/MUSE survey. The filled blue square represents a galaxy found to be very close in velocity (within $\pm1000$\,\kms) to the BLAs at $z\approx0.18787$ and $z\approx0.18919$. This galaxy corresponds to the object with ID $14$ in the Table~\ref{tab:muse_ids}. At $z\approx0.04021$ and $z\approx0.04866$ we have an impact parameter coverage of $\sim40$ kpc and $\sim50$ kpc respectively. This is small compared to the size suggested for the circum-galactic medium, on the order of a few hundred kpc \cite[e.g., ][]{Prochaska2011, Tumlinson2017}. Therefore, we have included SDSS data in our survey to look for nearby galaxies beyond the VLT/MUSE FoV.

The diamonds in the bottom panel of Fig.~\ref{fig:muse_vimos_ipz} represent these data from the SDSS DR16 spectroscopic catalog. We have additionally searched for intervening sources in the SDSS photometric redshift catalog, but the large uncertainties in their redshift measurement make it difficult to obtain an appropriate measurement of their impact parameter to the QSO sightline, and therefore, we do not include these sources in the figure. 

This search revealed one additional galaxy that lies close (in redshift and impact parameter) to the BLA at $z\approx0.04021$ and two more galaxies near the BLA at $z\approx0.04866$. However, no additional galaxies are found near any other BLA in our sample. In the following sections, we examine closely each one of the absorbing systems, together with their potential relation to galaxies near the QSO sightline. We stress that we use the SDSS data only to look for galaxies near the QSO sightline that could be directly associated with a BLA (same as our VLT/MUSE survey), and not to infer an overdensity of galaxies at a given redshift, due to the complex selection function of the SDSS spectroscopic survey. On the other hand, our understanding of the VLT/VIMOS survey completeness allows us to use our VLT/VIMOS survey to infer the completeness-corrected number of galaxies in a given cosmic volume. Thus, galaxies from the SDSS data set are only shown in the bottom panel of Fig.~\ref{fig:muse_vimos_ipz}.

\subsection{BLA - overdensity of galaxies at $z\approx0.04021$}
\label{sec:sys_at_004}

\begin{figure*}
        \begin{minipage}[b]{0.48\textwidth}
                \includegraphics[width=\columnwidth]{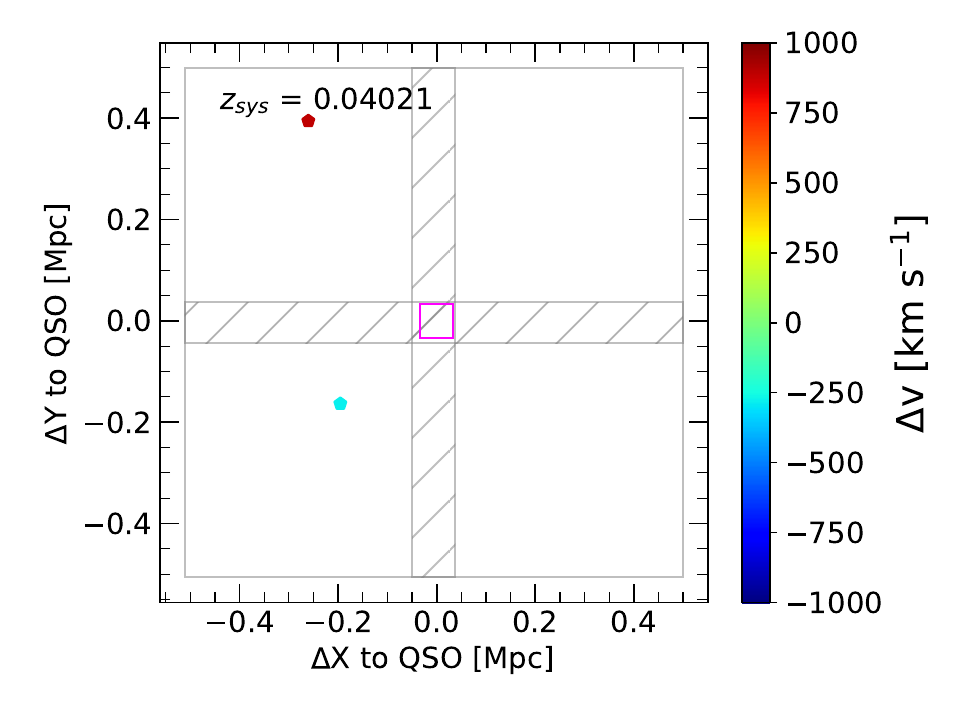}

        \end{minipage}%
        \begin{minipage}[b]{0.48\textwidth}
                \includegraphics[width=\columnwidth]{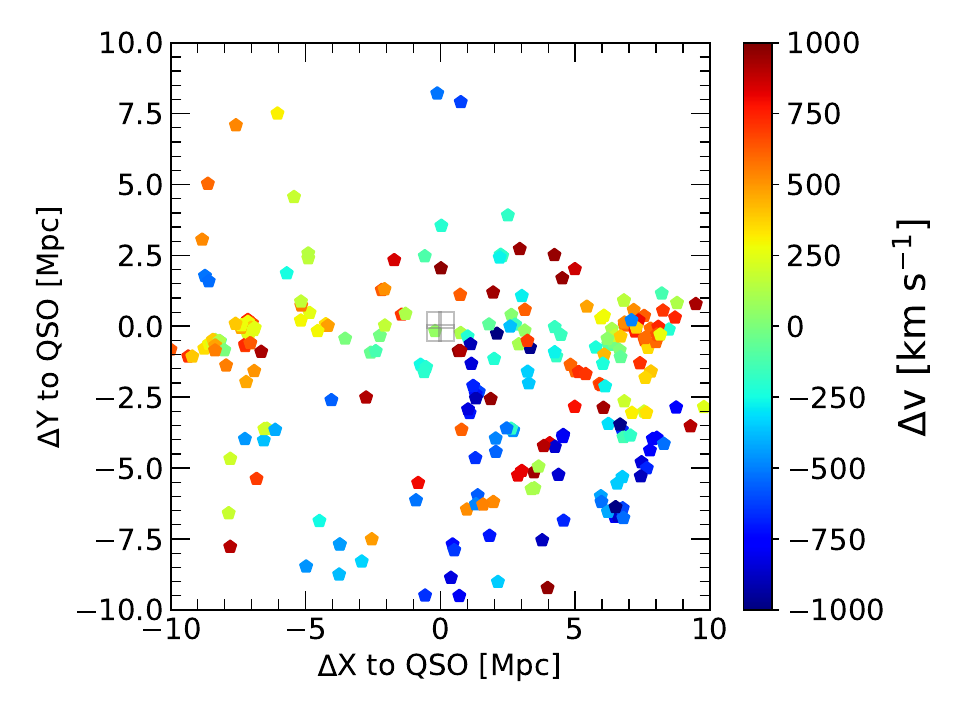}

        \end{minipage}%

        \caption{\label{fig:gal_dist_004}  Spatial distribution of galaxies with a redshift within $z_{\mathrm{sys}}\pm1000$ \kms, for the BLA at $z = 0.04021$. \textit{Left}: Full VLT/VIMOS FoV around SDSSJ161940.56+254323.0 at $z \approx 0.04021$. The dashed cross represents the separation between the four VLT/VIMOS quadrants, indicated in grey. The central magenta square represents the VLT/MUSE FoV. The VLT/VIMOS FoV spans a square of $\sim 1.0$ Mpc side at this redshift. The VLT/MUSE FoV is significantly smaller, reaching impact parameters of $\sim 40$ kpc from the QSO sightline. Two galaxies in our VLT/VIMOS survey lie within $1000$ \kms\ from the BLA.  \textit{Right}: Zoom-out of the left panel, using spectroscopic data from the SDSS DR16 survey. The separation between the VLT/VIMOS quadrants is kept for reference. The colorbar of both panels indicates the velocity offset of each galaxy with respect to $z_{\mathrm{sys}}$.}
\end{figure*}

Figure~\ref{fig:bla_excess} shows a clear excess of galaxies close to the BLA at $z\approx0.04021$, of nearly $\sim7$ times over the average number of galaxies (once completeness effects are accounted for). Figure~\ref{fig:gal_dist_004} shows the spatial distribution of these galaxies around the QSO sightline, with a redshift within a velocity window of $\pm1000$\,\kms\ from the BLA, for the galaxies in our VLT/VIMOS survey (left) and the galaxies found in the SDSS data with spectroscopic redshift (right). Note that while the SDSS spectroscopic data is helpful to investigate the environment around the QSO sightline on ever larger scales than the VLT/VIMOS data, its completeness is lower inside VLT/VIMOS FoV, where the existence of a large-scale cosmic filament would have a more substantial impact in the local density of galaxies. Furthermore, our understanding of the completeness of our VLT/VIMOS survey (see Sec.~\ref{sect:vimos_compl}) makes the latter suitable for evaluating whether there is an under- or an over-density of galaxies at the redshift of the BLAs, compared to the random expectation. Although the number of galaxies identified in our VIMOS survey might seem low (2), this number is large considering the relatively small FoV of VLT/VIMOS at this redshift ($\sim1.0$ Mpc$^{2}$) and the completeness fraction of our VLT/VIMOS survey. Recall that the lower luminosity limit at lower redshifts has been taken into account when computing the average number of galaxies expected. In addition to these two galaxies, we find one more additional galaxy near the QSO sightline, at an impact parameter of $\approx250$ kpc, nearly at the same redshift of the BLA in the SDSS spectroscopic data (see bottom panel of Fig.~\ref{fig:muse_vimos_ipz}). We then followed \citet{Taylor2011} to infer the stellar masses of these three galaxies from their SDSS photometry:

\begin{equation}
    \log M_{*} [\mathrm{M}_{\odot}] = 1.15 + 0.7 (g - i) - 0.4 M_{i}
\end{equation}
Where $M_{i}$ corresponds to the absolute magnitude in the restframe i-band. This relation can be used to estimate the stellar mass with a $\sim 0.10$ dex scatter \citep{Taylor2011}. We assumed the bijective relation between total halo mass and stellar mass presented in \citet{Moster2010} to estimate their virial masses ($M_{\mathrm{vir}}$). Hydrodynamical simulations show that the scatter in the halo-stellar mass varies as a function of redshift and halo mass, and it is expected to be of the order of $\sim0.1-0.2$ dex at $z = 0.1$ \citep{Matthee2017}.

Finally, we estimated their virial radii ($R_{200}$) and velocity dispersion ($\sigma_{\rm vir}$) following:

\begin{equation}
R_{200} = \bigg(\frac{M_{\rm vir}}{\frac{4}{3}\pi200\rho_{\rm c}(z)}\bigg)^{1/3}
\end{equation}

\noindent and

\begin{equation}
\sigma_{\rm vir} = \sqrt[]{\frac{GM_{\rm vir}}{R_{200}}}
\end{equation}

\noindent where $R_{200}$ is defined as the radius of the spherical volume where $M_{\rm vir}$ is contained at $200$ times the critical density of the Universe at a given redshift, $\rho_{\rm c}(z)$. Comparing the physical distance of these galaxies to the QSO sightline with $R_{200}$, and their velocity offset respect to the BLA ($\Delta v$) with $\sigma_{\rm vir}$, allow us to evaluate if a BLA could be produced by hot gas in the halo of a galaxy, as opposed to only be tracing the large-scale filamentary structure of the Universe \citep{Pessa2018}. Generally, if the product between $\Delta v$, in units of $\sigma_{\rm vir}$, and the impact parameter of a galaxy to the QSO sightline, in units of $R_{200}$ ($\eta \equiv \frac{\mathrm{i.p.}}{R_{200}}\frac{\Delta v}{\sigma_{\rm vir}}$) is $\gtrsim 2$, it could be indicative of a not gravitationally bounded system \citep{Shen2017}. Indeed, \citet{Wilde2023} find evidence that suggests that the CGM of galaxies reaches up to two times their virial radii. We expect, however, that while BLA-galaxy systems with $\eta\geq 2$ are most likely not gravitationally bound, a system with $\eta<2$ does not necessarily correspond to virialized gas associated with a galaxy, for instance, if the impact parameter is much larger than the virial radius of the galaxy, but the $\Delta v$ is $\sim0$, $\eta$ would be $\ll2$, and still be likely that the absorption is not associated to the galaxy. Thus, we discuss possible BLA-galaxy associations case by case.

In this regard, we find that the inferred velocity dispersion and virial radii of these three galaxies are significantly smaller than their velocity offset to the BLA and impact parameter to the QSO sightline, respectively (by a factor of $\sim 1.5 - 30$ for velocity dispersion, and a factor of $\sim5 - 15$ in the case of the virial radii), that lead to $\eta$ values in the range $8-500$. This suggests that the absorption feature is likely tracing intergalactic gas, rather than being associated with a specific galaxy. Furthermore, the right panel of Fig.~\ref{fig:gal_dist_004} shows a possible filament of galaxies intersected by the QSO sightline (red to yellow dots at velocities $200-700$\,\kms), which would further support this idea. Finally, we find several NLAs at a similar redshift ($\Delta v \sim 400$ \kms), and for the same reasons discussed above, we do not find it likely that the NLAs are associated with a particular galaxy, meaning that these absorption features are the imprint of warm-hot intergalactic gas near to a colder component.

\subsection{BLA - overdensity of galaxies at $z\approx0.04866$}
\label{sec:sys_at_005}

\begin{figure*}
        \begin{minipage}[b]{0.48\textwidth}
                \includegraphics[width=\columnwidth]{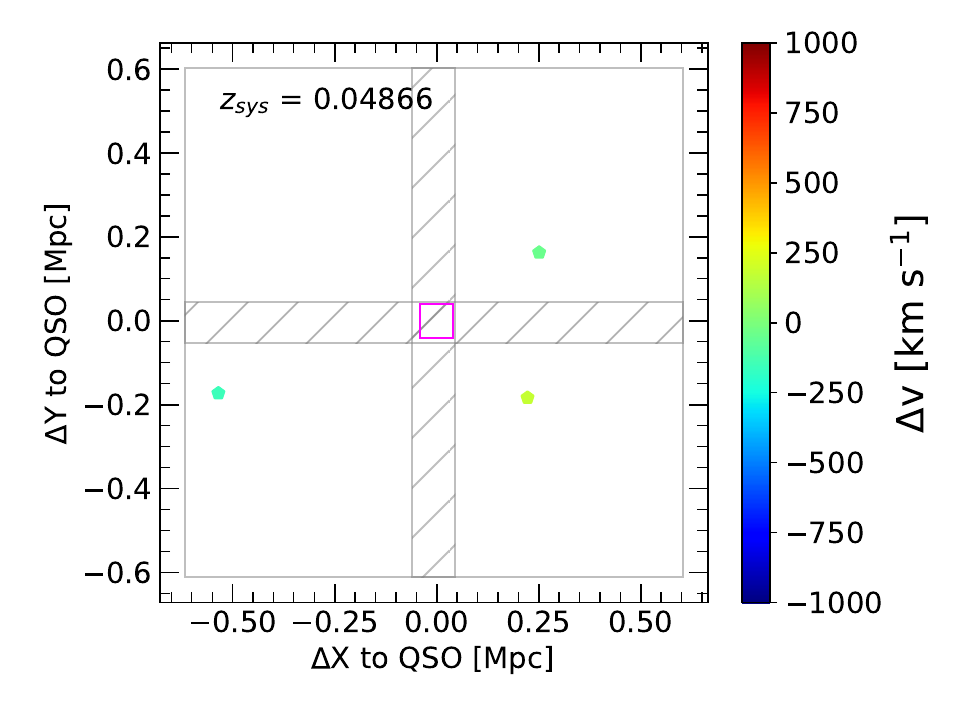}

        \end{minipage}%
        \begin{minipage}[b]{0.48\textwidth}
                \includegraphics[width=\columnwidth]{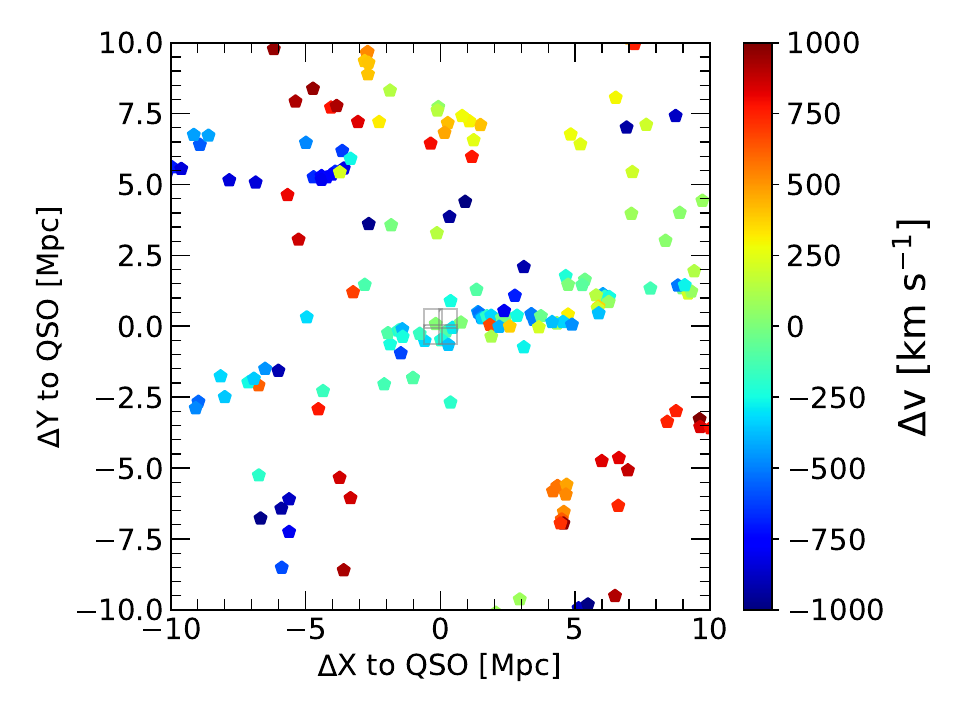}

        \end{minipage}%

        \caption{\label{fig:gal_dist_005}  Spatial distribution of galaxies with a redshift within $z_{\mathrm{sys}}\pm1000$ \kms, for the BLA at $z = 0.04866$. \textit{Left}: Full VLT/VIMOS FoV around SDSSJ161940.56+254323.0 at $z \approx 0.04866$. The dashed cross represents the separation between the four VLT/VIMOS quadrants, indicated in grey. The central magenta square represents the VLT/MUSE FoV. The VLT/VIMOS FoV spans a square of $\sim 1.2$ Mpc side at this redshift. The VLT/MUSE FoV is significantly smaller, reaching impact parameters of $\sim 50$ kpc from the QSO sightline. Three galaxies in our VLT/VIMOS survey lie within $1000$ \kms\ from the BLA.  \textit{Right}: Zoom-out of the left panel, using spectroscopic data from the SDSS DR16 survey. The separation between the VLT/VIMOS quadrants is kept for reference. The colorbar of both panels indicates the velocity offset of each galaxy with respect to $z_{\mathrm{sys}}$.}
\end{figure*}

Similar to the previous case, Fig.~\ref{fig:bla_excess} also shows a clear excess of galaxies at the redshift $z\sim0.04866$, of nearly $\sim9$ times over the average density of galaxies.

Figure~\ref{fig:gal_dist_005} shows the spatial distribution of these galaxies around the QSO sightline. Again, three (uncorrected by incompleteness) galaxies might seem like a small number, but this number is large considering the small FoV of VLT/VIMOS at this redshift ($\sim1.2$ Mpc$^{2}$). A closer inspection shows that the galaxy in the bottom-right quadrant of the VLT/VIMOS FoV is the same as that found in the SDSS data at an impact parameter of $\approx270$ kpc (see bottom panel of Fig.~\ref{fig:muse_vimos_ipz}). This leaves us with four galaxies near this BLA in the impact parameter - redshift space, three in the VLT/VIMOS FoV plus one extra galaxy in the SDSS data at an impact parameter of $\sim180$ kpc.

We proceed in the same way as for the BLA at $z\approx0.04021$ to evaluate if this absorption could be associated with a nearby galaxy, and we find that while the inferred velocity dispersion of these four galaxies is generally comparable to the $\Delta v$ respect to the BLA, their virial radii are significantly smaller (at least a factor 3) than their physical separation to the QSO sightline. For the galaxies in our VLT/VIMOS survey, we calculate $\eta$ values in the range $8-40$. For the SDSS galaxy, on the other hand, we get $\eta = 1.6$. Despite that this is lower than our reference value of 2, this value is mostly driven by the low velocity separation between the BLA and the galaxy, since the galaxy is located at an impact parameter of $\sim3$ times its virial radius. Thus, we find it unlikely that the BLA originates from virialized gas associated with this galaxy. Altogether, this suggests that the absorption feature is tracing intergalactic gas, rather than being associated with any of the galaxies in our survey. 

Moreover, when examining the distribution of galaxies at larger spatial scales using the SDSS galaxies with spectroscopically determined redshifts (right panel of Fig.~\ref{fig:gal_dist_005}), we find a coherent structure that is intersected by the QSO sightline, supporting the idea that the BLA at $z\approx0.04866$ is indeed tracing the filamentary large-scale structure of the Universe. Unfortunately, the GMBCG cluster catalog does not reliably cover redshifts lower than $0.1$, thus, we are unable to infer the presence of an inter-cluster axis close to the QSO sightline at this redshift. Nevertheless, the presence of metals at this redshift (see Table~\ref{tab:blas_params}) suggests that this gas could have been processed by a nearby galaxy at some point and escaped to the IGM \citep[see, e.g.,][]{Tripp2006}, or, alternatively, that it is indeed bound to a galaxy missed in our surveys.

\subsection{BLA - overdensity of galaxies at $z\approx0.18853$}
\label{sec:sys_at_019}
\begin{figure*}
        \begin{minipage}[b]{0.48\textwidth}
                \includegraphics[width=\columnwidth]{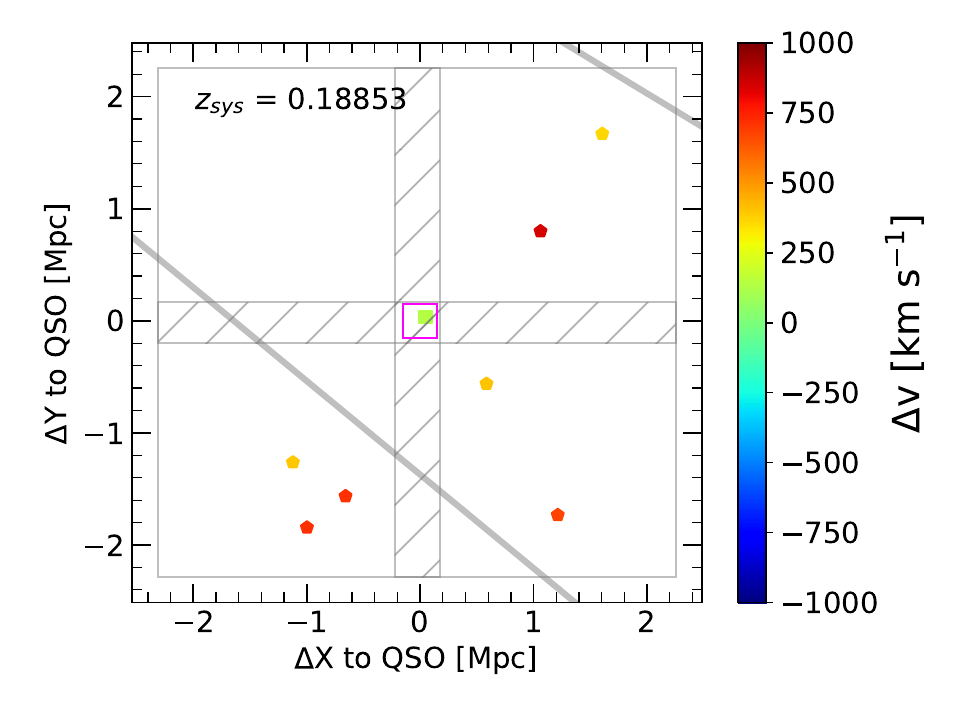}

        \end{minipage}%
        \begin{minipage}[b]{0.48\textwidth}
                \includegraphics[width=\columnwidth]{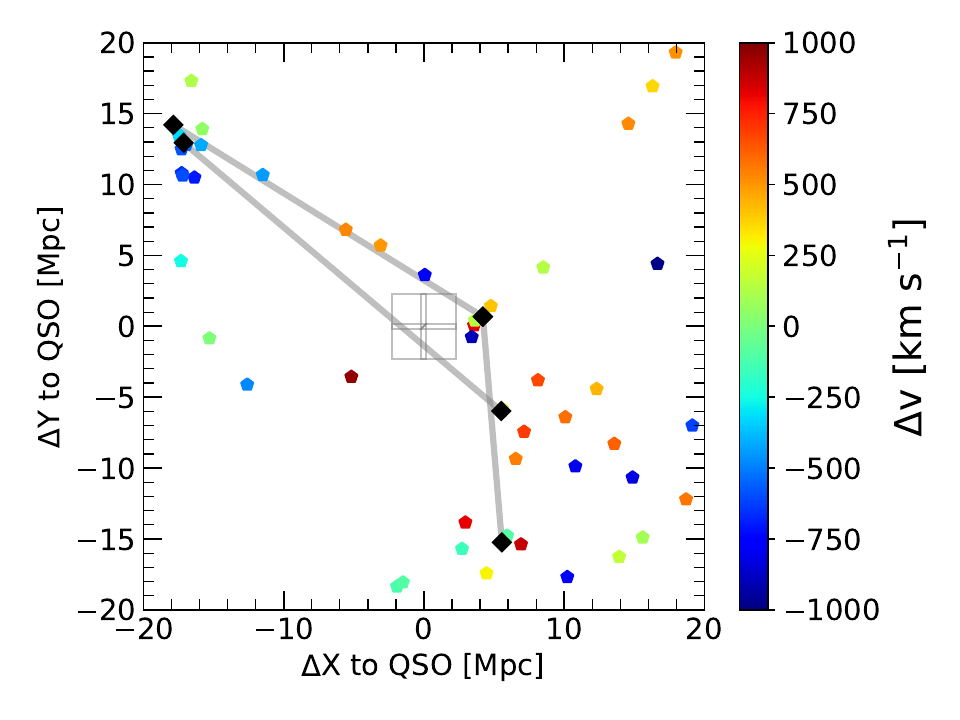}

        \end{minipage}%

        \caption{\label{fig:gal_dist_019} Same as Fig.~\ref{fig:gal_dist_004}, for the two BLAs at (average) $z \approx 0.18853$. The grey lines in the right panel indicate the inter-cluster axes connecting the cluster pairs described in Sec.~\ref{sec:gal_clust}, indicated with black diamonds. The VLT/VIMOS FoV spans a square of $\sim 4.4$ Mpc side at this redshift. The VLT/MUSE FoV is smaller, reaching impact parameters of $\sim 140$ kpc from the QSO sightline. Seven galaxies in our VLT/VIMOS survey and one galaxy from our VLT/MUSE survey lie within $1000$ \kms\ from the BLA. The right panel shows a zoom-out of the left panel, using spectroscopic data from the SDSS DR16 survey.}
\end{figure*}

Since we have found 2 BLAs at similar redshifts ($z_{1} = 0.18787$ and $z_{2} = 0.18919$), we have defined $z_{\mathrm{sys}} = 0.18853$ as the average of the two.
\\
Similar to the previous figures, Fig.~\ref{fig:gal_dist_019} shows the spatial distribution of galaxies with a redshift within $z_{\mathrm{sys}}\pm1000$\,\kms\ in our VLT/VIMOS and VLT/MUSE survey and the SDSS data. Both panels show a high number of galaxies close to the QSO sightline. Figure~\ref{fig:bla_excess} shows that this elevated number clearly represents an excess with respect to the expected number of galaxies. Furthermore, we also find three cataloged inter-cluster axes close to the QSO at this redshift (see Table~\ref{tab:cluster_pairs_props}), although a clear filamentary structure is not visible in the spatial distribution of SDSS galaxies in Fig.~\ref{fig:gal_dist_019}.

Although this high number of galaxies and the presence of close cluster pairs provide a first evidence of a large-scale structure, the presence of a galaxy close to the BLA redshift in the VLT/MUSE FoV (see bottom panel of Fig.~\ref{fig:muse_vimos_ipz} and left panel of Fig.~\ref{fig:gal_dist_019}) could indicate that the BLAs found at this redshift arise from hot coronal gas ~\citep{Williams2013} belonging to the galaxy halo, rather than from warm-hot intergalactic gas. To evaluate this possibility, we have proceeded in an analogous fashion to Sec.~\ref{sec:sys_at_005}, using SDSS photometric data to determine the $R_{\mathrm{vir}}$ and $\sigma_{\rm vir}$ of the galaxy, and comparing it with the impact parameter to the QSO sightline and the $\Delta v$ with respect to the BLAs.

We find that its $R_{\mathrm{vir}}$ is comparable, even slightly larger than the impact parameter to the sightline ($116$ kpc and $102$ kpc, respectively). However, its $\Delta v$ to the BLAs are $11$\,\kms\ and $-322$\,\kms. While $11$\,\kms\ is a seemingly low value, significantly smaller than its estimated $\sigma_{\rm vir}$ of $\sim90$\,\kms\ ($\eta\approx0.1$), a $\Delta v$ of $322$\,\kms\ is considerably larger, and represents a much larger separation between the galaxy and the BLA ($\eta\approx3.1$). We have additionally detected \ion{O}{VI} absorption at the redshift of one of these BLAs (although it is close in velocity to the BLA at $z = 0.18787$, see Table~\ref{tab:blas_params}), which further suggests a galaxy-BLA association (although in a recent work, using a simple 1$D$ model, \citealt{Bromberg2024} report that the IGM can actually contribute the majority of the observed \ion{O}{VI} column densities around the CGM of galaxies). Moreover, the detection of several NLAs at this redshift (only one with reliability `a', $\Delta v \sim 150$ \kms), suggests the presence of an additional colder component near the warm-hot gas. Altogether, we conclude that while the BLA at $z = 0.18919$ could be associated with the circumgalactic medium of a galaxy identified in our VLT/MUSE survey, this seems to not be the case for the BLA at $z = 0.18787$ \citep[although we can not rule out the possibility that this gas is indeed an outflow from the galaxy, as it high $\Delta v$ value is also consistent with outflowing gas, see, e.g.,][ and we caution the reader about this potential caveat]{Schroetter2019}, for which the absence of nearby galaxy suggest it is originated by intergalactic warm-hot gas, tracing a large scale structure with a clear excess of galaxies identified in our VLT/VIMOS survey. 

\subsection{BLAs not associated with a robust overdensity of galaxies}
\label{sec:not_overdensity_BLAs}
For some of the BLAs, we find a mild overdensity of nearby galaxies, but given the uncertainties, they are also consistent with the field expectation (i.e., those at $z = 0.06805$, $0.13831$, $0.15417$, and $0.25294$). For the BLAs at $z = 0.17576$, $0.17649$, $0.17729$, and $0.17813$, we find a number of galaxies roughly consistent with the average expected number (see Fig.~\ref{fig:bla_excess}) and for the BLA at $z=0.21316$, we do not find any nearby galaxy, neither in our VLT/VIMOS nor in our VLT/MUSE survey. 

In the case of the BLA at $z=0.06805$, our VLT/MUSE survey shows no nearby galaxies up to impact parameters of $\sim60$ kpc. Similarly, we do not find any nearby galaxy in the SDSS dataset either. On the other hand, our VLT/VIMOS survey shows one galaxy at the same redshift, at an impact parameter of $\sim300$ kpc. However, proceeding in the same way as in Sec.~\ref{sec:sys_at_004}, we find that the virial radii and velocity dispersion of this galaxy are considerably smaller than its impact parameter, and velocity offset to the BLA (by a factor $\sim10$, leading to $\eta\approx180$), making it unlikely that it is associated with the BLA. Thus, while it is possible that we are missing a nearby galaxy due to the small FoV of our VLT/MUSE survey, our analysis suggests that the BLA at $z=0.06805$ is probably not associated with any nearby galaxy, and it is truly tracing intergalactic medium.

Finally, given that the BLAs at $z = 0.13831$, $0.15417$, $0.17576$, $0.17649$, $0.17729$, $0.17813$, $0.21316$, and $0.25294$ do not present any nearby (impact parameter to the sightline  $<500$ kpc) galaxy in any of our datasets, and that our VLT/MUSE FoV covers impact parameters higher than $\sim120$ kpc at those redshifts, and it is complete up to $r \sim 24$ mag (see Sec.~\ref{sec:MUSE_survey}), we find unlikely that they could be associated to any particular galaxy. For the BLA at $z\approx0.17576$, we also find a nearby inter-cluster filament, with an impact parameter of $3.5$ Mpc to the QSO sightline (see Table~\ref{tab:cluster_pairs_props} and Figure~\ref{fig:gal_dist_017}). Altogether, we conclude that the BLAs at $z=0.17576$, $0.17649$, $0.17729$, and $0.17813$ are likely truly tracing intergalactic medium (naturally, there is the caveat that we could be missing a nearby galaxy in our surveys, but we do not find evidence that this is the case).

For completeness, in Appendix~\ref{sec:appendix_plot_no_overdensity}, we show the spatial distribution of sources at the redshift of the BLAs discussed in this subsection, found in our VLT/MUSE and VLT/VIMOS surveys, as well as in the SDSS dataset.

\subsection{A potential BLA in a void of galaxies?}
\label{sec:bla_in_void}

Figure~\ref{fig:gal_dist_021} shows the absence of galaxies close to the BLA at $z\approx0.21316$ in our data, as well as in the SDSS dataset. Remarkably, the top panel of Fig.~\ref{fig:muse_vimos_ipz} shows the presence of an obvious overdense region at a slightly higher redshift, beyond our imposed velocity window threshold of $1000$\,\kms. Under the assumption that BLAs trace warm-hot gas residing in large-scale galaxy filaments \citep[][]{Richter2006a, Richter2006b, Tripp2006, Danforth2010, Wakker2015}, the detection of a BLA at this redshift is intriguing.

One possibility is that this BLA is tracing warm-hot gas that does not belong to a galaxy filament. In this line, \citet{Tejos2012} and \citet{Watson2022} have reported the existence of a population of \lya\ absorbers within cosmic voids. A second possibility is that this BLA arises artificially, due to the blending of multiple narrower components, instead of actually tracing warm-hot gas. As discussed in Sec.~\ref{sec:blas_caveats}, this limitation is intrinsic to the absorption-line technique. We also refer the reader to \citet{Tejos2016} for a more comprehensive discussion of this issue. 

In both these scenarios, the BLA at $z\approx0.21316$ would not be tracing a large-scale filament of galaxies. Alternatively, the lack of galaxies could be due to the sightline probing an underdense region of a filament. A final possibility is that there is indeed an overdense region at this redshift, but due to the imperfect statistical completeness of our VLT/VIMOS survey, we do not detect any of its galaxies. However, given the completeness and size of our survey, we have estimated a probability $\lesssim3\%$ of this occurring (with lower probability for overdensities larger than our threshold of 2.5 times the expected number of field galaxies). Thus, we rule out this latter possibility.

In the following sections we infer the physical properties of the warm-hot gas assuming that this BLA is real, and we opt by not removing it from our sample, although we caution the reader about this potential caveat, as we can not rule out that this BLA emerges from the blending of two narrower components or from one noisy narrower component.


\section{Discussion}
\label{sec:discussion}

\subsection{Inter-cluster axes and BLAs connection}
As detailed in Sec.~\ref{sec:selection}, the sightline toward the QSO SDSSJ161940.56+254323.0 was chosen to maximize the number of inter-cluster axes (as proxies of inter-cluster filaments) intersected, in the same fashion as SDSSJ141038.39+230447.1 \citep{Tejos2016,Pessa2018}. However, while for SDSSJ141038.39+230447.1 the authors find a good agreement between the redshift of the BLAs and that of the cluster pairs (all detected BLAs are located within $800$\,\kms\ of a cluster pair, with a mean velocity offset of $500$\,\kms), that is not the case for SDSSJ161940.56+254323.0. Figure~\ref{fig:bla_excess} shows that only the BLAs at average $z\approx0.18853$ and the BLA at $z\approx0.17576$ lie within $1000$\,\kms\ from the closest cluster pair. The rest of them (even those at $z>0.1$, where clusters in the GMBCG catalog are available) do not present any nearby (in redshift) inter-cluster axis (see Fig.~\ref{fig:bla_excess}). This either implies that the cluster pairs that do not present a nearby BLA are not actually connected by a large-scale filament that contains WHIM, or that there is a WHIM filament, but it produces a shallow absorption feature that is not detectable in our data. A third possibility could be that there is a WHIM filament, but with a morphology that evades the QSO sightline \citep[we note that in contrast to the sample of][here most of the impact parameters to the inter-cluster axes are larger than 3 Mpc]{Tejos2016}. In any case, this difference in the inter-cluster axes and BLAs connection for both QSOs is intriguing and deserves further exploration to assess the impact of the number of intersected inter-cluster axes in the observed number of BLAs, as a function of impact parameters, using a larger sample of QSOs with a known number of intersected inter-cluster axes. Nevertheless, the sightline of SDSSJ161940.56+254323.0 still presents a large number of BLAs, compared to the random expectation \citep{Danforth2010}, despite that they may not be connected to the inter-cluster axes (rather simplistic) definition used to select this specific sightline.

\subsection{Overdensity of galaxies towards larger physical scales}
\label{sec:overdensity_larger_scales}
In previous sections, we have established the existence of a strong overdensity of galaxies close to the QSO sightline coinciding with the redshift of four of the BLAs in our sample, in addition to milder overdensities at the redshift of other four BLAs. We acknowledge, however, that due to our overall low number statistics of at most a few galaxies at the redshift of a given BLA, the reported overdensities of galaxies are subject to large Poissonian errors. Indeed, Fig.~\ref{fig:bla_excess} makes clear that the four strongest overdensities are only $\sim1-\sigma$ above the overdensity threshold of a factor of 2.5 more galaxies than the field expectation.  Nevertheless, keeping this caveat in mind, in this section we explore the extent of these overdensities in both, velocity and impact parameter space.

\begin{figure}
\centering
\includegraphics[width = \columnwidth]{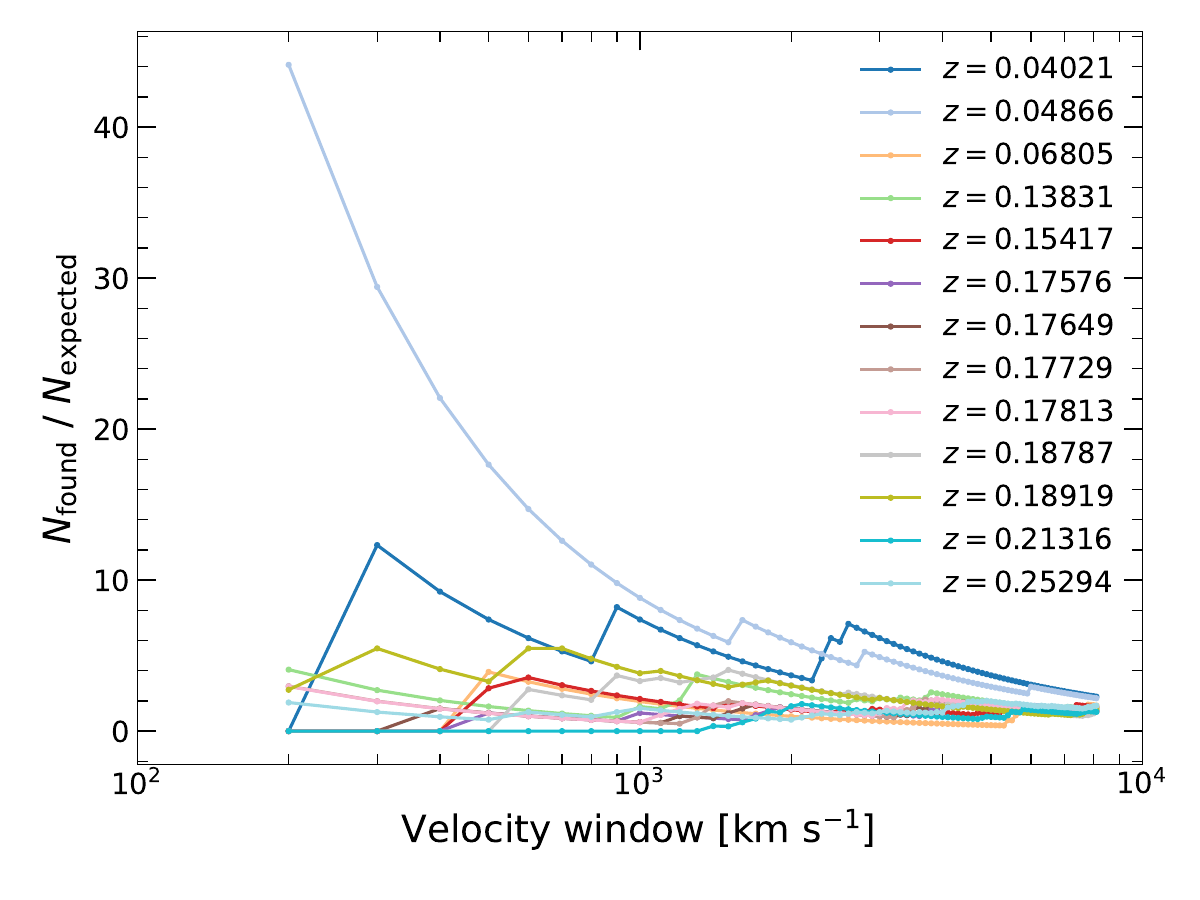}
\caption{\label{fig:velocity_window_function} Ratio of completeness-corrected found and expected number of galaxies in our VLT/VIMOS survey (considering the full FoV) at the redshift of the BLAs, as a function of the size of the velocity window used to define the depth of the cosmic volume subtended by the VLT/VIMOS FoV, from $200$\,\kms\ to 8200 km $s^{-1}$. Each line is colored by the redshift of the BLA, as indicated in the top-right part of the panel. Although 8000 km $^{-1}$ is considerably higher than the $\Delta v$ expected for a large-scale galaxy filament, we included large velocity windows to show how the local excess of galaxies converges towards the cosmic average in a large enough volume. }     
\end{figure}

Figure~\ref{fig:velocity_window_function} shows the ratio between the (completeness-corrected) inferred and expected number of galaxies in the full VLT/VIMOS FoV, at the redshift of the thirteen BLAs in our sample, considering a velocity window that ranges from $\pm200$\,\kms\ up to $\pm8000$\,\kms. It is clear that the BLA at $z\approx0.04866$ shows the most extreme excess of galaxies, over a factor $\sim40$ on the smallest velocity windows. That is, our VLT/VIMOS survey shows an abnormally high number of galaxies close to the BLA in a relatively small cosmic volume (although none of these galaxies seem to be directly associated with the BLA, as discussed in Sec.~\ref{sec:sys_at_005}). This excess of galaxies decays rapidly when a larger velocity window is considered. At $\pm2000$\,\kms, the excess is drastically smaller, although still significant ($\sim6$ times higher than the average expected value), pointing to a local overdensity of galaxies, rather than, for example, the possibility that our modeling is underestimating the number of expected galaxies in a given volume.

The BLA at $z\approx0.04021$ also shows a significant excess of nearby galaxies, peaking at an excess of a factor $\sim12$ relative to the cosmic average at a velocity window of $\pm300$\,\kms. This excess is still relevant towards larger velocity windows. At $\pm1000$\,\kms, the number of galaxies is still a factor $\sim7$ larger than the expected cosmic average. On the other hand, The two BLAs at average $z\approx0.18853$ show a peak overdensity of factor 4 and 6 with respect to the cosmic average on velocity windows of $700-1000$\,\kms, which is not as extreme as for the BLAs at $z\approx 0.04021$ and $0.04866$, but still well above the threshold of factor 2.5 used to define overdense regions of the Universe (see Sec.~\ref{sec:overdensity_blas}).

The rest of the BLAs that coincide with an overdensity of galaxies present a more modest excess that peaks at absolute velocity windows $\lesssim1000$\,\kms, a scale typically used to represent the extent in the velocity space of large-scale filaments \citep[e.g.,][]{Tejos2016, Bouma2021}. The BLAs at $z\approx0.06805$ and $z=0.15417$ exhibit an excess of $\sim$4 times the expected number of galaxies at a velocity window of $\pm500$\,\kms, with a steady decline towards larger velocity windows. The BLA at $z=0.13831$ also presents a modest peak of a factor of $3-4$ at velocity windows of about $\pm1000$\,\kms. 

The BLA at $z\sim0.25294$ shows a first peak of factor $\sim2$ at the smallest velocity window, followed by a peak of only a factor of $\sim1.5$ at a velocity window of $\pm1000$\,\kms. However, it also shows an additional peak at velocities $> 4000$\,\kms, due to the presence of a large number of galaxies found at a slightly higher redshift (see the top panel of Fig.~\ref{fig:muse_vimos_ipz}). Although this excess of galaxies at higher velocities could, in principle, be consistent with a galactic filament, it is beyond the scope of this paper to focus on the overdensities of galaxies that do not present a BLA. In this regard, the lack of a BLA at the redshift of an existing filament is not necessarily unexpected, as their detection is often difficult due to the high temperature of the gas, leading to an extremely low fraction ($<10^{-5}$) of neutral hydrogen and a large thermal broadening. Thus, the detection of gas at higher temperatures requires higher quality data \citep[see, e.g.,][]{Danforth2010} or a different technique (e.g. using fast radio bursts). 

For completeness, we have also included in the figure those BLAs that do not present an excess of nearby galaxies. The BLAs at $z=0.17576$, $0.17649$, $0.17729$, $0.17813$, and $0.21316$ do not present any clear peak above our threshold of $2.5$ times the average density.

The fact that the ratio of galaxies found and expected, at the redshift of all the different absorption systems, converges towards large volumes reflects the consistency between our model and our galaxy survey (although the ratio of found and expected galaxies converges towards a slightly higher number $\sim2$ for those systems that present an overdensity of galaxies). A relevant aspect of this analysis is that at different redshifts, we probe different maximum impact parameters with the full VLT/VIMOS FoV, being significantly smaller towards lower redshifts, implying that purely using a velocity criteria (as in Fig.~\ref{fig:velocity_window_function}) to compare the systems do not provide a complete picture.

\begin{figure}
\centering
\includegraphics[width = \columnwidth]{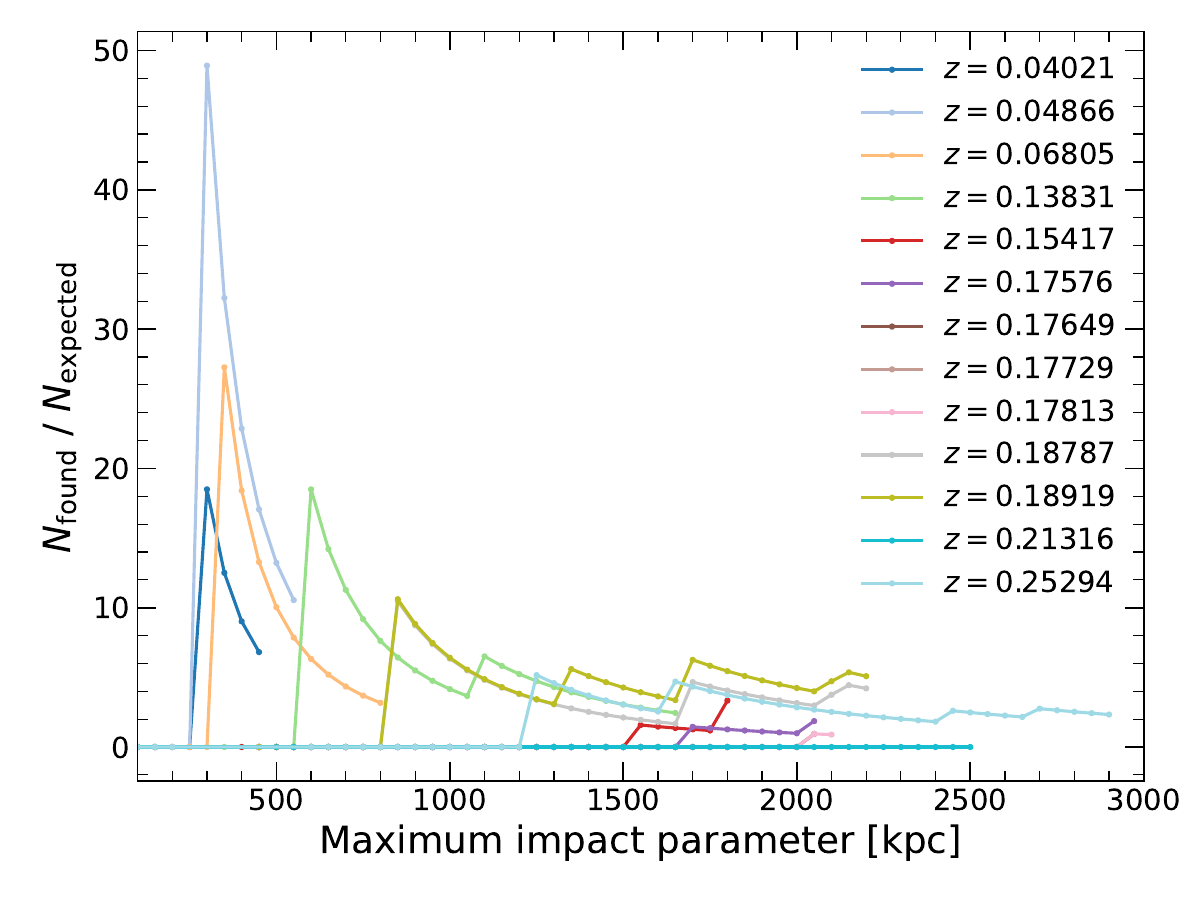}
\caption{\label{fig:impact_parameter_function} Ratio of completeness-corrected found and expected number of galaxies in our VLT/VIMOS survey within $1000$\,\kms\ from the redshift of the BLAs, as a function of the maximum impact parameter used to define the area of the cosmic volume subtended by VLT/VIMOS, from 100 kpc to 3.2 Mpc. Each line is colored by the redshift of the BLA, as indicated in the top-right part of the panel. Note that for the BLAs at low redshift (blue and orange lines), we can only probe impact parameters up to $\sim500$ kpc.}     
\end{figure}

Figure~\ref{fig:impact_parameter_function} makes this difference clearer. It shows the found to expected galaxies ratio close to each BLA, within a fixed velocity window of $\pm1000$\,\kms, considering different maximum impact parameters, from $100$ kpc, up to the maximum impact parameter covered by VLT/VIMOS at each redshift.

The systems at $z=0.04866$ and $0.06805$ present the strongest excess of galaxies, with an excess of a factor $\sim50$ and $\sim25$, respectively. In general, all the systems peak at maximum impact parameters lower than $1.5$ Mpc (except for the system at $z=0.15417$, since the galaxies at this redshift are located relatively far away from the central sightline), which is well within the expected width of cosmic filaments, according to cosmological simulations \citep{GonzalezPadilla2010}. Additionally, their ratio of found to expected galaxies (within $\pm1000$\,\kms) is notoriously larger when we consider a smaller fraction of the FoV. This is because when one or more galaxies start being accounted for within a very small cosmic volume, parametrized by a small maximum impact parameter, it represents a strong excess that is then averaged when a larger cosmic volume is considered. Hence, at small impact parameters, measurements are statistically less robust, as a single galaxy can drastically increase the ratio of found to expected galaxies (this naturally also occurs when only a small velocity window is considered).

In general, we conclude that the overdensities of galaxies at the redshift of the BLAs peak within impact parameters and velocity windows comparable to or smaller than the expected size of the large-scale structure of the Universe, being indeed local in the velocity space, and representing a true excess of galaxies with respect to the average density of galaxies.

\subsection{Correlation between intergalactic medium conditions and excess of galaxies in our VLT/VIMOS survey}
\label{sec:correlation_galaxies_BLA_gas}
\begin{figure}
        \begin{minipage}[b]{\columnwidth}
                \includegraphics[width=\columnwidth]{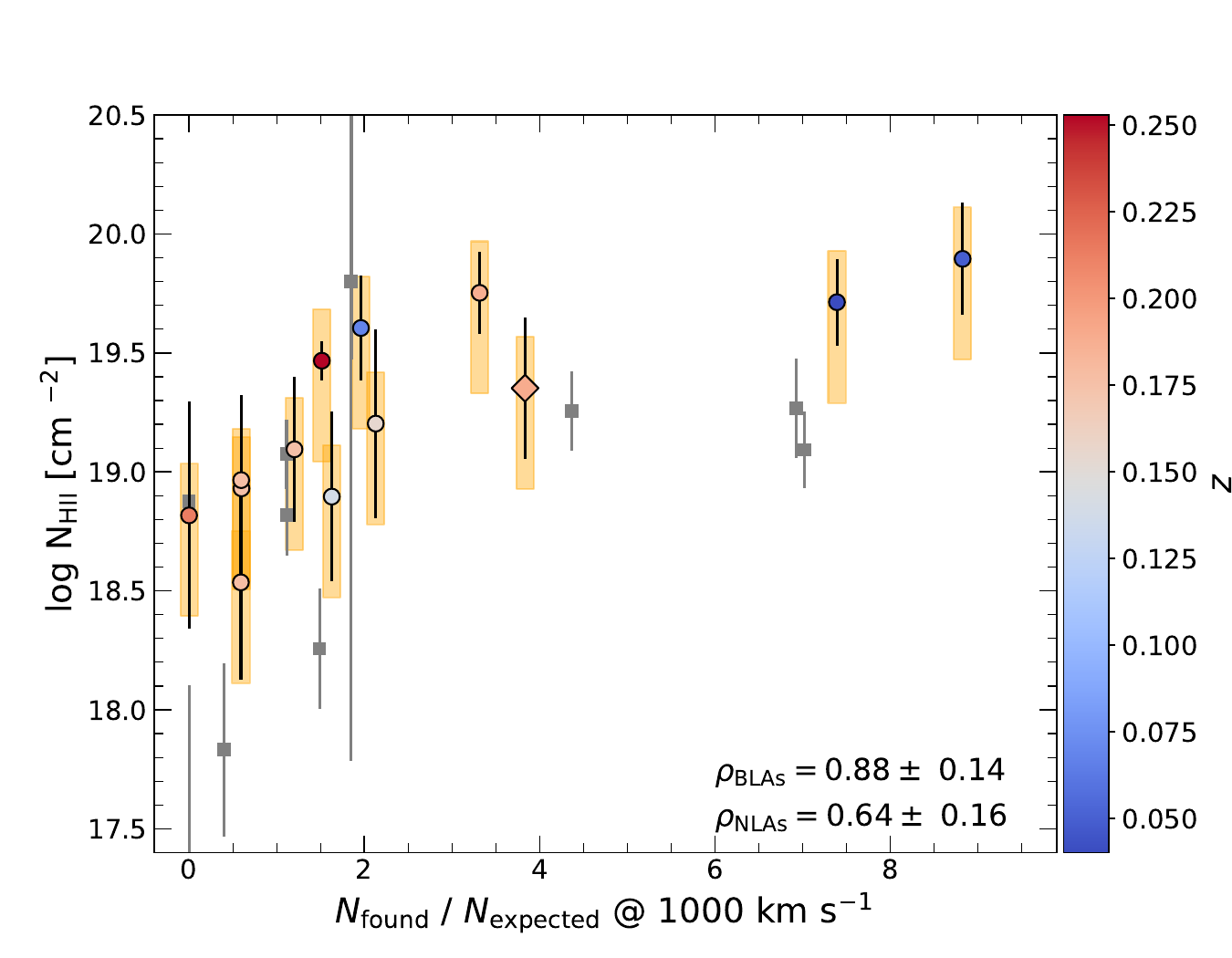}

        \end{minipage}%
        \\
        \begin{minipage}[b]{\columnwidth}
                \includegraphics[width=\columnwidth]{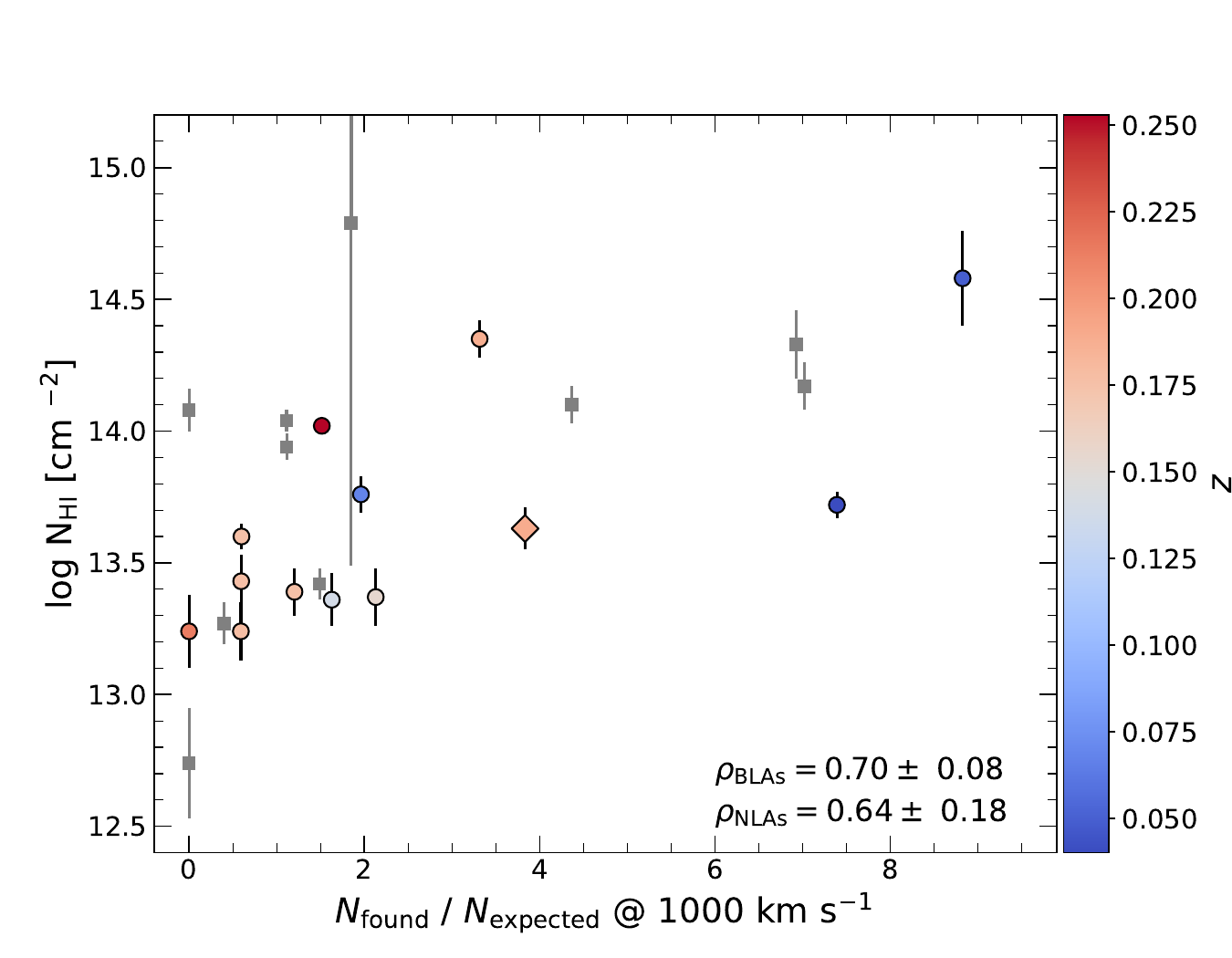}

        \end{minipage}%
        \\
        \begin{minipage}[b]{\columnwidth}
                \includegraphics[width=\columnwidth]{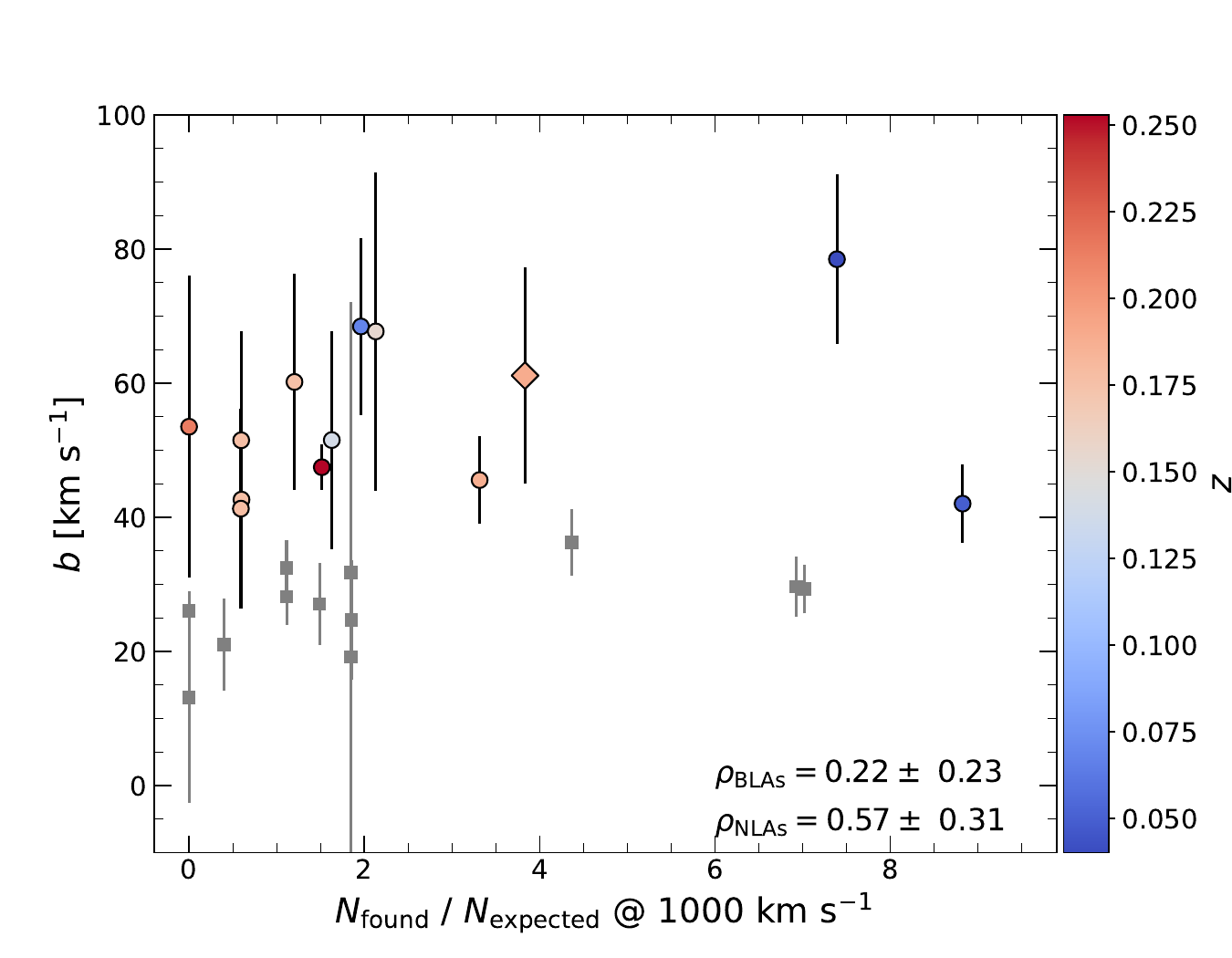}

        \end{minipage}%

        \caption{\label{fig:column_doppler_vs_excess_galaxies} Correlation between the total ionized hydrogen column density (\textit{top}), neutral hydrogen column density (\textit{middle}), and Doppler parameter (\textit{bottom}) measured for each absorption feature and the completeness-corrected number of galaxies found in our VLT/VIMOS survey at the same redshift, divided by the cosmic average expected. The redshift of each BLA is indicated by the colorbar. The BLA close to a nearby galaxy from our VLT/MUSE survey is indicated with a diamond in the three panels (i.e., that at $z\approx 0.18919$). The grey squares correspond to our NLA sample. The orange bars in the top panel quantify the range of variation in the inferred ionized hydrogen column density between the $\alpha = 0$ and $\alpha = 1.5$ scenarios (see Sec.~\ref{sec:blas_survey}).}
\end{figure}

Our VLT/VIMOS survey gives us the opportunity to study the number and distribution of galaxies at the redshift of the BLAs. In this section, we use the BLAs to infer the physical conditions of the intergalactic gas, and correlate gas properties with the galaxy distribution (For completeness and comparison purposes, we also include the NLAs in our analyses). The top panel of Fig.~\ref{fig:column_doppler_vs_excess_galaxies} shows the derived column density of ionized gas (N$_{\mathrm{HII}}$, which comprises the bulk of the gas mass) for each absorption (see Tables~\ref{tab:blas_params} and~\ref{tab:nlas_params}) as a function of the ratio between the (completeness-corrected) inferred and expected number of galaxies found in our VLT/VIMOS survey, within velocity windows of $\pm1000$\,\kms\ considering the full VLT/VIMOS FoV (note that we now treat the excess of galaxies as the independent variable shown in the $x$-axis). The figure shows a tentative correlation between both quantities, for both, the BLA and the NLA samples. The only notable exceptions are the three NLAs that lie close (in velocity and impact parameter, $\eta\approx0.4$) to a galaxy from our VLT/MUSE survey at $z\approx0.125$. For two of them, we infer a column density N$_{\mathrm{HII}} > 10^{21}$ cm$^{-2}$, and are not shown in the figure (the error bars of the NLAs at this redshift are large because the absorptions are blended and heavily saturated). We also identify several metal transitions at the same redshift of these NLAs (\ion{C}{II}, \ion{C}{IV}, \ion{Si}{II}, \ion{Si}{III}, \ion{O}{VI}). The NLA at $z\approx0.01702$ is also an outlier to some degree (on the low- N$_{\mathrm{HII}}$ side), but unfortunately, the small FoV of VLT/MUSE at this redshift does not allow for a meaningful search of nearby galaxies at the same velocity. 

After excluding these outlier absorptions, since they are either very likely tracing the circumgalactic medium of a galaxy, rather than being truly intergalactic (in the case of the NLAs at $z\approx0.125$), or do not allow for an appropriate exploration of their origin (as for the NLA at $z\approx0.01702$), it is interesting that both BLAs and NLAs follow a similar correlation. We compute the Spearman correlation coefficient $\rho$ for the BLAs, after excluding the BLA found to be near a galaxy in our VLT/MUSE survey ($z\approx 0.18919$), and find $\rho = 0.88 \pm 0.14$, a value typically interpreted as a strong correlation between both quantities. This value is higher, albeit consistent with the $\rho$ measured for the NLAs (after excluding the outlier NLAs, a detailed analysis of the environment at the redshift of each NLA is beyond the scope of this paper, and none of the other NLAs are found to be an outlier in any aspect) of  $\rho = 0.64 \pm 0.16$. This similarity further suggests that the density of both, cold and warm-hot phases of the IGM correlates with the underlying overdensity of galaxies \citep[e.g., see][]{Tejos2014}. However, one aspect that separates both samples is that while about half of the BLAs coincide with ratios of found-to-expected number of galaxies $\gtrsim2$, most of the NLAs coincide with lower densities of galaxies, closer to the expected average, indicating that BLAs seem to coincide preferentially with more overdense environments, compared to NLAs \citep[see also][]{Tejos2016}. We estimated the uncertainty in $\rho$ by performing 100 measurements of $\rho$, perturbing the data according to its uncertainty in column density in each iteration, and then using the standard deviation of these 100 measurements as the uncertainty in the Spearman coefficient.

In principle, this correlation is similar to the previously explored in the literature correlation between impact parameter to a filament of galaxies and strength of an absorption feature \citep[see, e.g,][]{Wakker2015,Bouma2021}, except that here we use the local overdensity of galaxies to trace properties of the galaxy filament, whose presence is initially inferred from the BLA, but its exact location is unknown. In this line, \citet{Wakker2015} used cosmological hydrodynamical simulations \citep[][]{Oppenheimer2010} to show that a higher density of galaxies is expected closer to the center of the filament (although differences between different filaments may exist). Thus, the top panel of Fig.~\ref{fig:column_doppler_vs_excess_galaxies} can be interpreted as the intergalactic medium having higher total gas densities either closer to the center of a filament, or in overall denser (in terms of number of galaxies) filaments. A similar approach was carried out by \citet{Burchett2020}, where they find a correlation between the modeled density of the cosmic web in a given volume and the strength of the \hi\ absorption feature in an average sightline passing through that volume. Furthermore, it is interesting that \citet{Burchett2020} report that the \hi\ absorption strength increases only up to a certain density threshold, after which the strength of the absorption saturates (or even decrease), and the scatter raises. This is because the \hi\ absorption is harder to detect in the hotter gas, with a lower neutral fraction, as expected in the inner and denser regions of the cosmic web filaments. 

Here, we see that the column density of ionized gas also flattens towards larger density of galaxies (i.e., the systems at $z=0.04021$ and $z=0.04866$), although this should be treated carefully, as it could also be a result of our statistically small sample and cosmic volume probed at these redshifts. For a more consistent comparison with the findings reported by \citet{Burchett2020}, the middle panel of Fig.~\ref{fig:column_doppler_vs_excess_galaxies} shows the same correlation for neutral hydrogen column density. Remarkably, the figure shows a correlation that is qualitatively consistent with the results from \citet{Burchett2020}, where the strongest outliers are again the NLAs at $z\approx0.125$, for which we found a nearby galaxy in our VLT/MUSE survey and the presence of metals at the same redshift (i.e., they are likely not produced by intergalactic gas). For this correlation we find a Spearman coefficient $\rho = 0.70\pm0.08$ for the BLA sample, and  $\rho = 0.64\pm0.18$ for the NLA sample, which is still significant, but lower than that for the correlation with ionized gas column density in the case of the BLAs (and the same for the NLAs).

It is also interesting that the BLA for which we do not find any nearby galaxy in our survey (i.e., the BLA at $z\approx0.21316$) presents nearly the lowest column density (of neutral and ionized hydrogen) among the full sample. This is consistent with the scenario where this BLA is indeed tracing diffuse low-density intergalactic gas in an overall underdense cosmic region. We have also explored the correlation between N$_{\mathrm{HII}}$ and the impact parameter to the inter-cluster axes. Unfortunately, we find a BLA coinciding with an inter-cluster axis only at two independent redshifts ($z\sim$ $0.18853$ and $0.17576$), and although for these two absorbing systems (with multiple BLAs nearby each), we do find higher column densities for the BLAs closer (in velocity) to an inter-cluster axis, we do not have enough inter-cluster axes-BLA systems to robustly determine a correlation.

The bottom panel of Fig~\ref{fig:column_doppler_vs_excess_galaxies} shows the Doppler parameter measured for each absorption feature (a tracer for the temperature of the gas) as a function of the magnitude of the galaxy overdensity. In this case, we do not see any correlation between both quantities for the BLA sample ($\rho = 0.22\pm0.23$). This is in agreement with previous findings in the literature. \citet{Bouma2021} reports a large scatter between $b$ and the impact parameter to the galaxy filament. On the other hand, \citet{Wakker2015} reports a tentative correlation between the same quantities, although this correlation disappears when only broad absorptions are considered, because the BLAs are located preferentially at smaller impact parameters to the filament, where the scatter in the correlation between both quantities is higher. Interestingly, we find a tentative correlation for the NLA sample, with a correlation coefficient of $\rho = 0.57\pm0.31$, where NLAs with a higher Doppler parameter tend to be in overall denser environments, in line with the findings from \citet{Wakker2015}.

\section{Summary and Conclusions}
\label{sec:summary}
In this paper, we have used a sample of $13$ BLAs identified on the HST/COS FUV spectra of the QSO SDSSJ161940.56+254323.0 to infer the physical properties of the WHIM. We have also further identified $12$ NLAs that we include in our analyses for comparison purposes. SDSSJ161940.56+254323.0 is targeted because its sightline maximizes the number of inter-cluster filaments intersected. Additionally, we have used VLT/MUSE and VLT/VIMOS data, together with available SDSS data, to infer the distribution of galaxies near the QSO sightline. The detection threshold of our VLT/MUSE and VLT/VIMOS galaxy surveys are to magnitude $r\sim24$\,mag and $r\sim22.5$\,mag, respectively. We study the presence of galaxies at the redshift of the BLAs to infer if the BLAs could potentially be produced by gas gravitationally bound to a galaxy, or, alternatively, if both, BLAs and galaxies are tracing the same large-scale filament of galaxies but are otherwise not directly associated. We also compare the number of galaxies present in our dataset at the redshift of the BLAs with predictions from $N$-body simulations to evaluate if a certain number of galaxies does represent an excess with respect to the cosmic average, and finally, we cross-correlate the gas properties inferred from the BLAs with the strength of the excess of galaxies. Our main results are the following:
\begin{itemize}
    \item We find an elevated number of galaxies, consistent with an overdense region of the Universe, at the redshift of four out of the thirteen BLAs in our sample. Furthermore, the two BLAs at average $z\approx0.18853$ are also close, in terms of velocity and projected distance, to three inter-cluster axes identified from the GMBCG catalog. Other four BLAs present a milder excess of galaxies, that could also be consistent with the expected cosmic average.
    \item Based on the distance of nearby galaxies to the BLAs, in terms of velocity offset and impact parameter to the QSO sightline, we conclude that one of the BLAs is likely associated with the circumgalactic medium of a nearby galaxy, rather than being really probing WHIM gas (BLA at $z=0.18919$). For the rest of the BLAs, we do not find evidence that suggests that they are gravitationally bound to any specific galaxy.

    \item Interestingly, the BLA at $z=0.21316$ does not show any nearby galaxy in either of our datasets, implying that it could be either tracing warm-hot gas within a cosmic void, or alternatively, could be an artificial BLA produced by the blending of multiple narrower components. However, the fact that this BLA presents nearly the lowest column density among our sample could suggest that it is tracing diffuse low-density intergalactic gas in an overall underdense cosmic region.

    \item We study how the excess of galaxies found in our VLT/VIMOS survey at the redshift of the different BLAs change when we consider a broader or narrower cosmic volume around the BLA, in terms of velocity windows and impact parameter, and  find that the overdensities of galaxies close to BLAs are indeed local, with all of them reaching their maxima within velocity windows $\leq 1000$\,\kms\ and being averaged up to the expected cosmic average number of galaxies towards larger cosmic volumes

    \item The strength of the excess of galaxies close to a BLA, parametrized by the ratio of completeness-corrected found-to-expected number of galaxies in our VLT/VIMOS survey, correlates well with the neutral- and ionized-hydrogen column density inferred for the BLAs, where BLAs with higher column density are generally associated with a stronger excess of galaxies. Our NLA sample also follows a similar trend, which suggests that the density of both, cold and warm-hot phases of the IGM correlates with the underlying overdensity of galaxies in a similar manner. The strongest outliers in this correlation correspond to the NLAs at $z\approx0.125$, which we believe are likely not probing intergalactic gas, given that they are very close to a galaxy in our VLT/MUSE survey. Furthermore, we found the presence of metal absorptions in the QSO spectra at the same redshift, providing further evidence of the galactic origin of these NLAs.

    \item For the BLA sample, the measured Doppler parameter does not correlate with the strength of the underlying excess of galaxies. However, for the NLA sample, we find a tentative correlation where broader NLAs lie preferentially in overall denser environments, in line with previous findings from the literature.
\end{itemize}

Altogether, this work shows that most of the BLAs identified in the FUV spectra of QSO SDSSJ161940.56+254323.0 are really tracing WHIM, and that WHIM properties correlate with the distribution of galaxies within its local volume, where denser intergalactic gas correlates with a stronger local excess of galaxies, relative to the cosmic expectancy. This provides an alternative observational perspective to similar correlations reported in the literature, derived using cosmological simulations, between properties of cosmic filaments and properties of the intergalactic gas residing in them.

\begin{acknowledgements}
    
    We sincerely thank the anonymous referee for their valuable comments and constructive suggestions, which have greatly improved the quality of this paper.
    We thank Simon L. Morris, Gabriel Altay, Charles Finn, Rich Bielby,  Neil H. Crighton, Tom Theuns and Nelson Padilla for their contribution to the early stages of this project.

This research is based on observations collected at the European Organisation for Astronomical Research in the Southern Hemisphere under ESO programmes:
096.A-0426(B), 
097.A-0560(B), and 
099.A-0197(B). 
This research is based on data obtained from the NASA/ESA Hubble Space Telescope under programme GO~13832, obtained at the Space Telescope Science Institute and from the Mikulski Archive for Space Telescopes (MAST). STScI is operated by the Association of Universities for Research in Astronomy, Inc., under NASA contract NAS5-26555.

I.P. acknowledges funding by the European Research Council through ERC-AdG SPECMAP-CGM, GA 101020943.
I.P. and N.T. acknowledge support from CONICYT PAI/82140055. K.M. acknowledges support from the National Agency for Research and Development (ANID)/Scholarship Program/Doctorado Nacional/2022-21220649. SL acknowledges support by FONDECYT grant number 1231187. N.T. and J.X.P. acknowledge support from NASA programme HST-GO-13832. J.X.P. acknowledges support from NSF grants AST-1911140, AST-1910471 and AST-2206490.  

We thank contributors to {\sc SciPy}, {\sc Matplotlib}, {\sc Astropy} \citep{Astropy, Astropy2018}  and the PYTHON programming language; the free and open-source community; and the NASA Astrophysics Data System for software and services. We also thank contributors to linetools \citep{Prochaska2016} and PyMUSE \citep{PyMUSE}.

\end{acknowledgements}

\bibliographystyle{aa}
\bibliography{aanda} 

\begin{appendix}

\section{Galaxies in our VLT/MUSE galaxy survey}
\label{sec:appendix_muse_sources}
Table~\ref{tab:muse_ids} lists the sources characterized in our VLT/MUSE survey.

\begin{table*}[h]
\centering
\begin{tabular}{lccccrcccll}
\hline
ID&Object&RA&DEC&\multicolumn{2}{c}{Impact Parameter}&$r_{\rm AB}$&$M_{r}$&$z$&class&reliability\\
          & & J2000 &J2000   &    (arcsec)&(kpc)                      &       &       & &     &\\
(1)  &(2)&(3)&  (4)   &  (5)                      & (6)   &  (7)  &(8)&(9)&(10)&(11)\\
\hline
1&J161938.98+254318.7&244.91242&25.72186&22&-&18.6&-&0.0000&Star&a\\
2&J161939.97+254330.2&244.91654&25.72506&9&-&17.7&-&0.0000&Star&a\\
3&J161939.31+254247.9&244.91379&25.71331&41&-&24.3&-&0.0000&Star&a\\
4&J161943.25+254351.5&244.93021&25.73097&45&-&22.0&-&0.0000&Star&a\\
5&J161941.26+254353.0&244.92192&25.73139&29&-&21.0&-&0.0000&Star&a\\
6&J161942.90+254324.2&244.92875&25.72339&32&-&23.1&-&0.0000&Star&a\\
7&J161942.22+254324.9&244.92592&25.72358&23&-&20.7&-&0.0000&Star&a\\
8&J161940.07+254315.6&244.91696&25.72100&12&-&21.5&-&0.0000&Star&a\\
9&J161937.79+254401.3&244.90746&25.73369&52&-&24.4&-&0.0000&Star&a\\
10&J161943.46+254344.9&244.93108&25.72914&44&-&19.0&-&0.0000&Star&a\\
11&J161937.77+254333.0&244.90737&25.72583&38&74&23.2&-15.5&0.1016&SF&b\\
12&J161939.24+254336.1&244.91350&25.72669&21&47&18.0&-21.2&0.1248&non-SF&a\\
13&J161937.79+254254.9&244.90746&25.71525&48&128&23.1&-16.1&0.1456&SF&b\\
14*&J161938.25+254321.0&244.90937&25.72250&31&103&18.8&-21.6&0.1891&non-SF&a\\
15&J161940.53+254325.6&244.91887&25.72378&0&0&16.3&-24.5&0.2685&QSO&a$^{\bot}$\\
16&J161941.18+254341.7&244.92158&25.72825&18&89&23.5&-18.4&0.3337&SF&a\\
17&J161941.92+254326.7&244.92467&25.72408&19&112&23.9&-19.4&0.4676&non-SF&a\\
18&J161942.91+254338.9&244.92879&25.72747&34&209&23.5&-18.6&0.4697&SF&a\\
19&J161941.11+254313.0&244.92129&25.72028&15&90&-&-&0.4697&SF&a\\
20&J161938.33+254300.3&244.90971&25.71675&39&240&20.5&-22.6&0.4704&SF&a\\
21&J161938.08+254246.0&244.90867&25.71278&52&340&22.7&-19.8&0.5409&SF&a\\
22&J161939.03+254251.3&244.91262&25.71425&40&293&22.4&-21.6&0.6870&SF&a\\
23&J161938.77+254254.1&244.91154&25.71503&40&291&23.1&-22.2&0.6899&non-SF&a\\
24&J161938.68+254256.5&244.91117&25.71569&39&289&24.8&-20.0&0.7261&SF&c$^{\bot}$\\
25&J161942.36+254311.8&244.92650&25.71994&28&214&-&-&0.7669&SF&b\\
26&J161940.85+254333.8&244.92021&25.72606&9&69&25.6&-19.1&0.7826&SF&a\\
27&J161941.28+254345.0&244.92200&25.72917&22&166&23.7&-21.0&0.7827&SF&a\\
28&J161940.53+254345.0&244.91887&25.72917&19&148&24.0&-22.1&0.7917&SF&a\\
29&J161940.50+254344.8&244.91875&25.72911&19&147&23.5&-22.6&0.7918&SF&a\\
30&J161940.62+254348.4&244.91925&25.73011&23&174&22.8&-23.2&0.7935&SF&a\\
31&J161941.14+254316.4&244.92142&25.72122&12&96&24.7&-20.4&0.8334&SF&a\\
32&J161941.59+254335.6&244.92329&25.72656&17&140&23.5&-21.6&0.9782&SF&b\\
33&J161942.47+254347.6&244.92696&25.72989&34&281&24.4&-22.1&1.0353&SF&b\\
34&J161942.79+254341.5&244.92829&25.72819&34&283&24.5&-19.6&1.0356&SF&b\\
35&J161941.77+254401.6&244.92404&25.73378&39&330&23.6&-29.0&1.0742&SF&a\\
36&J161940.47+254306.8&244.91862&25.71856&19&161&24.3&-20.5&1.1629&SF&b\\
37&J161940.60+254255.0&244.91917&25.71528&31&261&25.0&-21.6&1.1649&SF&b\\
38&J161938.16+254308.7&244.90900&25.71908&37&310&23.4&-19.8&1.1653&SF&b$^{\bot}$\\
39&J161941.82+254316.3&244.92425&25.72119&20&170&23.9&-21.8&1.4179&SF&b$^{\bot}$\\
40&J161940.05+254258.5&244.91687&25.71625&28&244&23.7&-21.6&1.4239&SF&b$^{\bot}$\\
\hline
\end{tabular}
\caption{\label{tab:muse_ids} Sources characterized with a redshift in our VLT/MUSE survey. ($^{\bot}$): Redmonster did not converge to a $z$ on these sources; redshifts were assigned by a visual inspection. (*): Nearby galaxy to a reported BLA. Sources were classified according to their spectral type. SF galaxies show strong emission lines and a blue continuum and non-SF galaxies show a strong red continuum and an absence of emission lines. Columns (5) and (6) shows the impact parameter of each source to the QSO sight-line, in arcsecs and kpc, respectively. Sources where $r$ is undefined were not detected by SExtractor and we used MUSELET to identify them and include them in the survey. The uncertainties in the redshift measurements in the column (9) are of the order of $\sim0.0001$.}
\end{table*}

\FloatBarrier

\section{Uncharacterized VLT/MUSE sources}
\label{sec:noidsources}

Table~\ref{tab:muse_noids} lists photometric sources identified in the VLT/MUSE dataset but for which no redshift solution was found.

\begin{table*}[h]
\centering
\begin{tabular}{lcccccr}
\hline
ID&Object&RA&DEC&Impact Parameter&$r_{\rm AB}$&reliability\\
          & & J2000 & J2000   &    (arcsec)&        &   \\
(1)  &(2)&(3)&  (4)   &  (5)                      & (6)   &  (7) \\
\hline
1&J161937.72+254351.2&244.90717&25.73089&45.9&22.2&d\\
2&J161937.88+254317.0&244.90783&25.72139&37.2&24.9&d\\
3&J161942.70+254305.3&244.92792&25.71814&35.5&24.4&d\\
4&J161939.77+254308.4&244.91571&25.71900&20.3&24.2&d\\
5&J161942.37+254321.0&244.92654&25.72250&25.0&23.2&d\\
6&J161938.98+254248.6&244.91242&25.71350&42.8&24.7&d\\
7&J161941.20+254329.0&244.92167&25.72472&9.3&24.0&d\\
8&J161940.11+254322.2&244.91712&25.72283&7.0&24.8&d\\
9&J161941.87+254308.7&244.92446&25.71908&24.7&23.7&d\\
10&J161939.07+254325.4&244.91279&25.72372&20.0&24.8&d\\
11&J161938.84+254309.8&244.91183&25.71939&28.1&-&d\\
\hline
\end{tabular}
\caption{\label{tab:muse_noids}List of the sources that could not be characterized with a redshift in our VLT/MUSE survey. The object where $r$ is undefined is faint and SExtractor did not compute its photometry.}
\end{table*}

\FloatBarrier

\section{Galaxies in our VLT/VIMOS galaxy survey}
Table~\ref{tab:VIMOS_survey_table} lists the sources characterized in our VLT/VIMOS survey. 

\begin{table*}
\centering
\begin{tabular}{cccccccccccc}

ID&Object&RA&DEC&\multicolumn{2}{c}{Impact Parameter}&$g_{\mathrm{SDSS}}$&$r_{\mathrm{SDSS}}$&$i_{\mathrm{SDSS}}$&$M_{i}$&$z$&reliability\\
    & & J2000 &J2000   &    (arcsec)&(kpc)                      &       &       & &     &&\\
(1)  &(2)&(3)&  (4)   &  (5)                      & (6)   &  (7)  &(8)&(9)&(10)&(11)&(12)\\
\hline

1 & J161931.89+254216.6 & 244.88286 & 25.70461 & 136 & - & 22.4 & 21.7 & 21.9 & - & 0.0000 & a \\
2 & J162005.40+253845.5 & 245.02251 & 25.64596 & 438 & - & 21.3 & 20.9 & 20.8 & - & 0.0000 & a \\
3 & J162018.06+254846.8 & 245.07525 & 25.81299 & 600 & - & 19.9 & 19.5 & 19.4 & - & 0.0000 & a \\
4 & J162011.76+254647.7 & 245.04902 & 25.77993 & 468 & - & 23.3 & 21.9 & 21.5 & - & 0.0000 & a \\
5 & J162006.25+254151.4 & 245.02604 & 25.69760 & 360 & - & 18.5 & 17.9 & 17.7 & - & 0.0000 & a \\
6 & J162017.42+253645.8 & 245.07258 & 25.61273 & 639 & - & 22.3 & 20.9 & 20.2 & - & 0.0000 & a \\
7 & J162014.68+253737.7 & 245.06115 & 25.62715 & 578 & - & 20.5 & 19.1 & 18.5 & - & 0.0000 & a \\
8 & J162013.00+253804.6 & 245.05416 & 25.63460 & 544 & - & 18.7 & 18.4 & 18.2 & - & 0.0000 & a \\
9 & J162018.12+254540.8 & 245.07551 & 25.76132 & 525 & - & 23.5 & 22.3 & 21.7 & - & 0.0000 & a \\
10 & J162013.47+253838.8 & 245.05613 & 25.64412 & 530 & - & 19.4 & 18.0 & 17.3 & - & 0.0000 & a \\
11 & J161900.65+253626.6 & 244.75272 & 25.60739 & 683 & - & 21.1 & 20.0 & 19.4 & - & 0.0000 & a \\
12 & J162000.16+254840.5 & 245.00066 & 25.81126 & 412 & - & 18.8 & 18.0 & 17.8 & - & 0.0000 & a \\
13 & J161910.81+253640.6 & 244.79505 & 25.61127 & 571 & - & 20.7 & 20.5 & 20.3 & - & 0.0000 & a \\
14 & J162004.96+254749.2 & 245.02067 & 25.79701 & 422 & - & 21.6 & 21.1 & 21.0 & - & 0.0000 & a \\
15 & J161908.79+253818.3 & 244.78663 & 25.63842 & 528 & - & 20.4 & 20.1 & 19.9 & - & 0.0000 & a \\
16 & J162014.01+254520.7 & 245.05836 & 25.75576 & 466 & - & 22.0 & 20.8 & 20.3 & - & 0.0000 & a \\
17 & J162001.62+254609.4 & 245.00675 & 25.76929 & 328 & - & 18.7 & 17.9 & 17.6 & - & 0.0000 & a \\
18 & J161902.61+254630.5 & 244.76087 & 25.77515 & 545 & - & 21.5 & 20.8 & 20.4 & - & 0.0000 & a \\
19 & J161904.55+254958.8 & 244.76896 & 25.83301 & 625 & - & 20.3 & 19.8 & 19.5 & - & 0.0000 & a \\
20 & J161928.91+254629.8 & 244.87044 & 25.77494 & 242 & - & 18.7 & 18.4 & 18.3 & - & 0.0000 & a \\
21 & J161916.24+255010.7 & 244.81765 & 25.83631 & 521 & - & 21.1 & 20.4 & 20.0 & - & 0.0000 & a \\
22 & J161932.31+254650.7 & 244.88463 & 25.78074 & 233 & - & 19.6 & 19.3 & 19.1 & - & 0.0000 & a \\
23 & J161929.94+254713.2 & 244.87473 & 25.78701 & 269 & - & 22.7 & 21.5 & 20.9 & - & 0.0000 & a \\
24 & J161926.65+254831.6 & 244.86103 & 25.80877 & 359 & - & 18.8 & 17.7 & 17.3 & - & 0.0000 & a \\
25 & J161916.91+254750.4 & 244.82044 & 25.79734 & 415 & - & 20.9 & 20.2 & 20.0 & - & 0.0000 & a \\
26 & J161957.45+254049.2 & 244.98936 & 25.68033 & 277 & - & 21.2 & 20.0 & 19.6 & - & 0.0000 & a \\
27 & J161916.76+254728.9 & 244.81985 & 25.79135 & 403 & - & 19.5 & 18.4 & 18.0 & - & 0.0000 & a \\
28 & J161928.03+254142.5 & 244.86678 & 25.69515 & 198 & - & 20.0 & 18.7 & 18.0 & - & 0.0000 & a \\
29 & J161914.66+254621.6 & 244.81110 & 25.77268 & 392 & - & 22.0 & 21.2 & 20.9 & - & 0.0000 & a \\
30 & J161956.09+253729.3 & 244.98372 & 25.62481 & 414 & - & 20.0 & 18.8 & 18.3 & - & 0.0000 & a \\
31 & J161952.02+253845.4 & 244.96674 & 25.64595 & 320 & - & 21.0 & 20.3 & 20.0 & - & 0.0000 & a \\
32 & J161959.82+254005.1 & 244.99927 & 25.66807 & 329 & - & 22.1 & 21.7 & 21.7 & - & 0.0000 & a \\
33 & J161915.48+254427.0 & 244.81452 & 25.74082 & 344 & - & 22.3 & 21.5 & 21.3 & - & 0.0000 & a \\
34 & J162000.95+253540.1 & 245.00397 & 25.59448 & 541 & - & 21.3 & 20.0 & 19.3 & - & 0.0000 & a \\
35 & J161905.17+253834.4 & 244.77155 & 25.64288 & 560 & - & 20.0 & 18.6 & 17.9 & - & 0.0000 & a \\
36 & J162014.23+253918.4 & 245.05930 & 25.65512 & 518 & - & 21.1 & 20.1 & 19.8 & - & 0.0000 & a \\
37 & J161921.11+254103.7 & 244.83794 & 25.68437 & 299 & - & 24.0 & 22.3 & 21.7 & - & 0.0000 & a \\
38 & J161950.73+254630.6 & 244.96138 & 25.77518 & 230 & - & 21.7 & 20.4 & 19.7 & - & 0.0000 & a \\
39 & J161916.90+253832.2 & 244.82043 & 25.64229 & 434 & - & 23.4 & 22.0 & 21.4 & - & 0.0000 & a \\
40 & J161953.84+255002.3 & 244.97435 & 25.83397 & 435 & - & 20.5 & 19.3 & 18.9 & - & 0.0000 & a \\
41 & J161951.30+254731.6 & 244.96377 & 25.79210 & 286 & - & 20.5 & 19.7 & 19.3 & - & 0.0000 & a \\
42 & J161917.66+253657.6 & 244.82359 & 25.61599 & 496 & - & 21.6 & 20.2 & 19.6 & - & 0.0000 & a \\
43 & J161930.52+253725.1 & 244.87717 & 25.62364 & 385 & - & 18.4 & 17.8 & 17.6 & - & 0.0000 & a \\
44 & J161955.27+254846.8 & 244.98031 & 25.81300 & 378 & - & 20.1 & 19.7 & 19.5 & - & 0.0000 & a \\
45 & J161950.80+255057.4 & 244.96165 & 25.84928 & 472 & - & 19.9 & 19.3 & 19.1 & - & 0.0000 & a \\
46 & J161951.78+254522.7 & 244.96574 & 25.75631 & 192 & - & 25.4 & 23.3 & 22.1 & - & 0.0000 & a \\
47 & J161949.81+254514.6 & 244.95754 & 25.75405 & 166 & - & 22.7 & 21.2 & 20.6 & - & 0.0000 & a \\
48 & J161933.65+253847.3 & 244.89022 & 25.64647 & 294 & - & 23.2 & 21.6 & 20.8 & - & 0.0000 & a \\
49 & J161905.74+253857.2 & 244.77390 & 25.64921 & 542 & - & 23.0 & 22.2 & 21.6 & - & 0.0000 & a \\
50 & J161948.92+254657.9 & 244.95384 & 25.78276 & 240 & - & 18.5 & 18.2 & 18.0 & - & 0.0000 & a \\
\hline
\end{tabular}
\caption{Sources characterized with a redshift in our VLT/VIMOS survey. ($^{\bot}$): {\sc Redmonster} did not converge to a $z$ on these sources. Redshift were calculated by a visual inspection on these cases. Columns. (5) and (6) shows the impact parameter of each source to the QSO sightline, in arcsecs and kpc, respectively. Columns (7), (8), and (9) show the SDSS photometry if available. Column (10) shows the rest-frame $i$-band absolute magnitude (For galaxies at $z<0.5$, the k-correction was calculated using the code available in \protect\url{http://kcor.sai.msu.ru}). The uncertainties in the redshift measurements in column (11) are of the order of $\sim0.0001$. The reliability categories in column (11) are described in Sec.~\ref{sec:MUSE_survey}.}

\label{tab:VIMOS_survey_table}

\end{table*}

\begin{table*}
\centering
\begin{tabular}{cccccccccccc}

ID&Object&RA&DEC&\multicolumn{2}{c}{Impact Parameter}&$g_{\mathrm{SDSS}}$&$r_{\mathrm{SDSS}}$&$i_{\mathrm{SDSS}}$&$M_{i}$&$z$&reliability\\
    & & J2000 &J2000   &    (arcsec)&(kpc)                      &       &       & &     &&\\
(1)  &(2)&(3)&  (4)   &  (5)                      & (6)   &  (7)  &(8)&(9)&(10)&(11)&(12)\\
\hline
51 & J161902.22+254024.7 & 244.75926 & 25.67354 & 549 & - & 23.1 & 21.5 & 21.0 & - & 0.0000 & a \\
52 & J161949.51+254841.1 & 244.95630 & 25.81143 & 338 & - & 18.7 & 18.2 & 18.0 & - & 0.0000 & a \\
53 & J161933.67+253706.9 & 244.89029 & 25.61857 & 390 & - & 18.2 & 17.6 & 17.4 & - & 0.0000 & a \\
54 & J161923.22+254009.5 & 244.84675 & 25.66931 & 306 & 255 & 21.8 & 22.6 & 22.4 & -13.7 & 0.0392 & a \\
55 & J161919.56+255034.6 & 244.83152 & 25.84294 & 514 & 473 & 21.7 & 21.2 & 20.9 & -15.6 & 0.0433 & b \\
56 & J161901.79+254036.3 & 244.75745 & 25.67676 & 551 & 562 & 19.7 & 19.0 & 18.7 & -18.1 & 0.0481 & b \\
57 & J161958.62+254603.8 & 244.99424 & 25.76771 & 291 & 299 & 20.9 & 20.2 & 19.8 & -16.9 & 0.0485 & a \\
58 & J161956.31+254030.3 & 244.98463 & 25.67509 & 276 & 288 & 17.3 & 16.9 & 16.6 & -20.2 & 0.0493 & a \\
59 & J161950.44+254611.0 & 244.96017 & 25.76971 & 212 & 313 & 23.3 & 22.5 & 22.4 & -15.1 & 0.0697 & a \\
60 & J161925.95+254943.9 & 244.85813 & 25.82886 & 426 & 849 & 20.7 & 20.3 & 19.9 & -18.3 & 0.0948 & b \\
61 & J161955.15+253658.1 & 244.97979 & 25.61615 & 435 & 877 & 19.1 & 18.3 & 17.9 & -20.4 & 0.0961 & b \\
62 & J161928.77+254615.9 & 244.86987 & 25.77107 & 233 & 485 & 18.0 & 17.3 & 16.9 & -21.5 & 0.0993 & b \\
63 & J161916.64+253915.5 & 244.81935 & 25.65430 & 409 & 956 & 21.6 & 21.4 & 20.8 & -17.7 & 0.1120 & a \\
64 & J161931.71+254518.4 & 244.88213 & 25.75511 & 164 & 431 & 22.1 & 21.2 & 21.0 & -17.9 & 0.1262 & b \\
65 & J161931.32+254505.1 & 244.88049 & 25.75142 & 160 & 422 & 22.7 & 21.7 & 22.0 & -16.8 & 0.1270 & a \\
66 & J162016.84+254943.1 & 245.07018 & 25.82864 & 619 & 1715 & 20.5 & 19.9 & 19.6 & -19.4 & 0.1334 & a \\
67 & J162015.32+254559.7 & 245.06385 & 25.76659 & 494 & 1371 & 22.9 & 22.1 & 21.7 & -17.3 & 0.1335 & b \\
68 & J162013.24+254755.1 & 245.05515 & 25.79863 & 517 & 1440 & 18.4 & 17.3 & 16.9 & -22.3 & 0.1340 & a \\
69 & J161953.03+253741.2 & 244.97097 & 25.62810 & 384 & 1074 & 21.8 & 21.4 & 21.6 & -17.2 & 0.1348 & a \\
70 & J161949.89+254040.8 & 244.95788 & 25.67801 & 208 & 598 & - & - & - & - & 0.1389 & a \\
71 & J161903.61+254159.1 & 244.76504 & 25.69974 & 507 & 1503 & 21.7 & 21.1 & 20.8 & -18.3 & 0.1431 & a$^{\bot}$ \\
72 & J161922.81+253848.9 & 244.84505 & 25.64692 & 366 & 1126 & 22.8 & 22.0 & 21.6 & -17.7 & 0.1486 & b \\
73 & J162003.82+255114.3 & 245.01592 & 25.85396 & 564 & 1775 & 22.6 & 22.4 & 22.0 & -17.1 & 0.1521 & a \\
74 & J161908.58+253714.7 & 244.78574 & 25.62074 & 570 & 1796 & 22.7 & 22.0 & 21.5 & -17.9 & 0.1525 & a \\
75 & J162005.51+254902.9 & 245.02297 & 25.81748 & 477 & 1504 & 20.7 & 20.1 & 19.8 & -19.5 & 0.1525 & a$^{\bot}$ \\
76 & J161933.49+253653.5 & 244.88953 & 25.61486 & 404 & 1362 & 22.4 & 21.7 & 21.3 & -18.2 & 0.1636 & c$^{\bot}$ \\
77 & J162014.22+254055.8 & 245.05923 & 25.68216 & 479 & 1696 & 23.7 & 23.1 & 22.6 & -17.0 & 0.1721 & a \\
78 & J161913.49+255018.7 & 244.80621 & 25.83854 & 552 & 2014 & 20.4 & 19.4 & 19.0 & -20.8 & 0.1777 & c$^{\bot}$ \\
79 & J162002.52+254523.3 & 245.01051 & 25.75646 & 319 & 1194 & 21.3 & 20.7 & 20.4 & -19.2 & 0.1821 & b \\
80 & J161903.12+254102.7 & 244.76300 & 25.68407 & 526 & 1971 & 19.0 & 17.8 & 17.3 & -22.6 & 0.1827 & b \\
81 & J161951.01+254911.8 & 244.96255 & 25.81994 & 374 & 1403 & 21.9 & 21.3 & 21.2 & -18.3 & 0.1829 & a \\
82 & J162011.11+255034.2 & 245.04629 & 25.84282 & 595 & 2315 & 23.5 & 22.3 & 21.9 & -18.0 & 0.1900 & b \\
83 & J161919.25+253801.2 & 244.83019 & 25.63368 & 434 & 1689 & 21.9 & 21.5 & 21.3 & -18.3 & 0.1901 & a \\
84 & J161951.66+254101.5 & 244.96525 & 25.68375 & 208 & 811 & 21.1 & 20.0 & 19.5 & -20.5 & 0.1901 & a \\
85 & J162003.45+253603.1 & 245.01436 & 25.60086 & 540 & 2115 & 21.5 & 21.1 & 20.8 & -18.8 & 0.1912 & a \\
86 & J161928.14+253646.5 & 244.86727 & 25.61291 & 433 & 1697 & 19.8 & 19.3 & 19.0 & -20.8 & 0.1914 & a \\
87 & J161921.73+253535.1 & 244.84053 & 25.59308 & 535 & 2096 & 22.4 & 22.6 & 23.7 & -17.2 & 0.1914 & a \\
88 & J162000.55+254648.9 & 245.00228 & 25.78026 & 338 & 1328 & 22.2 & 21.1 & 20.9 & -19.1 & 0.1919 & b \\
89 & J161953.11+254459.8 & 244.97129 & 25.74995 & 194 & 769 & 20.1 & 19.4 & 19.0 & -20.9 & 0.1935 & c$^{\bot}$ \\
90 & J161922.22+253544.1 & 244.84258 & 25.59559 & 524 & 2327 & 22.5 & 21.9 & 21.7 & -18.3 & 0.2185 & c$^{\bot}$ \\
91 & J162015.60+254010.5 & 245.06501 & 25.66959 & 512 & 2287 & 23.1 & 21.9 & 21.5 & -18.8 & 0.2195 & a \\
92 & J162009.43+254031.7 & 245.03929 & 25.67547 & 427 & 1910 & 20.8 & 19.4 & 18.9 & -21.5 & 0.2199 & a \\
93 & J162016.40+253957.6 & 245.06835 & 25.66599 & 527 & 2360 & 20.4 & 19.2 & 18.6 & -21.8 & 0.2201 & a \\
94 & J162011.75+253731.7 & 245.04897 & 25.62547 & 551 & 2464 & 22.6 & 22.1 & 21.8 & -18.3 & 0.2201 & a \\
95 & J162010.26+254128.2 & 245.04273 & 25.69116 & 418 & 1880 & 21.7 & 20.9 & 20.7 & -19.4 & 0.2211 & a \\
96 & J161903.56+253544.6 & 244.76483 & 25.59573 & 680 & 3057 & 20.9 & 20.6 & 20.3 & -19.7 & 0.2211 & b \\
97 & J161904.05+254525.2 & 244.76689 & 25.75701 & 507 & 2282 & 19.7 & 18.7 & 18.2 & -22.2 & 0.2214 & b \\
98 & J161931.95+254645.5 & 244.88314 & 25.77930 & 231 & 1089 & 22.0 & 21.2 & 20.9 & -19.4 & 0.2327 & c$^{\bot}$ \\
99 & J162003.37+253919.5 & 245.01403 & 25.65541 & 395 & 1877 & 21.4 & 20.4 & 19.9 & -20.5 & 0.2349 & a \\
100 & J161954.41+253532.8 & 244.97669 & 25.59245 & 509 & 2503 & 20.7 & 19.3 & 18.7 & -22.0 & 0.2436 & a \\
\hline
\end{tabular}
\caption{Continuation of Table~\ref{tab:VIMOS_survey_table}.}
\end{table*}

\begin{table*}
\centering
\begin{tabular}{cccccccccccc}

ID&Object&RA&DEC&\multicolumn{2}{c}{Impact Parameter}&$g_{\mathrm{SDSS}}$&$r_{\mathrm{SDSS}}$&$i_{\mathrm{SDSS}}$&$M_{i}$&$z$&reliability\\
    & & J2000 &J2000   &    (arcsec)&(kpc)                      &       &       & &     &&\\
(1)  &(2)&(3)&  (4)   &  (5)                      & (6)   &  (7)  &(8)&(9)&(10)&(11)&(12)\\                  & (6)   &  (7)  &(8)&(9)&(10)&(11)\\
\hline
101 & J161904.67+253955.5 & 244.76945 & 25.66541 & 529 & 2693 & 23.4 & 22.6 & 22.3 & -18.2 & 0.2527 & b \\
102 & J161953.01+253605.0 & 244.97086 & 25.60138 & 472 & 2427 & - & - & - & - & 0.2553 & b \\
103 & J161926.76+254053.4 & 244.86150 & 25.68150 & 241 & 1243 & 21.7 & 20.7 & 20.5 & -20.0 & 0.2564 & b \\
104 & J161955.76+254722.3 & 244.98234 & 25.78954 & 313 & 1621 & 25.1 & 23.2 & 21.7 & -18.2 & 0.2569 & a \\
105 & J162016.30+253911.0 & 245.06790 & 25.65306 & 546 & 2874 & 23.2 & 22.0 & 21.9 & -18.7 & 0.2616 & a \\
106 & J161905.39+254046.7 & 244.77246 & 25.67965 & 501 & 2641 & 22.0 & 21.7 & 21.5 & -18.9 & 0.2621 & a \\
107 & J161951.41+255044.2 & 244.96420 & 25.84560 & 462 & 2448 & 22.0 & 20.5 & 20.0 & -20.9 & 0.2635 & b \\
108 & J161951.32+255033.6 & 244.96385 & 25.84267 & 452 & 2397 & 21.2 & 20.2 & 19.9 & -20.8 & 0.2639 & b \\
109 & J161921.18+254844.7 & 244.83826 & 25.81241 & 412 & 2198 & 22.7 & 21.8 & 22.1 & -18.4 & 0.2652 & a \\
110 & J161917.54+254657.3 & 244.82307 & 25.78259 & 376 & 2008 & 22.4 & 20.8 & 20.0 & -20.9 & 0.2658 & b \\
111 & J161922.01+253856.1 & 244.84170 & 25.64893 & 368 & 1978 & 22.8 & 22.5 & 22.1 & -18.4 & 0.2677 & a \\
112 & J161909.04+253902.1 & 244.78768 & 25.65058 & 501 & 2705 & 21.8 & 21.0 & 20.7 & -19.9 & 0.2691 & a \\
113 & J161950.19+254447.4 & 244.95914 & 25.74649 & 154 & 833 & 22.8 & 21.9 & 21.7 & -18.9 & 0.2699 & a \\
114 & J161921.13+254650.7 & 244.83804 & 25.78076 & 333 & 1804 & 21.4 & 20.8 & 20.5 & -20.0 & 0.2700 & a \\
115 & J162003.45+254807.4 & 245.01438 & 25.80206 & 418 & 2267 & 22.5 & 22.0 & 21.5 & -19.1 & 0.2700 & a \\
116 & J162009.32+254857.1 & 245.03885 & 25.81585 & 511 & 2768 & 21.5 & 20.9 & 20.8 & -19.7 & 0.2701 & a \\
117 & J162009.41+254923.9 & 245.03919 & 25.82331 & 529 & 2874 & 21.9 & 21.1 & 20.8 & -19.9 & 0.2706 & a \\
118 & J161953.88+254148.9 & 244.97452 & 25.69692 & 205 & 1111 & 23.3 & 22.8 & 22.4 & -18.2 & 0.2708 & b \\
119 & J162006.50+254555.2 & 245.02707 & 25.76533 & 381 & 2075 & 21.1 & 20.2 & 20.0 & -20.7 & 0.2715 & a \\
120 & J161923.08+254435.5 & 244.84616 & 25.74319 & 246 & 1341 & 22.6 & 22.4 & 23.5 & -18.1 & 0.2716 & a \\
121 & J161950.10+254652.1 & 244.95873 & 25.78114 & 243 & 1327 & 23.6 & 22.7 & 22.7 & -17.9 & 0.2718 & a \\
122 & J161918.59+255024.6 & 244.82746 & 25.84017 & 513 & 2802 & 22.4 & 21.6 & 21.3 & -19.3 & 0.2722 & b \\
123 & J161953.22+254640.9 & 244.97175 & 25.77802 & 260 & 1420 & 22.7 & 22.2 & 21.6 & -19.0 & 0.2728 & a \\
124 & J161904.94+254120.3 & 244.77060 & 25.68896 & 497 & 2754 & 22.7 & 21.7 & 21.2 & -19.7 & 0.2764 & a \\
125 & J161903.52+253602.8 & 244.76465 & 25.60077 & 668 & 3738 & 20.7 & 20.1 & 19.9 & -20.7 & 0.2793 & a$^{\bot}$ \\
126 & J162002.64+254429.1 & 245.01098 & 25.74141 & 305 & 1713 & 22.0 & 21.2 & 21.1 & -19.5 & 0.2806 & a \\
127 & J161903.22+254613.7 & 244.76342 & 25.77048 & 532 & 3007 & 22.2 & 20.9 & 20.3 & -20.7 & 0.2828 & c$^{\bot}$ \\
128 & J161930.97+253623.8 & 244.87903 & 25.60662 & 441 & 2536 & 21.4 & 20.7 & 20.6 & -20.0 & 0.2877 & a \\
129 & J161930.84+253633.5 & 244.87850 & 25.60931 & 433 & 2490 & 22.3 & 21.9 & 21.8 & -18.8 & 0.2883 & a \\
130 & J161929.40+254136.0 & 244.87249 & 25.69333 & 186 & 1079 & 22.0 & 20.9 & 20.4 & -20.6 & 0.2900 & a \\
131 & J161958.43+253807.0 & 244.99345 & 25.63527 & 400 & 2412 & 23.4 & 22.0 & 21.8 & -19.3 & 0.3031 & a \\
132 & J162005.88+254438.4 & 245.02450 & 25.74400 & 350 & 2178 & 21.5 & 20.7 & 20.3 & -20.7 & 0.3139 & a \\
133 & J162002.58+254037.3 & 245.01073 & 25.67703 & 342 & 2137 & 21.3 & 20.5 & 20.3 & -20.7 & 0.3151 & a \\
134 & J162008.09+254445.2 & 245.03372 & 25.74589 & 381 & 2378 & 22.1 & 20.7 & 20.1 & -21.2 & 0.3151 & b \\
135 & J161928.87+254911.4 & 244.87030 & 25.81984 & 380 & 2404 & 23.5 & 21.8 & 21.4 & -20.0 & 0.3194 & a \\
136 & J162017.65+254903.8 & 245.07354 & 25.81772 & 604 & 3900 & 23.0 & 22.4 & 22.9 & -18.3 & 0.3264 & a \\
137 & J161959.86+255025.5 & 244.99941 & 25.84041 & 494 & 3297 & 22.0 & 21.3 & 20.9 & -20.3 & 0.3386 & b \\
138 & J161925.76+255018.5 & 244.85732 & 25.83847 & 459 & 3066 & 20.7 & 19.5 & 19.0 & -22.5 & 0.3394 & b \\
139 & J161922.21+253753.1 & 244.84253 & 25.63141 & 415 & 2776 & 23.4 & 22.4 & 22.6 & -18.6 & 0.3397 & a \\
140 & J162018.38+254717.2 & 245.07658 & 25.78812 & 561 & 3862 & 23.5 & 22.8 & 22.6 & -18.6 & 0.3504 & a \\
141 & J161918.78+253705.2 & 244.82827 & 25.61810 & 481 & 3315 & 21.8 & 20.2 & 19.7 & -22.0 & 0.3509 & a \\
142 & J161924.98+253825.0 & 244.85409 & 25.64027 & 367 & 2532 & 24.0 & 21.2 & 20.7 & -20.7 & 0.3513 & a \\
143 & J162016.90+254807.1 & 245.07043 & 25.80197 & 566 & 3915 & 23.1 & 22.1 & 22.3 & -18.9 & 0.3523 & a \\
144 & J161950.70+254814.4 & 244.96125 & 25.80400 & 320 & 2215 & 22.7 & 21.8 & 21.6 & -19.7 & 0.3531 & a \\
145 & J161953.50+254942.2 & 244.97290 & 25.82839 & 415 & 2883 & 25.2 & 24.6 & 24.6 & -16.7 & 0.3539 & a \\
146 & J161930.98+254743.7 & 244.87908 & 25.79548 & 289 & 2005 & 22.5 & 21.3 & 21.5 & -19.8 & 0.3540 & a \\
147 & J161950.35+254431.1 & 244.95979 & 25.74198 & 148 & 1026 & 22.4 & 21.5 & 21.5 & -19.7 & 0.3541 & a \\
148 & J161923.58+254534.9 & 244.84826 & 25.75969 & 263 & 1832 & 22.9 & 22.2 & 22.2 & -19.1 & 0.3548 & b \\
149 & J161931.01+253734.8 & 244.87921 & 25.62633 & 374 & 2610 & 22.3 & 21.9 & 21.8 & -19.5 & 0.3559 & a \\
150 & J162001.46+254516.3 & 245.00609 & 25.75454 & 304 & 2123 & 23.6 & 22.5 & 22.3 & -19.2 & 0.3567 & a \\
\hline
\end{tabular}
\caption{Continuation of Table~\ref{tab:VIMOS_survey_table}.}
\end{table*}

\begin{table*}
\centering
\begin{tabular}{cccccccccccc}

ID&Object&RA&DEC&\multicolumn{2}{c}{Impact Parameter}&$g_{\mathrm{SDSS}}$&$r_{\mathrm{SDSS}}$&$i_{\mathrm{SDSS}}$&$M_{i}$&$z$&reliability\\
    & & J2000 &J2000   &    (arcsec)&(kpc)                      &       &       & &     &&\\
(1)  &(2)&(3)&  (4)   &  (5)                      & (6)   &  (7)  &(8)&(9)&(10)&(11)&(12)\\                  & (6)   &  (7)  &(8)&(9)&(10)&(11)\\
\hline
151 & J162007.45+255053.0 & 245.03103 & 25.84805 & 576 & 4096 & 22.8 & 21.8 & 21.3 & -20.2 & 0.3632 & a \\
152 & J161903.89+253849.3 & 244.76620 & 25.64704 & 567 & 4113 & 22.2 & 21.0 & 20.6 & -21.0 & 0.3711 & a \\
153 & J161921.77+254512.7 & 244.84070 & 25.75354 & 275 & 2083 & 21.2 & 20.1 & 19.8 & -21.9 & 0.3891 & c \\
154 & J161925.51+253923.5 & 244.85629 & 25.65654 & 316 & 2397 & 23.3 & 21.9 & 21.6 & -20.2 & 0.3902 & a \\
155 & J161915.24+254053.8 & 244.81352 & 25.68160 & 374 & 2910 & 23.2 & 22.3 & 21.7 & -20.1 & 0.4013 & b \\
156 & J161957.00+254202.5 & 244.98752 & 25.70070 & 237 & 1850 & 22.5 & 20.8 & 20.1 & -22.0 & 0.4023 & b \\
157 & J161930.65+255033.9 & 244.87769 & 25.84274 & 449 & 3504 & 23.4 & 21.9 & 21.6 & -20.3 & 0.4033 & c \\
158 & J161913.12+253746.7 & 244.80467 & 25.62963 & 502 & 3933 & 21.5 & 19.9 & 19.3 & -22.7 & 0.4043 & b \\
159 & J161908.08+253648.2 & 244.78367 & 25.61338 & 592 & 4636 & 23.7 & 22.1 & 21.5 & -20.5 & 0.4044 & b \\
160 & J161903.61+253533.7 & 244.76505 & 25.59269 & 687 & 5381 & 24.1 & 23.2 & 22.1 & -19.9 & 0.4044 & a \\
161 & J161915.39+253739.8 & 244.81413 & 25.62771 & 485 & 3809 & 21.6 & 20.6 & 20.3 & -21.5 & 0.4057 & a \\
162 & J161923.22+254009.5 & 244.84675 & 25.66931 & 306 & 2438 & 21.8 & 22.6 & 22.4 & -20.4 & 0.4132 & c$^{\bot}$ \\
163 & J162002.48+253706.5 & 245.01032 & 25.61846 & 481 & 3855 & 22.7 & 21.0 & 20.3 & -21.9 & 0.4148 & a \\
164 & J161929.36+253813.5 & 244.87233 & 25.63708 & 347 & 2781 & 23.1 & 22.1 & 22.0 & -19.8 & 0.4152 & c$^{\bot}$ \\
165 & J161919.28+254039.0 & 244.83032 & 25.67750 & 332 & 2666 & 21.9 & 21.0 & 20.8 & -20.9 & 0.4155 & b \\
166 & J161957.46+254054.9 & 244.98940 & 25.68192 & 274 & 2196 & 23.3 & 22.3 & 21.9 & -19.9 & 0.4155 & b \\
167 & J162016.36+254107.1 & 245.06816 & 25.68531 & 503 & 4043 & 24.7 & 23.2 & 22.5 & -19.7 & 0.4161 & c \\
168 & J161929.53+254702.5 & 244.87303 & 25.78404 & 263 & 2113 & 23.5 & 22.8 & 22.1 & -19.8 & 0.4163 & b \\
169 & J161951.73+253946.0 & 244.96555 & 25.66277 & 267 & 2143 & 22.3 & 20.9 & 20.5 & -21.4 & 0.4164 & b \\
170 & J161931.42+254850.7 & 244.88092 & 25.81407 & 348 & 2799 & 23.0 & 21.5 & 20.8 & -21.3 & 0.4174 & c \\
171 & J161906.20+254942.9 & 244.77583 & 25.82859 & 598 & 4820 & 22.2 & 21.3 & 20.9 & -21.0 & 0.4179 & a \\
172 & J161919.26+254522.9 & 244.83025 & 25.75636 & 311 & 2521 & 22.9 & 21.7 & 21.4 & -20.5 & 0.4212 & c$^{\bot}$ \\
173 & J161932.37+255026.5 & 244.88486 & 25.84069 & 435 & 3542 & 21.2 & 19.7 & 19.2 & -22.9 & 0.4226 & b \\
174 & J161925.65+254812.6 & 244.85687 & 25.80351 & 350 & 2854 & 23.2 & 22.0 & 21.8 & -20.1 & 0.4227 & b \\
175 & J161955.84+254827.0 & 244.98267 & 25.80751 & 365 & 2984 & 23.6 & 21.9 & 21.2 & -21.1 & 0.4242 & b \\
176 & J161929.35+254557.5 & 244.87229 & 25.76598 & 214 & 1761 & 23.3 & 22.1 & 21.6 & -20.4 & 0.4269 & b \\
177 & J161931.42+254850.7 & 244.88092 & 25.81407 & 348 & 2858 & 23.0 & 21.5 & 20.8 & -21.4 & 0.4273 & b \\
178 & J161931.44+253911.3 & 244.88098 & 25.65315 & 283 & 2326 & 24.3 & 22.3 & 21.9 & -20.3 & 0.4278 & b \\
179 & J161912.23+254021.2 & 244.80095 & 25.67256 & 425 & 3498 & 22.7 & 21.9 & 22.1 & -19.7 & 0.4279 & b \\
180 & J162000.18+253932.3 & 245.00075 & 25.65898 & 353 & 2910 & 23.1 & 22.3 & 22.2 & -19.6 & 0.4280 & b \\
181 & J161956.77+254114.2 & 244.98654 & 25.68729 & 256 & 2105 & 23.7 & 22.0 & 21.6 & -20.6 & 0.4280 & b \\
182 & J162015.90+253902.6 & 245.06623 & 25.65072 & 546 & 4492 & 23.3 & 22.4 & 21.9 & -20.0 & 0.4281 & b \\
183 & J162008.31+254459.3 & 245.03464 & 25.74981 & 387 & 3187 & 23.0 & 22.1 & 22.0 & -19.9 & 0.4286 & c \\
184 & J161922.60+255116.6 & 244.84417 & 25.85460 & 530 & 4372 & 22.8 & 21.9 & 21.3 & -20.7 & 0.4292 & b \\
185 & J161909.79+254503.7 & 244.79078 & 25.75103 & 427 & 3528 & 23.1 & 22.3 & 21.7 & -20.2 & 0.4296 & c \\
186 & J162013.72+254203.6 & 245.05717 & 25.70101 & 456 & 3773 & 23.2 & 22.1 & 22.2 & -19.7 & 0.4306 & a \\
187 & J161908.67+254911.3 & 244.78611 & 25.81981 & 552 & 4575 & 24.0 & 22.0 & 21.7 & -20.6 & 0.4311 & b \\
188 & J161905.83+254923.5 & 244.77428 & 25.82319 & 590 & 4899 & 23.4 & 22.6 & 22.3 & -19.6 & 0.4322 & c$^{\bot}$ \\
189 & J161927.31+254111.4 & 244.86381 & 25.68650 & 224 & 1867 & 23.6 & 22.8 & 22.3 & -19.6 & 0.4347 & b \\
190 & J161917.95+254906.2 & 244.82479 & 25.81839 & 457 & 3819 & 22.9 & 22.1 & 22.0 & -19.9 & 0.4348 & b \\
191 & J162008.56+255007.7 & 245.03567 & 25.83547 & 552 & 4616 & 21.3 & 20.9 & 20.8 & -21.1 & 0.4356 & c$^{\bot}$ \\
192 & J162020.03+254655.7 & 245.08346 & 25.78215 & 573 & 4919 & 22.7 & 21.8 & 21.5 & -20.5 & 0.4485 & b \\
193 & J161917.83+253720.1 & 244.82428 & 25.62225 & 478 & 4113 & 22.4 & 21.3 & 21.1 & -20.9 & 0.4504 & a \\
194 & J162018.64+254638.9 & 245.07767 & 25.77748 & 550 & 4750 & 21.9 & 21.0 & 20.6 & -21.4 & 0.4521 & b \\
195 & J162012.83+254613.7 & 245.05345 & 25.77046 & 467 & 4048 & 23.3 & 22.5 & 22.0 & -20.1 & 0.4534 & c \\
196 & J161919.85+254215.7 & 244.83272 & 25.70436 & 288 & 2544 & 21.6 & 20.6 & 20.2 & -21.9 & 0.4630 & a \\
197 & J161949.44+254703.5 & 244.95602 & 25.78431 & 249 & 2224 & 21.5 & 20.2 & 19.6 & -22.8 & 0.4701 & b \\
198 & J161912.74+254633.6 & 244.80310 & 25.77599 & 420 & 3800 & 21.8 & 20.9 & 20.6 & -21.6 & 0.4765 & b \\
199 & J162016.04+253543.1 & 245.06685 & 25.59531 & 667 & 6165 & 22.8 & 21.9 & 21.8 & -20.3 & 0.4888 & a \\
200 & J161901.92+253730.5 & 244.75800 & 25.62514 & 632 & 5852 & 23.8 & 22.6 & 22.3 & -20.0 & 0.4898 & c \\
\hline
\end{tabular}
\caption{Continuation of Table~\ref{tab:VIMOS_survey_table}.}
\end{table*}

\begin{table*}
\centering
\begin{tabular}{cccccccccccc}

ID&Object&RA&DEC&\multicolumn{2}{c}{Impact Parameter}&$g_{\mathrm{SDSS}}$&$r_{\mathrm{SDSS}}$&$i_{\mathrm{SDSS}}$&$M_{i}$&$z$&reliability\\
    & & J2000 &J2000   &    (arcsec)&(kpc)                      &       &       & &     &&\\
(1)  &(2)&(3)&  (4)   &  (5)                      & (6)   &  (7)  &(8)&(9)&(10)&(11)&(12)\\                    & (6)   &  (7)  &(8)&(9)&(10)&(11)\\
\hline
201 & J161907.43+254848.1 & 244.78094 & 25.81336 & 552 & 5242 & 22.8 & 22.2 & 22.0 & - & 0.5044 & b \\
202 & J162010.49+253747.3 & 245.04370 & 25.62980 & 528 & 5029 & 23.3 & 22.1 & 21.5 & - & 0.5062 & b \\
203 & J161919.61+254642.4 & 244.83169 & 25.77844 & 345 & 3308 & 22.5 & 22.1 & 22.5 & - & 0.5103 & c$^{\bot}$ \\
204 & J162007.14+254625.8 & 245.02975 & 25.77383 & 402 & 3858 & 22.8 & 22.5 & 21.9 & - & 0.5104 & c \\
205 & J161955.26+254922.2 & 244.98025 & 25.82282 & 408 & 3919 & 22.5 & 21.3 & 21.0 & - & 0.5106 & a \\
206 & J161927.24+253830.7 & 244.86352 & 25.64186 & 346 & 3322 & 22.6 & 21.8 & 21.4 & - & 0.5112 & b \\
207 & J161932.70+253601.6 & 244.88623 & 25.60044 & 457 & 4515 & 22.8 & 21.1 & 20.4 & - & 0.5281 & a \\
208 & J161915.34+254828.8 & 244.81391 & 25.80799 & 456 & 4508 & 21.9 & 21.2 & 21.0 & - & 0.5282 & b \\
209 & J162003.99+253625.1 & 245.01662 & 25.60696 & 527 & 5213 & 22.6 & 21.4 & 21.0 & - & 0.5288 & a \\
210 & J161951.64+253624.3 & 244.96517 & 25.60674 & 447 & 4434 & 23.8 & 22.3 & 22.6 & - & 0.5298 & c \\
211 & J161933.86+253801.8 & 244.89110 & 25.63382 & 336 & 3395 & 23.4 & 22.6 & 21.3 & - & 0.5411 & c \\
212 & J161957.55+253720.2 & 244.98978 & 25.62227 & 432 & 4362 & 23.3 & 22.2 & 21.3 & - & 0.5417 & b \\
213 & J161923.41+253903.2 & 244.84754 & 25.65088 & 350 & 3547 & 21.9 & 21.2 & 20.9 & - & 0.5433 & a \\
214 & J161932.79+253543.9 & 244.88662 & 25.59552 & 474 & 4798 & 22.6 & 21.8 & 21.4 & - & 0.5434 & c \\
215 & J162007.32+254910.1 & 245.03049 & 25.81946 & 499 & 5258 & 19.3 & 17.9 & 17.3 & - & 0.5688 & c$^{\bot}$ \\
216 & J161910.55+253752.5 & 244.79395 & 25.63126 & 525 & 5703 & 22.6 & 21.5 & 21.1 & - & 0.5903 & b \\
217 & J161913.56+255057.4 & 244.80648 & 25.84928 & 580 & 6328 & 23.4 & 22.2 & 21.8 & - & 0.5926 & b \\
218 & J161951.73+254546.0 & 244.96554 & 25.76278 & 206 & 2252 & 23.4 & 22.4 & 21.9 & - & 0.5940 & b \\
219 & J162015.70+253651.3 & 245.06543 & 25.61426 & 618 & 6929 & 21.6 & 20.7 & 19.9 & - & 0.6135 & c$^{\bot}$ \\
220 & J161917.04+254937.7 & 244.82098 & 25.82714 & 489 & 5493 & 23.0 & 22.4 & 21.7 & - & 0.6141 & c$^{\bot}$ \\
221 & J162013.50+253606.8 & 245.05624 & 25.60189 & 625 & 7024 & 23.1 & 22.1 & 21.8 & - & 0.6143 & a \\
222 & J161932.28+254123.5 & 244.88450 & 25.68986 & 166 & 1860 & 22.0 & 21.2 & 20.6 & - & 0.6145 & b \\
223 & J161956.06+253757.6 & 244.98358 & 25.63268 & 389 & 4383 & 22.5 & 21.7 & 21.3 & - & 0.6161 & a \\
224 & J162002.63+255106.7 & 245.01097 & 25.85186 & 549 & 6208 & 22.9 & 22.1 & 21.7 & - & 0.6192 & b \\
225 & J161953.06+253749.6 & 244.97110 & 25.63045 & 376 & 4257 & 23.0 & 21.6 & 20.9 & - & 0.6197 & b \\
226 & J161958.49+253924.8 & 244.99369 & 25.65689 & 342 & 3872 & 22.6 & 21.7 & 21.3 & - & 0.6207 & b \\
227 & J161929.10+254002.8 & 244.87123 & 25.66744 & 255 & 2892 & 19.8 & 19.3 & 19.1 & - & 0.6209 & c \\
228 & J162015.41+253812.7 & 245.06421 & 25.63687 & 566 & 6464 & 24.3 & 22.4 & 21.7 & - & 0.6271 & a \\
229 & J162017.34+253717.1 & 245.07224 & 25.62142 & 619 & 7177 & 23.8 & 22.8 & 21.7 & - & 0.6383 & c \\
230 & J161949.75+253820.0 & 244.95731 & 25.63890 & 330 & 3834 & 22.1 & 21.5 & 20.9 & - & 0.6399 & a \\
231 & J162007.26+254115.9 & 245.03023 & 25.68774 & 384 & 4601 & 23.2 & 22.0 & 21.8 & - & 0.6653 & b \\
232 & J162005.77+254949.0 & 245.02404 & 25.83027 & 513 & 6161 & 22.8 & 22.2 & 21.2 & - & 0.6667 & b \\
233 & J161902.86+254808.9 & 244.76192 & 25.80248 & 583 & 7037 & 24.6 & 23.2 & 23.7 & - & 0.6709 & c \\
234 & J161915.77+253949.8 & 244.81571 & 25.66382 & 398 & 4895 & 24.2 & 22.9 & 22.5 & - & 0.6850 & b \\
235 & J162003.08+254108.4 & 245.01285 & 25.68568 & 334 & 4173 & 22.8 & 22.1 & 21.3 & - & 0.6994 & c \\
236 & J161921.03+253605.5 & 244.83764 & 25.60153 & 513 & 6545 & 25.2 & 24.5 & 24.9 & - & 0.7175 & c \\
237 & J161917.33+254047.3 & 244.82221 & 25.67981 & 352 & 4558 & 23.0 & 22.0 & 21.7 & - & 0.7325 & c \\
238 & J161916.42+254554.3 & 244.81842 & 25.76509 & 358 & 4650 & 22.3 & 21.9 & 21.4 & - & 0.7334 & c \\
239 & J161901.51+253618.6 & 244.75628 & 25.60516 & 679 & 8923 & 20.6 & 19.7 & 19.3 & - & 0.7449 & c$^{\bot}$ \\
240 & J161950.44+254611.0 & 244.96017 & 25.76971 & 212 & 2931 & 23.3 & 22.5 & 22.4 & - & 0.7919 & c \\
241 & J161901.99+254134.4 & 244.75829 & 25.69288 & 533 & 9199 & 19.7 & 19.4 & 19.4 & - & 1.0674 & c \\
242 & J162018.12+254540.8 & 245.07551 & 25.76132 & 525 & 13978 & 23.5 & 22.3 & 21.7 & - & 2.1237 & b \\
243 & J162009.01+254712.0 & 245.03756 & 25.78668 & 446 & 12195 & 21.9 & 21.4 & 21.1 & - & 2.2331 & a \\
244 & J162001.92+254545.4 & 245.00799 & 25.76262 & 321 & 9910 & 19.3 & 19.1 & 19.1 & - & 2.8639 & a \\
\hline
\end{tabular}
\caption{Continuation of Table~\ref{tab:VIMOS_survey_table}.}
\end{table*}

\FloatBarrier
\section{Spatial distribution of galaxies at the redshift of BLAs not associated with a robust overdensity}
\label{sec:appendix_plot_no_overdensity}

For completeness, in this section, we show the spatial distribution of sources at the redshift of the BLAs that are not associated with an overdensity of galaxies, discussed in Sec.~\ref{sec:not_overdensity_BLAs}.

\begin{figure*}
        \begin{minipage}[b]{0.48\textwidth}
                \includegraphics[width=\columnwidth]{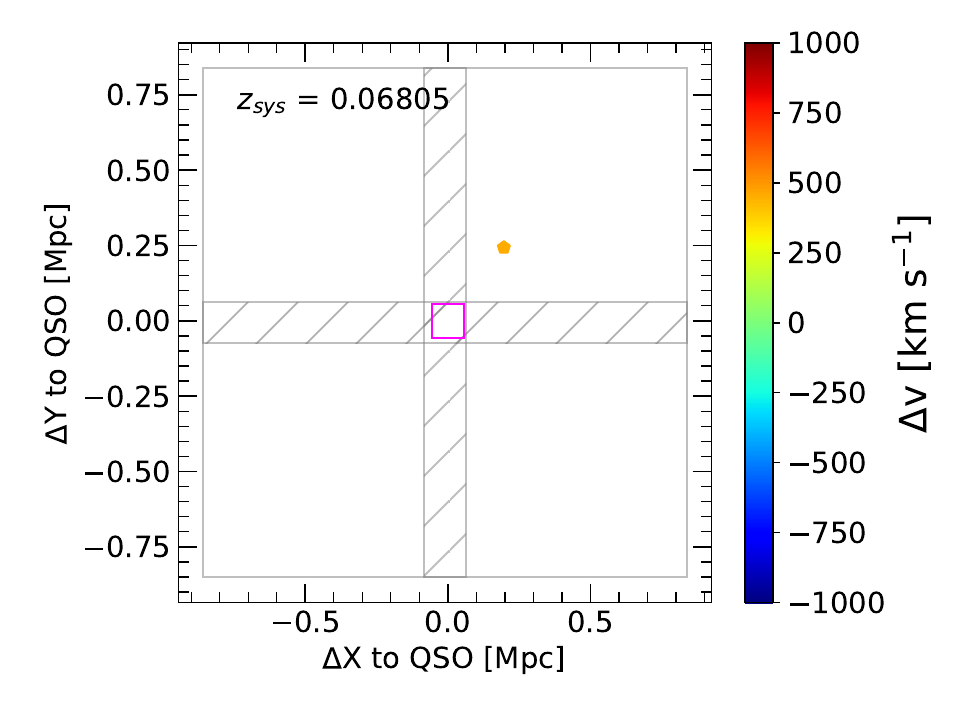}

        \end{minipage}%
        \begin{minipage}[b]{0.48\textwidth}
                \includegraphics[width=\columnwidth]{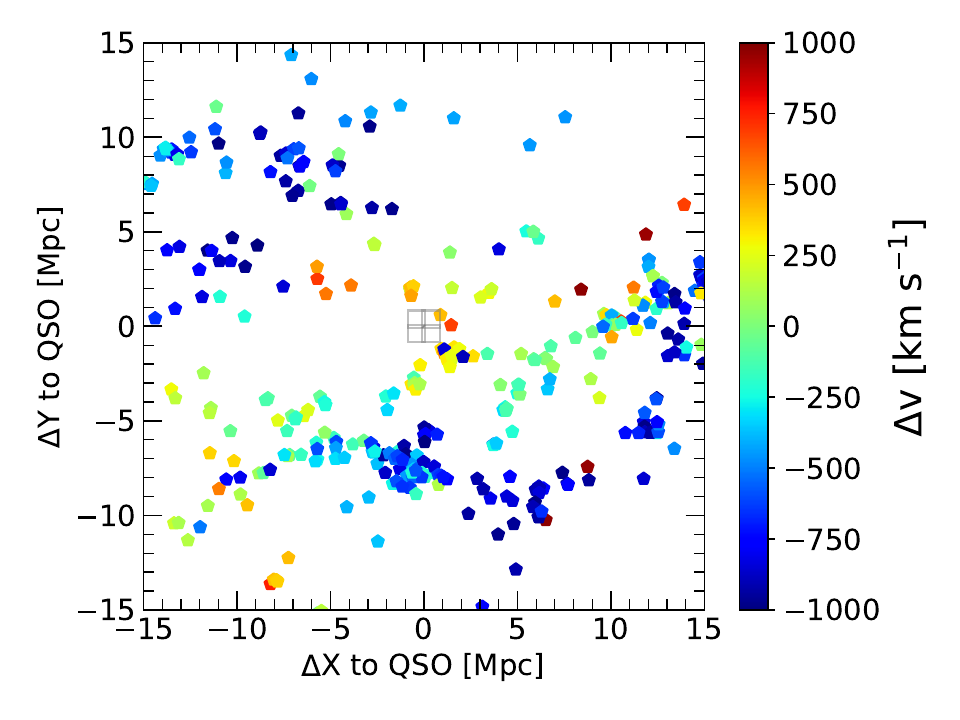}

        \end{minipage}%

        \caption{\label{fig:gal_dist_006} Same as Fig.~\ref{fig:gal_dist_004}, for the BLA at $z = 0.06805$. The VLT/VIMOS FoV spans a square of $\sim 1.6$ Mpc side at this redshift. The VLT/MUSE FoV is smaller, reaching impact parameters of $\sim 60$ kpc from the QSO sightline. One galaxy in our VLT/VIMOS survey lies within $1000$\,\kms\ from the BLA. The right panel shows a zoom-out of the left panel, using spectroscopic data from the SDSS DR16 survey.}
\end{figure*}

\begin{figure*}
        \begin{minipage}[b]{0.48\textwidth}
                \includegraphics[width=\columnwidth]{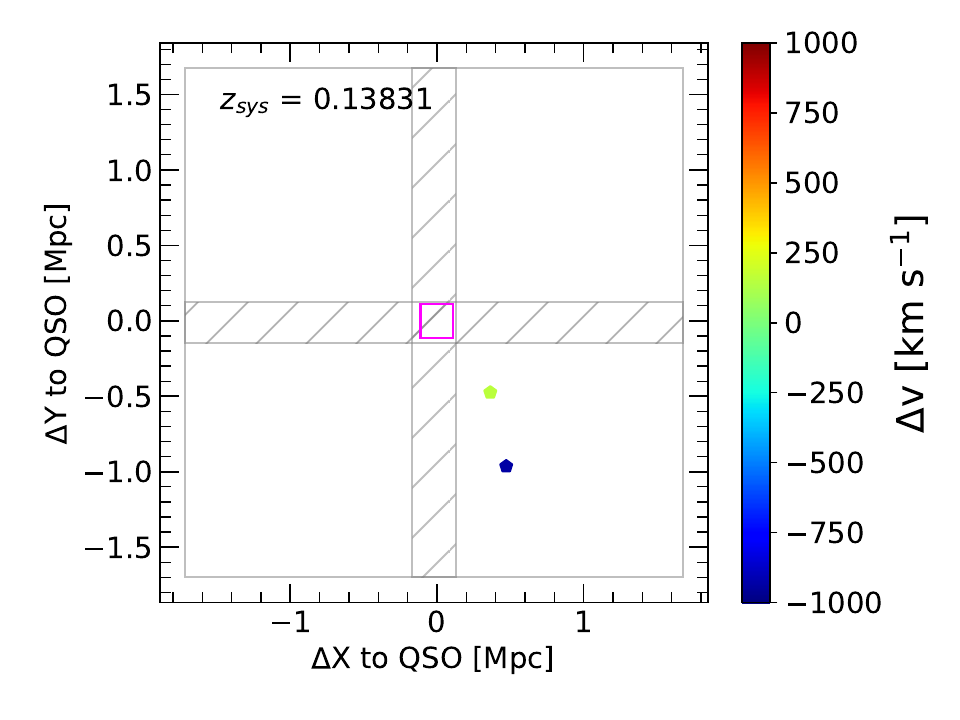}

        \end{minipage}%
        \begin{minipage}[b]{0.48\textwidth}
                \includegraphics[width=\columnwidth]{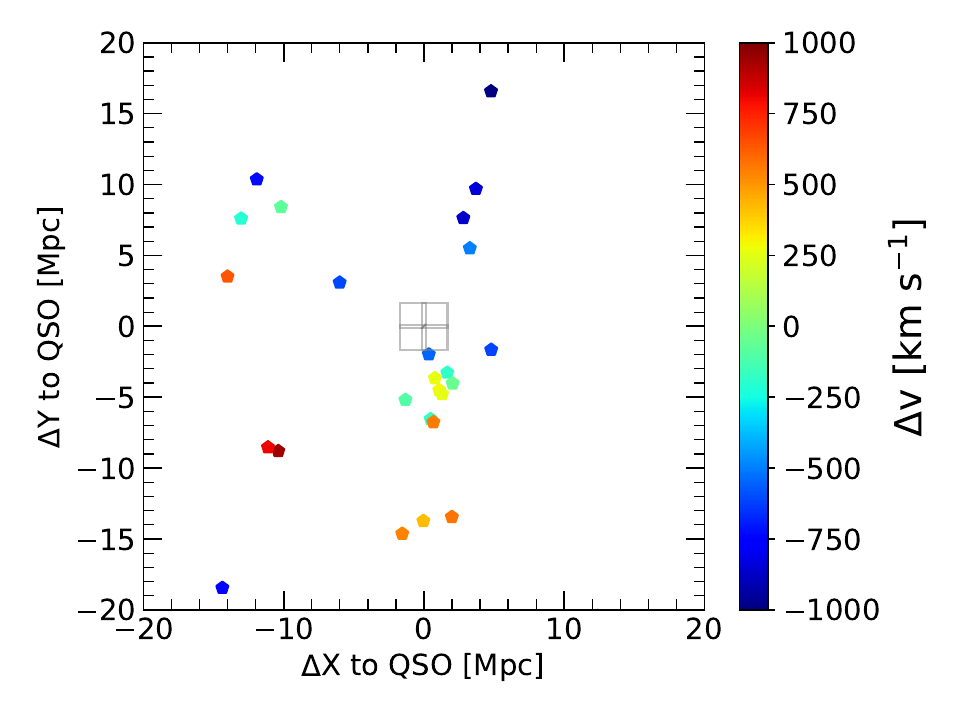}

        \end{minipage}%

        \caption{\label{fig:gal_dist_013} Same as Fig.~\ref{fig:gal_dist_004}, for the BLA at $z = 0.13831$. The VLT/VIMOS FoV spans a square of $\sim 3.4$ Mpc side at this redshift. The VLT/MUSE FoV is smaller, reaching impact parameters of $\sim 120$ kpc from the QSO sightline. Two galaxies in our VLT/VIMOS survey lie within $1000$\,\kms\ from the BLA. The right panel shows a zoom-out of the left panel, using spectroscopic data from the SDSS DR16 survey.}
\end{figure*}

\begin{figure*}
        \begin{minipage}[b]{0.48\textwidth}
                \includegraphics[width=\columnwidth]{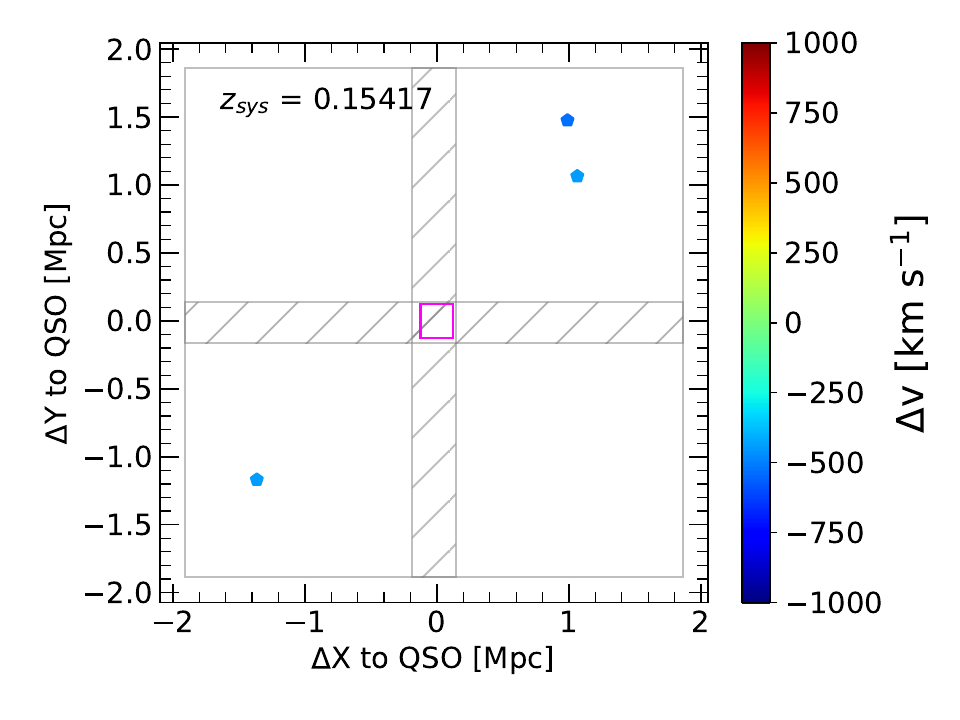}

        \end{minipage}%
        \begin{minipage}[b]{0.48\textwidth}
                \includegraphics[width=\columnwidth]{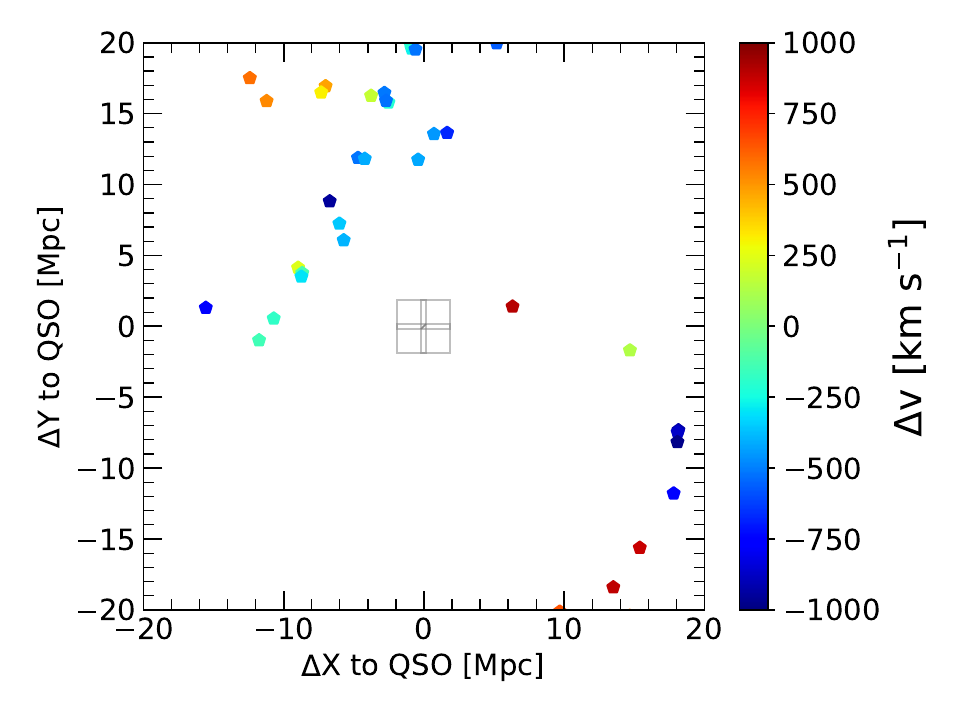}

        \end{minipage}%

        \caption{\label{fig:gal_dist_015} Same as Fig.~\ref{fig:gal_dist_004}, for the BLA at $z = 0.15417$. The VLT/VIMOS FoV spans a square of $\sim 3.8$ Mpc side at this redshift. The VLT/MUSE FoV is smaller, reaching impact parameters of $\sim 130$ kpc from the QSO sightline. Three galaxies in our VLT/VIMOS survey lie within $1000$\,\kms\ from the BLA. The right panel shows a zoom-out of the left panel, using spectroscopic data from the SDSS DR16 survey.}
\end{figure*}

\begin{figure*}
        \begin{minipage}[b]{0.48\textwidth}
                \includegraphics[width=\columnwidth]{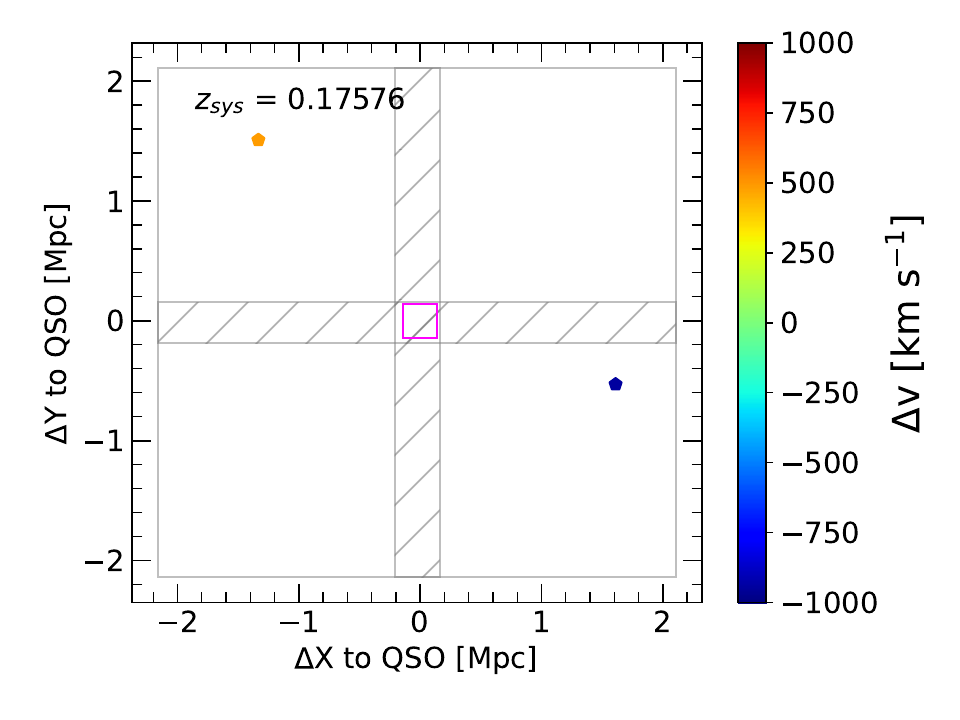}

        \end{minipage}%
        \begin{minipage}[b]{0.48\textwidth}
                \includegraphics[width=\columnwidth]{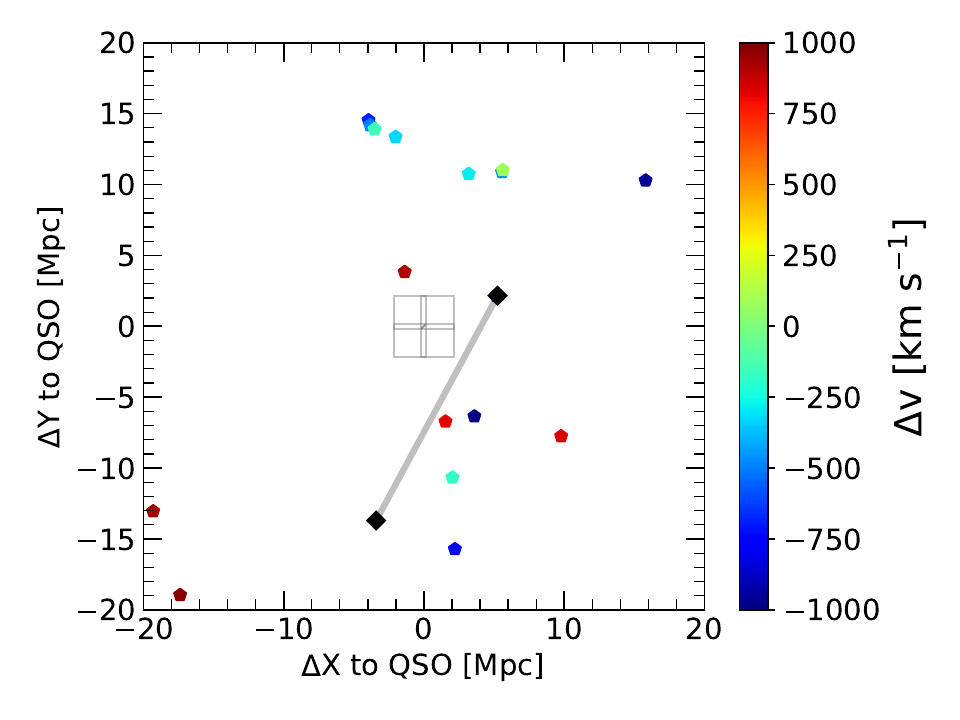}

        \end{minipage}%

        \caption{\label{fig:gal_dist_017} Same as Fig.~\ref{fig:gal_dist_004}, for the BLA at $z \sim 0.17576$. The VLT/VIMOS FoV spans a square of $\sim 4.2$ Mpc side at this redshift. The VLT/MUSE FoV is smaller, reaching impact parameters of $\sim 140$ kpc from the QSO sightline. The grey line in the right panel indicates the inter-cluster axis connecting the cluster pairs described in Sec.~\ref{sec:gal_clust}, indicated with black diamonds. Two galaxies in our VLT/VIMOS survey lie within $1000$\,\kms\ from the BLA. The right panel shows a zoom-out of the left panel, using spectroscopic data from the SDSS DR16 survey.}
\end{figure*}

Figures.~\ref{fig:gal_dist_006},~\ref{fig:gal_dist_013}, \ref{fig:gal_dist_015},~\ref{fig:gal_dist_017},~\ref{fig:gal_dist_021}, and~\ref{fig:gal_dist_025} show the spatial distribution of sources at the redshift of these subset of BLAs.

\begin{figure*}
        \begin{minipage}[b]{0.48\textwidth}
                \includegraphics[width=\columnwidth]{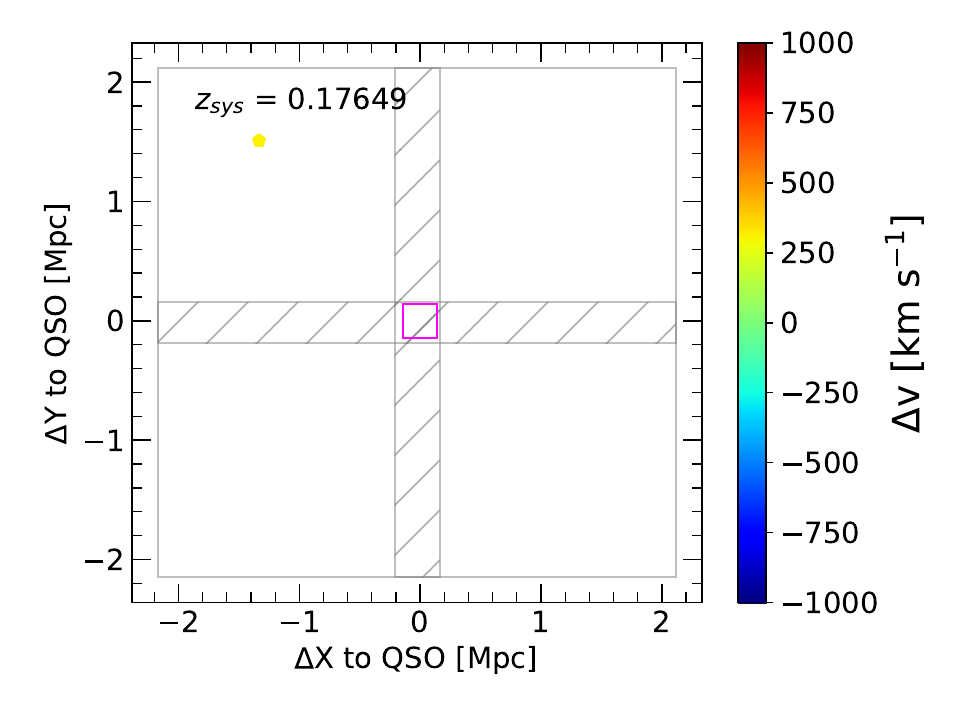}

        \end{minipage}%
        \begin{minipage}[b]{0.48\textwidth}
                \includegraphics[width=\columnwidth]{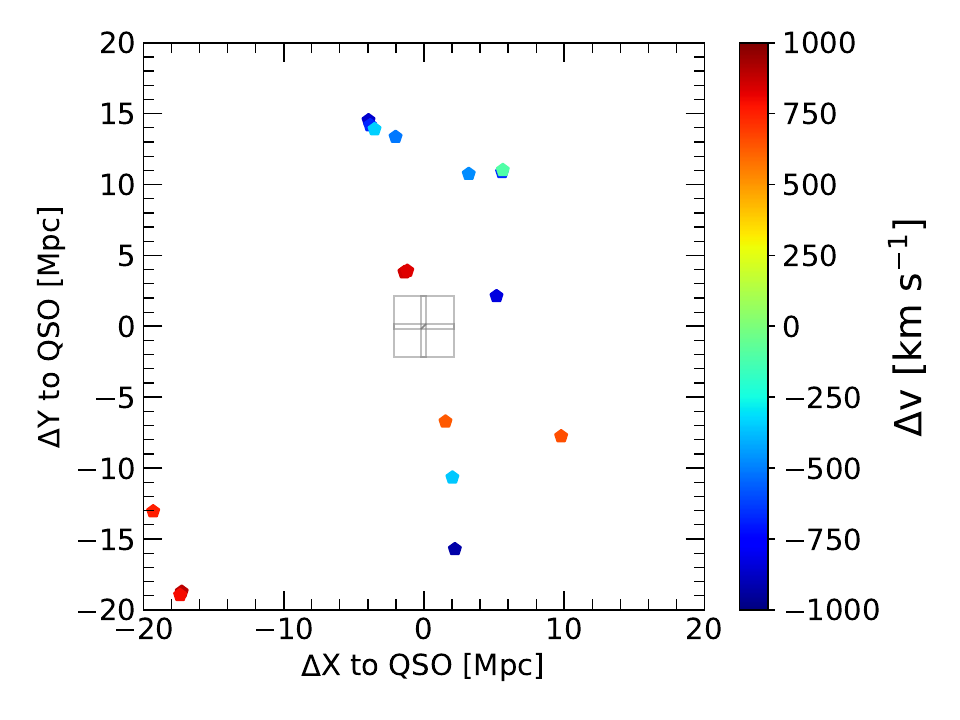}

        \end{minipage}%

        \caption{\label{fig:gal_dist_017649} Same as Fig.~\ref{fig:gal_dist_004}, for the BLA at $z \sim 0.17649$. The BLAs at $z \sim 0.17729$ and $z\sim 0.17813$  present a similar distribution of nearby galaxies. The VLT/VIMOS FoV spans a square of $\sim 4.2$ Mpc side at this redshift. The VLT/MUSE FoV is smaller, reaching impact parameters of $\sim 140$ kpc from the QSO sightline. One galaxy in our VLT/VIMOS survey lie within $1000$\,\kms\ from the BLAs at $z\sim$ $0.17649$, $0.17729$, and $0.17813$. The right panel shows a zoom-out of the left panel, using spectroscopic data from the SDSS DR16 survey.}
\end{figure*}

\begin{figure*}
        \begin{minipage}[b]{0.48\textwidth}
                \includegraphics[width=\columnwidth]{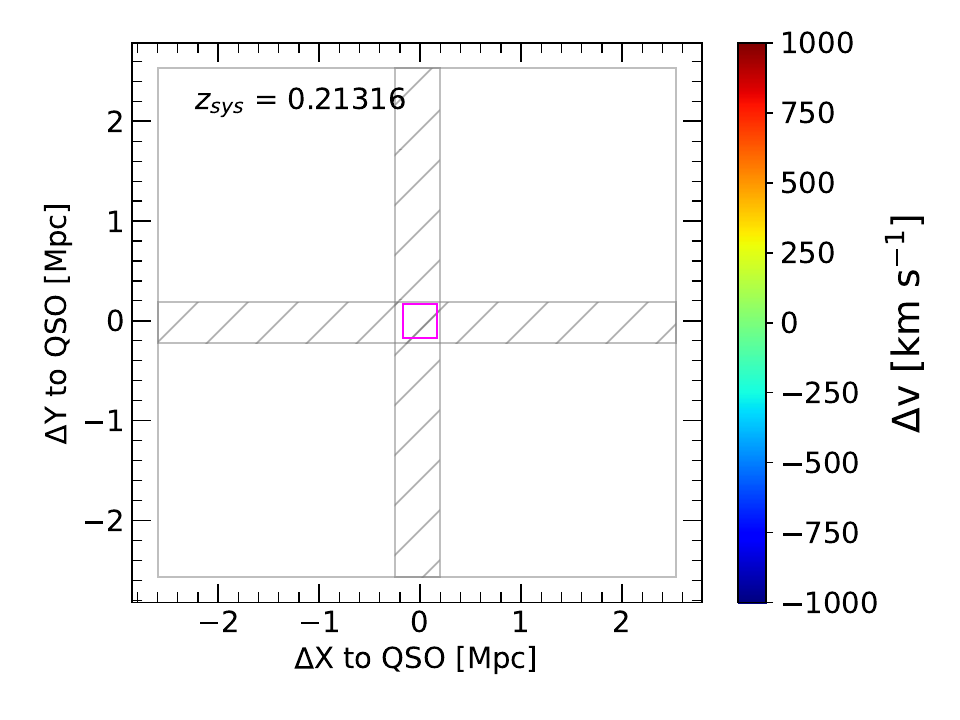}

        \end{minipage}%
        \begin{minipage}[b]{0.48\textwidth}
                \includegraphics[width=\columnwidth]{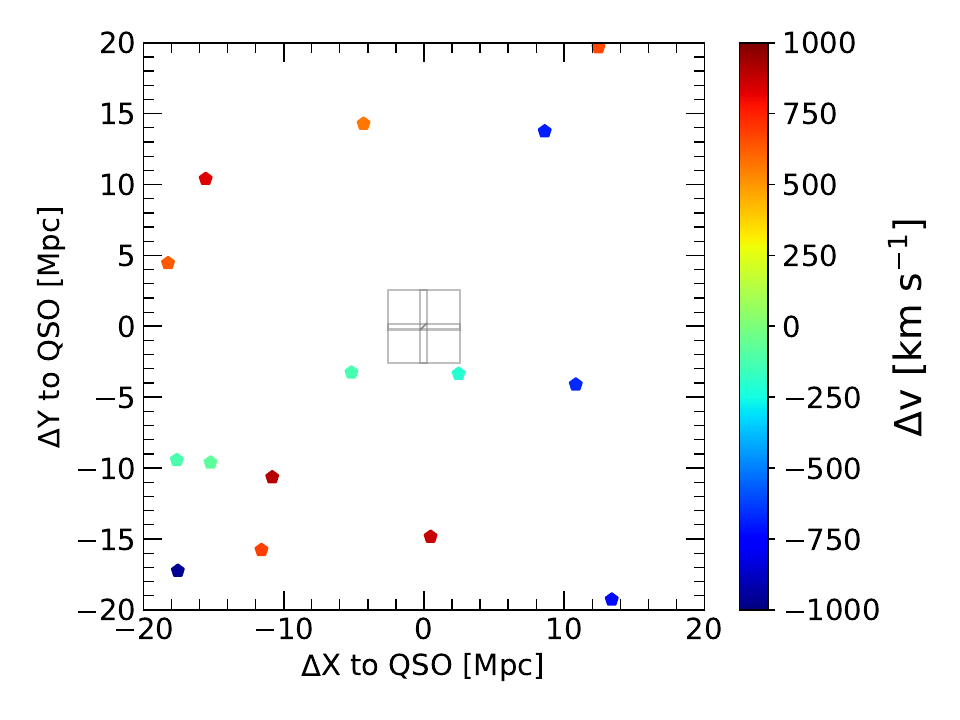}

        \end{minipage}%

        \caption{\label{fig:gal_dist_021} Same as Fig.~\ref{fig:gal_dist_004}, for the BLA at $z = 0.21316$. The VLT/VIMOS FoV spans a square of $\sim 4.8$ Mpc side at this redshift. The VLT/MUSE FoV is smaller, reaching impact parameters of $\sim 170$ kpc from the QSO sightline. We do not find any galaxy in our VLT/VIMOS survey within $1000$\,\kms\ from the BLA. he right panel shows a zoom-out of the left panel, using spectroscopic data from the SDSS DR16 survey.}
\end{figure*}

\begin{figure*}
        \begin{minipage}[b]{0.48\textwidth}
                \includegraphics[width=\columnwidth]{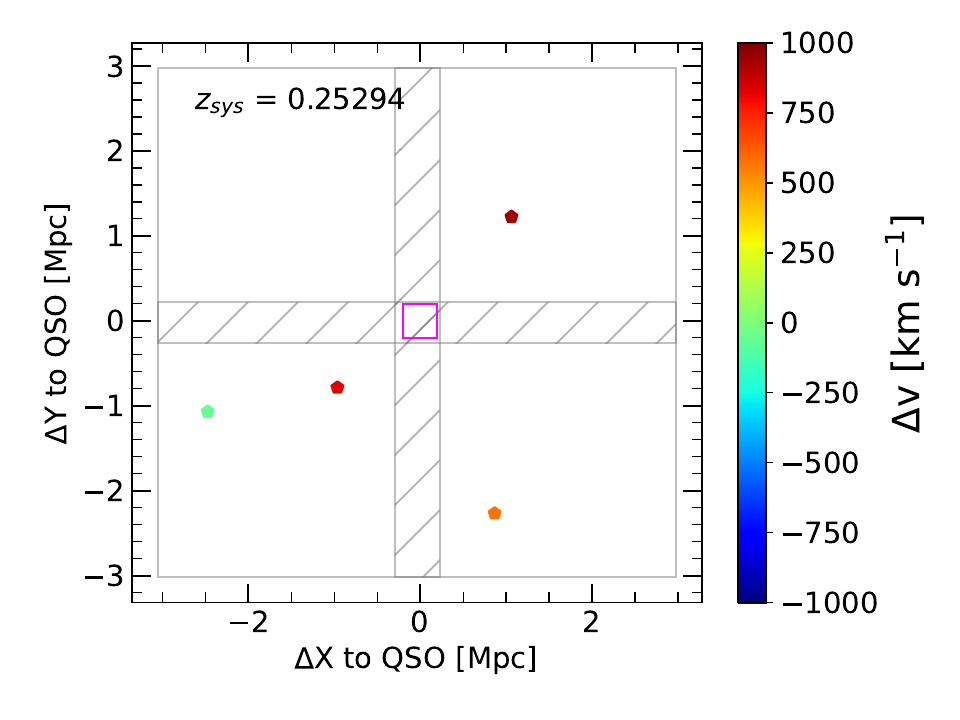}

        \end{minipage}%
        \begin{minipage}[b]{0.48\textwidth}
                \includegraphics[width=\columnwidth]{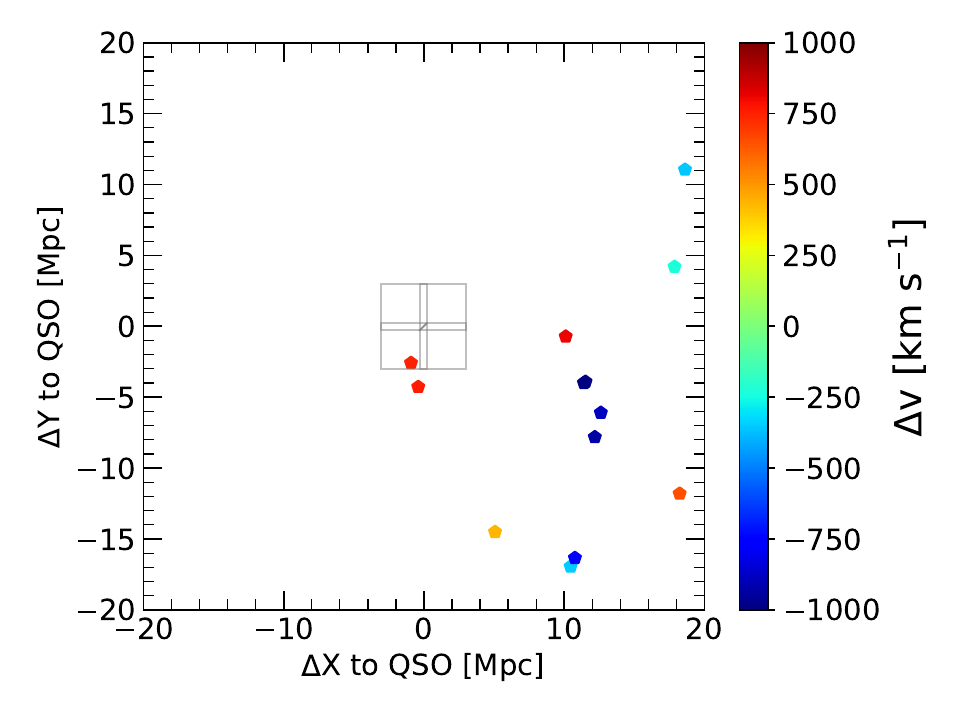}

        \end{minipage}%

        \caption{\label{fig:gal_dist_025} Same as Fig.~\ref{fig:gal_dist_004}, for the BLA at $z = 0.25294$. The VLT/VIMOS FoV spans a square of $\sim 6.0$ Mpc side at this redshift. The VLT/MUSE FoV is smaller, reaching impact parameters of $\sim 190$ kpc from the QSO sightline. Four galaxies in our VLT/VIMOS survey lie within $1000$\,\kms\ from the BLA. The right panel shows a zoom-out of the left panel, using spectroscopic data from the SDSS DR16 survey.}
\end{figure*}
\FloatBarrier
\section{Saturated BLAs and associated metals}
\label{sec:appendix_plots_SaturatedBLAs_metals}
For completeness, the plots of the BLAs are presented, which are shown saturated together with the metals associated with their detection. 

\begin{figure}
\centering
\includegraphics[width = \columnwidth]{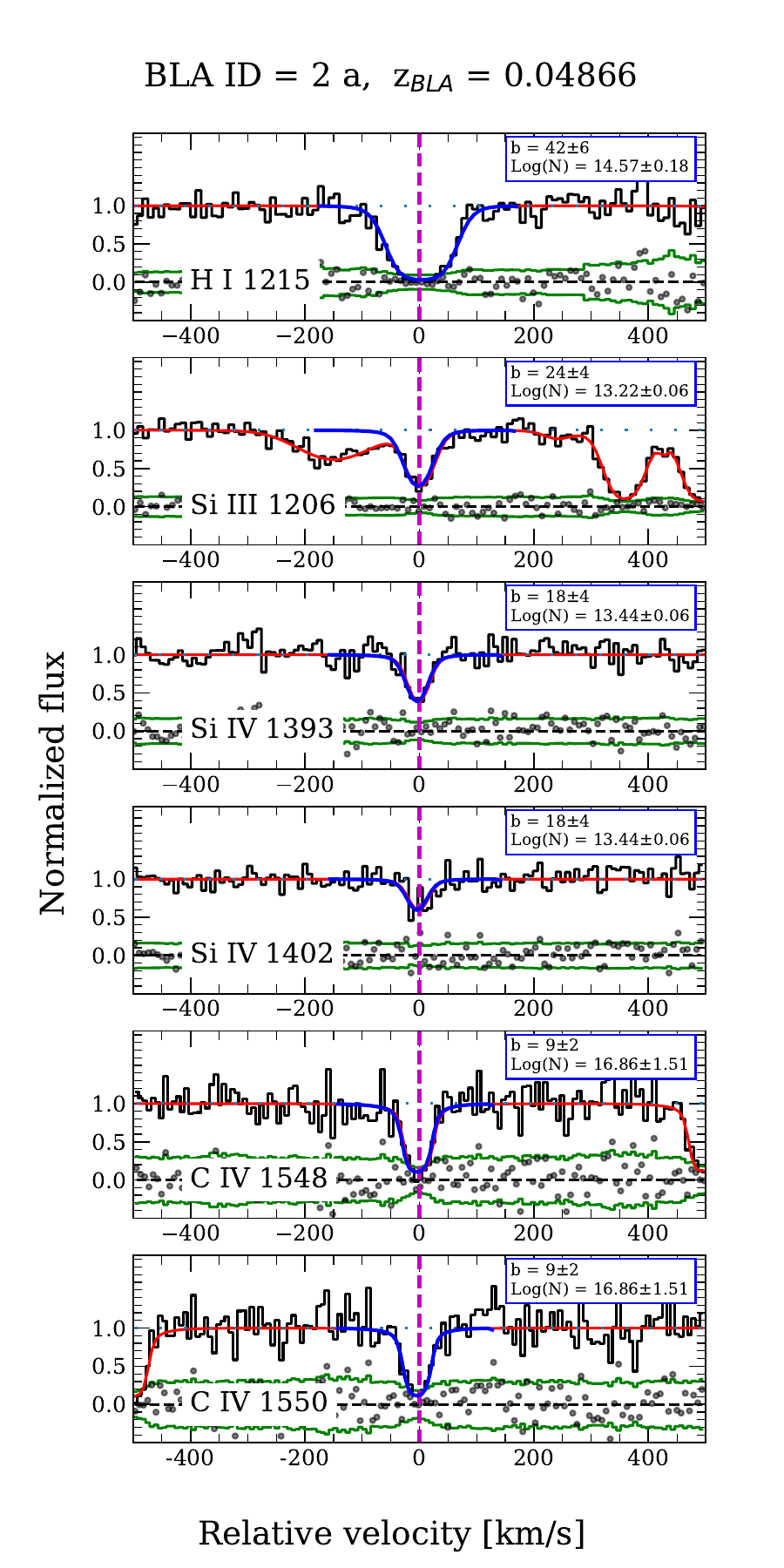}
\caption{\label{fig:blas_metals_2} \hi\ absorption feature for ID 2 with z=0.0486. The metals detected in this \hi\ absorption associated kinematically aligned are shown under the BLA (only transitions with reliability `a' and `b' are shown, see Sec.~\ref{sec:blas_survey} for details of the reliability categories). The spectrum is shown in black, and the continuum is shown in blue dots. The best Voigt best profile fit is shown in blue. The observational parameters of the Dopper parameter in \kms\ and column density log(N/cm$^{-2}$) are shown in the upper right part of the panels.}     
\end{figure}

\begin{figure}
\centering
\includegraphics[width = \columnwidth]{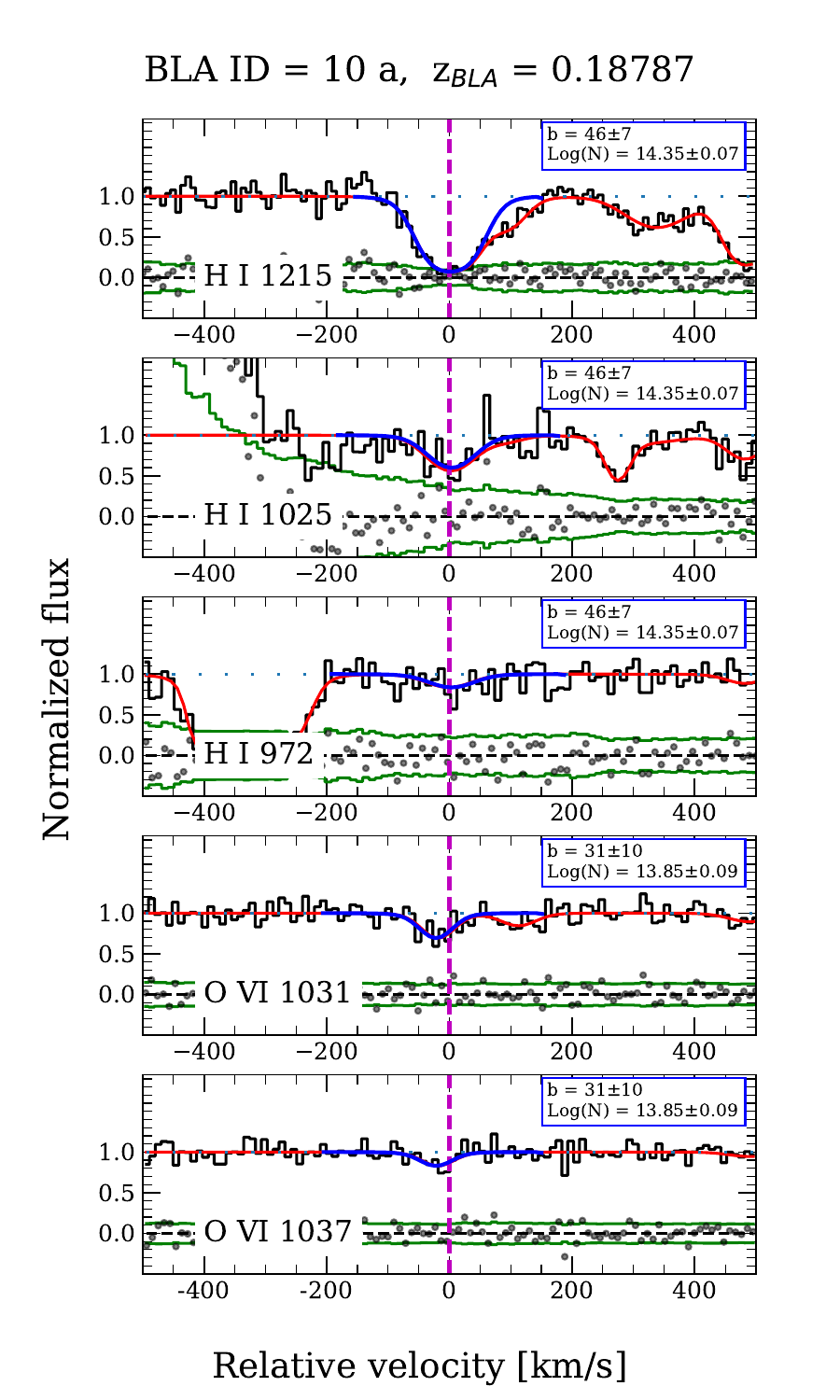}
\caption{\label{fig:blas_metals_10} Same as Fig.~\ref{fig:blas_metals_2}, but for ID 10 with z=0.18787.}     
\end{figure}

\begin{figure}
\centering
\includegraphics[width = \columnwidth]{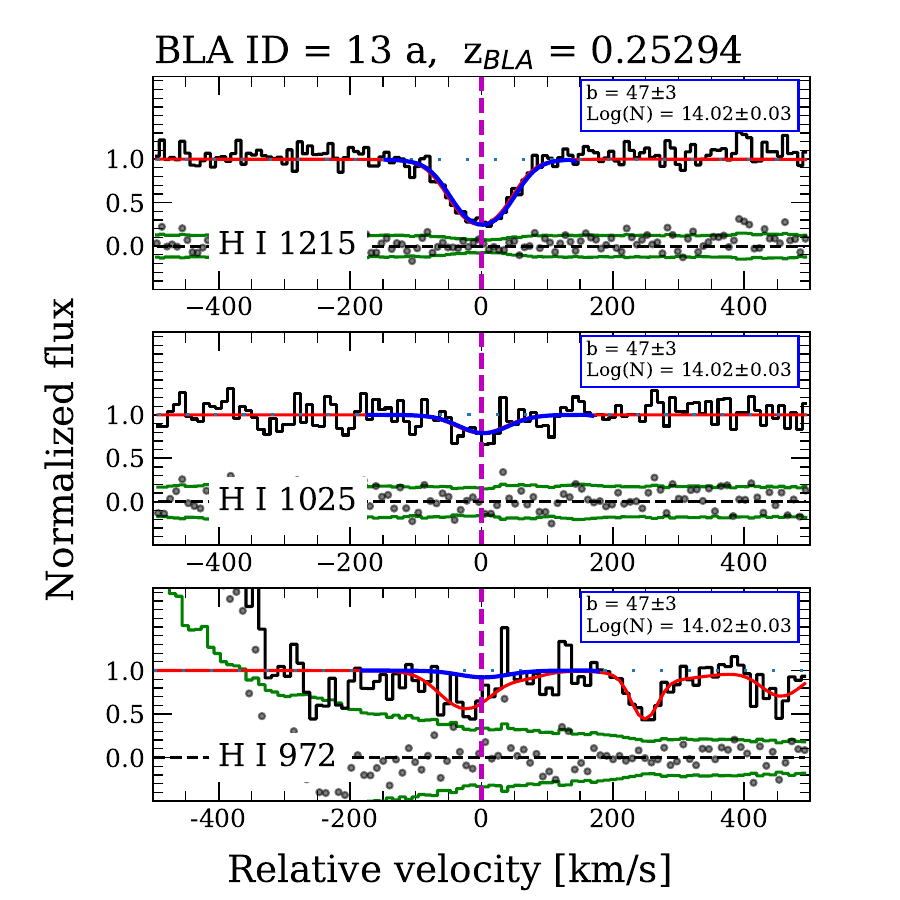}
\caption{\label{fig:blas_metals_13} Same as Fig.~\ref{fig:blas_metals_2}, but for ID 13 with z=0.25294.}     
\end{figure}

The Figures.~\ref{fig:blas_metals_2},~\ref{fig:blas_metals_10}, and ~\ref{fig:blas_metals_13} show the BLAs that present saturation and their associated metals.

\end{appendix}
\end{document}